\documentclass[11pt,a4paper]{article}
\usepackage[T1]{fontenc}
\usepackage[utf8]{inputenc}
\usepackage{lmodern}
\usepackage[margin=2.5cm]{geometry}
\usepackage[numbers,sort&compress]{natbib}
\usepackage{caption}
\usepackage{graphicx}%
\usepackage{multirow}%
\usepackage{amsmath,amssymb,amsfonts}%
\usepackage{amsthm}%
\usepackage{mathrsfs}%
\usepackage[title]{appendix}%
\usepackage{xcolor}%
\usepackage{textcomp}%
\usepackage{booktabs}
\usepackage{tabularx}
\usepackage{array}
\usepackage{enumitem}
\usepackage{microtype}
\usepackage{tikz}
\usetikzlibrary{arrows.meta,fit,backgrounds,positioning}

\theoremstyle{plain}
\newtheorem{theorem}{Theorem}
\newtheorem{proposition}{Proposition}
\newtheorem{corollary}{Corollary}
\theoremstyle{definition}
\newtheorem{definition}{Definition}
\theoremstyle{remark}
\newtheorem{remark}{Remark}

\DeclareMathOperator{\clip}{clip}
\newcommand{\R}{\mathbb{R}}
\newcommand{\one}{\mathbf{1}}

\newcommand{\constraintbox}[2]{\par\medskip\noindent
  \fbox{\parbox{\dimexpr\linewidth-2\fboxsep-2\fboxrule\relax}{\textbf{#1}\quad #2}}\par\medskip}

\newcolumntype{C}{>{\centering\arraybackslash}p{0.9cm}}
\newcolumntype{L}{>{\raggedright\arraybackslash\hspace{0pt}}X}

\usepackage[hidelinks]{hyperref}

\begin{document}

\title{Manipulation-resistant reputation for peer-to-peer scientific publishing: a combination-therapy approach}
\author{Fr\'ed\'eric Prost\\[2pt]
\small Inria, INSA Lyon, CITI, UR3720, 69621 Villeurbanne, France\\
\small\texttt{frederic.prost@insa-lyon.fr}}
\date{November 2026}
\maketitle

\begin{abstract}
In open peer-to-peer networks, where publishing is nearly free and identities are
cheap, reputation systems decide what deserves attention, and their metrics become
targets for manipulation. We study how to choose and combine reputation signals so
that manipulation does not pay while genuine contribution does. Inspired by
combination therapies in medicine, we describe each gaming action by its effects on
all signals. For aggregation by the minimum of the signals, we prove that, in
a global linear model, either some mix of gaming actions raises every signal by more
than it costs, or the net gain of gaming is bounded by the imbalance of the
attacker's honest profile; locally, gaming brings no first-order gain exactly when an
admissible weighting is supported on the attacker's weakest signals.
We then specify, without implementing it, MagMell, a two-tier system for peer-to-peer
scientific publication: a protocol that orders signed events by Byzantine
fault-tolerant consensus, enforces budgets and settles staked assessments, and open
reputation algorithms that any node can run. Simulations on synthetic data
show that minimum aggregation makes networks of fake identities, collusion rings and
floods of generated papers lose rank beyond small intensities, at the price of a
lower correlation between rank and quality for honest accounts. They also show that
criticism is rarely settled, because the evidence needed is seldom available. The replicated log should grow by about 26 kilobytes per identity and
year (analytic estimate), and a full recomputation of reputations takes 1.0 to 1.6
seconds for 100,000 identities (measured).

\end{abstract}

\section{Introduction}
\label{sec:introduction}

Digital networks have removed the marginal cost of publication, and generative
models have made the production of plausible text nearly free. The main difficulty is
therefore no longer to vet information before publication but to allocate attention
after it, since a wealth of information creates a poverty of
attention~\cite{simon1971designing}. Ranking systems address this problem through the
reputation attached to items and to their authors, from PageRank~\cite{Page1999ThePC}
to the likes and reposts of social media. In this paper, reputation denotes the trust
attached to an author, which summarizes their track record, including the reactions
of others to their past publications. Once a reputation metric is in use, however,
participants have an incentive to manipulate it.

Scientific publishing illustrates this tension. Pre-publication peer review is a
filter designed for scarcity. Under abundance, it could be complemented, or replaced,
by post-publication assessment on an open peer-to-peer (P2P) network, provided that
the reputation system that guides attention resists manipulation. In such a network,
identities are cheap, coalitions are easy to form and content can be generated at
scale. Existing defenses typically target one attack, for instance through trust
propagation~\cite{kamvar2003}, Sybil
detection~\cite{douceur2002sybil,yu2006sybilguard} or incentive-compatible
elicitation~\cite{miller2005eliciting}, and can be circumvented by combining attacks;
moreover, no single metric is immune to Goodhart's
law~\cite{manheim2018categorizing}.

This paper addresses the following question: \emph{how should the signals of a
reputation system be chosen and combined so that, in an open P2P network where
identities are cheap, manipulation does not pay while genuine contribution does?} We
decompose it into three research questions.
\begin{description}[leftmargin=2.5em,labelwidth=2em,labelsep=0.5em]
  \item[RQ1] Which constraints must a reputation system satisfy in such a network,
             and how do they interact?
  \item[RQ2] Under which conditions does a set of signals, together with an
             aggregation rule, make every combination of manipulation strategies
             unprofitable?
  \item[RQ3] Which mechanisms and incentive rules implement these constraints in a
             P2P network for scientific publication?
\end{description}

Our approach borrows from combination therapies. A pathogen exposed to several drugs
is caught in an evolutionary vise when resistance to one drug does not confer
resistance to the others, or even increases sensitivity to
them~\cite{pal2015collateral,imamovic2013use}. By analogy, the signals of a reputation
system should be chosen so that inflating one of them exposes the attacker on another.
This paper is a theoretical and design study: it develops a formal framework,
instantiates it in a system called MagMell, and uses synthetic numerical analyses to
test the internal consistency of the design. The empirical effectiveness of the
system remains to be established.

The main contributions of this paper are summarized as follows.
\begin{itemize}
  \item We state six design constraints and relate them to the literature and to an
        explicit threat model (Section~\ref{sec:constraints}). Rate limits and the
        asymmetry between gains and losses are effective only if identities have a
        cost, and impossibility results on Sybil-proofness require locality.
  \item We develop a combination-therapy theory of reputation signals
        (Section~\ref{sec:theory}). In a global linear model, either a cost-adjusted
        ``multi-resistant'' mix of gaming actions exists, or the net gain of gaming
        is bounded by the spread of the attacker's honest profile. Locally, gaming
        brings no first-order gain exactly when an admissible weighting is supported
        on the attacker's weakest signals. Conjunctive aggregation exploits
        admissible weightings without knowing them but cannot create them, and we
        give a robust test of their existence.
  \item We specify MagMell, a two-tier reputation system for a P2P scientific
        publication network (Section~\ref{sec:multirep}). Its protocol is specified as
        a distributed system (node roles, trust and failure model, operations,
        guarantees that hold whatever the reputation algorithm, and estimated costs)
        but not implemented; its reputation layer is specified from observable events
        to sub-indicators, aggregates, staked assessments, settlement and governance
        rules.
  \item We evaluate the design on synthetic data (Section~\ref{sec:numerics}): the
        robustness and false positives of the reputation pipeline, an estimated
        gaming matrix and a comparison of aggregation rules on the same attacks, the
        incentives of assessors with different risk attitudes, a dynamic simulation
        with endogenous outcomes, and the distributed properties of the protocol
        (dissemination under churn, equivocation detection, estimated costs and
        measured recomputation times up to $10^6$ identities, and rankings on partial
        replicas). Consensus is neither implemented nor benchmarked. The evaluation
        led to eight corrections of the design and exposed a remaining weakness, the
        settlement of criticism, which we trace to the availability of evidence in
        the network.
\end{itemize}
The theory is the main contribution. Its point is not that a non-compensatory rule
resists manipulation better than an average, but that local robustness holds exactly
when an admissible weighting is supported on the attacker's weakest signals, a
property of the signals that can be designed and tested. MagMell is one instantiation
of this theory, and the evaluation delimits the conditions under which this
instantiation works and the trade-offs it imposes.

The remainder of this paper is organized as follows. Section~\ref{sec:stateoftheart}
reviews related work. Section~\ref{sec:constraints} presents the threat model and the
design constraints. Section~\ref{sec:theory} develops the theory, and
Section~\ref{sec:multirep} specifies MagMell. Section~\ref{sec:numerics} reports the
evaluation on synthetic data. Section~\ref{sec:discussion} discusses what the
evaluation establishes, the limitations of this work and the protocol of a full
evaluation. Section~\ref{sec:conclusion} concludes the paper, and
Appendix~\ref{app:ga} details the methodology of the planned evolutionary simulation.

\section{Related work}
\label{sec:stateoftheart}

Reputation systems sit at the crossroads of several literatures: distributed
systems and security, economics, social psychology and, for our case study,
science studies. We review them by theme, then compare the works closest to ours.

\subsection{Definitions and surveys}

Resnick et al.\ define a reputation system as a system that collects,
distributes and aggregates feedback about the past behavior of participants,
and identify early the difficulties of eliciting, distributing and aggregating
such feedback~\cite{resnick2000}. The economic foundations are older: in
repeated games with incomplete information, an agent may find it profitable to
build a reputation for being of a type it is not~\cite{kreps1982reputation}.
Dellarocas reviews online feedback mechanisms as a digitization of word of
mouth and discusses their strategic vulnerabilities~\cite{dellarocas2003digitization}.
The survey of J{\o}sang et al.\ distinguishes trust from reputation and
classifies reputation computation engines into summation, Bayesian, discrete,
belief, fuzzy and flow models~\cite{josang2007survey}; Fan et al.\ survey
decentralized trust management with an emphasis on risk analysis and trust
aggregation~\cite{fan2020}. From the security standpoint, Hoffman et al.\
decompose reputation systems into components and classify attacks according to
the component they target~\cite{hoffman2009survey}, and Friedman, Resnick and
Sami give a game-theoretic overview of manipulation-resistant reputation
systems~\cite{friedman2007manipulation}.

\subsection{Aggregation models}

\paragraph{Summation and Bayesian models.} eBay's feedback forum aggregates
positive and negative ratings by summation; its empirical analysis revealed that
feedback is overwhelmingly positive and largely reciprocal~\cite{resnick2002trust}.
The beta reputation system models positive and negative evidence with a beta
distribution and introduces a forgetting factor~\cite{josang2002beta}; subjective
logic adds an explicit uncertainty component~\cite{josang2016subjective}. Sporas
and Histos~\cite{zacharia2000collaborative} illustrate the global/personalized
divide: in Sporas, newcomers start at the minimum reputation, so that changing
identity does not pay, and updates are damped for highly reputed users; Histos
computes a personalized reputation from the graph of ratings seen from the
querying user.

\paragraph{Flow models.} PageRank ranks web pages by the stationary distribution
of a random surfer~\cite{Page1999ThePC}. EigenTrust transposes the idea to P2P
networks: each peer normalizes its local trust values, global trust is the
principal eigenvector of the resulting matrix, and a set of pre-trusted peers
is used to break up collectives of malicious peers~\cite{kamvar2003}. Later
work improves the resilience of this scheme to attacks~\cite{FanXin2012} and to
collusion~\cite{cotronei2024}. PeerTrust weights feedback by the credibility of
its issuer, the number of transactions, and transaction and community context
factors~\cite{xiong2004peertrust}. TrustGuard addresses three vulnerabilities of
decentralized reputation: strategic oscillation, flooding with fake
transactions, and dishonest feedback~\cite{srivatsa2005trustguard}. The
Advogato metric computes trust as a network flow from a seed, which bounds the
number of malicious identities that can be accepted~\cite{levien2009attack}.
Trust measures have also been used to self-organize federations of
utility-computing infrastructures~\cite{DBLP:journals/fgcs/MessinaPRSS16}.

\paragraph{Local versus global metrics.} Massa and Avesani show on Epinions
data that a significant fraction of users are controversial, which calls for
local trust metrics, and frame the design space as a balance between the
tyranny of the majority and echo chambers~\cite{massa2007trust}. Bridging-based
ranking, deployed in X's Community Notes, selects annotations that are rated
helpful across heterogeneous groups of users, using a matrix factorization that
separates viewpoint from shared appreciation~\cite{wojcik2022birdwatch}.

\subsection{Identity and Sybil resistance}

Douceur showed that, without a logically centralized authority vouching for
identities, Sybil attacks cannot be prevented except under unrealistic
assumptions on resources and coordination~\cite{douceur2002sybil}. Friedman and
Resnick analyze the social cost of cheap pseudonyms: newcomers must be
distrusted, which is inefficient, unless pseudonyms are free but unreplaceable,
which they show how to implement with standard cryptography~\cite{friedman2001social}.
Cheng and Friedman prove that no symmetric reputation function is sybilproof,
while some asymmetric ones are~\cite{cheng2005sybilproof}. Graph-based defenses
such as SybilGuard~\cite{yu2006sybilguard} and SybilRank~\cite{cao2012aiding}
assume that the honest region of the social graph is well connected and linked
to the Sybil region by few attack edges, and propagate trust from honest seeds
by short random walks; Schoenebeck et al.\ show that attacks at the periphery of
the network defeat conductance-based methods and propose a detection scheme
based on latent network structure~\cite{schoenebeck2016sybil}. Other approaches
raise the cost of identities directly, through proof of personhood, e.g.\
physical ``pseudonym parties''~\cite{borge2017proof}, or through non-transferable
``soulbound'' tokens recording the commitments, credentials and affiliations of
an identity~\cite{ohlhaver2022decentralized}.

\subsection{Eliciting honest and informative feedback}

Dellarocas studies unfair ratings and discriminatory behavior and proposes
controlled anonymity and filtering techniques against them~\cite{dellarocas2000immunizing}.
When no ground truth is available, peer-prediction mechanisms score a rating
against the ratings of others with proper scoring rules, making honest reporting
an equilibrium~\cite{miller2005eliciting}, and the Bayesian truth serum rewards
answers that are more common than collectively predicted~\cite{prelec2004bayesian}.
Market scoring rules reward a forecaster for the improvement she brings over the
current consensus forecast~\cite{hanson2003combinatorial}. Fear of retaliation
biases public feedback~\cite{resnick2002trust}, and silence itself carries
information that can be exploited to correct reporting
bias~\cite{dellarocas2008sound}. In practice, Slashdot combines distributed
moderation with meta-moderation of the moderators~\cite{lampe2004slash}, and Stack
Overflow makes down-votes on answers cost the voter one reputation point, to
discourage unwarranted down-voting~\cite{stackoverflow-downvote}.

\subsection{Blockchain-based reputation and social media}

Public ledgers~\cite{nakamoto2008bitcoin,gilad2017algorand} make it possible to
record reputation events without a central operator. Arshad et al.\ propose a
decentralized reputation system for blockchain-based ecosystems~\cite{arshad2022reputable}.
Steemit, the largest blockchain-based social media platform, rewards both
content creation and curation with a cryptocurrency~\cite{li2019incentivized};
the analysis of four years of its data shows that a significant number of bot
accounts actively down-vote content~\cite{sun2023exploring}. Ohlhaver, Weyl and
Buterin criticize the focus of Web3 on transferable, financialized assets and
argue that relations of trust should be encoded by non-transferable
tokens~\cite{ohlhaver2022decentralized}.

\subsection{Reputation in science}

Scientific evaluation is a well-documented instance of Goodhart's law~\cite{strathern1997improving,manheim2018categorizing}:
citation cartels can be detected in citation networks~\cite{fister2016toward},
bibliometric indicators have been inflated by fabricated documents indexed by
Google Scholar~\cite{lopez2014google}, and the manipulation of research metrics
is now studied in its own right~\cite{biagioli2020gaming}. Reputation in science
is also subject to cumulative advantage, the Matthew effect~\cite{merton1968matthew}.
Wang et al.\ show that highly novel papers are more likely to become top-cited in
the long run, are more cited in other fields than in their home field, and are
recognized with a delay, so that indicators based on short citation windows are
biased against novelty~\cite{wang2017bias}. Open peer review covers a family of
practices, from open identities to open participation and open platforms,
reviewed by Ross-Hellauer~\cite{ross2017open}. Closest to our case study,
Tenorio-Forn{\'e}s et al.\ propose a decentralized publication system for open
science based on a blockchain and the InterPlanetary File System (IPFS), with a distributed reviewer reputation
system, open access by design, transparent governance, and support for both
anonymous and accountable reviews~\cite{tenorio2019towards}, later extended
towards an interoperable open peer review ecosystem and validated with a
prototype and interviews with editors~\cite{tenorio2021decentralizing}.

\subsection{Cross-disciplinary inspirations}

The strategic classification literature models the interaction between an
evaluator and agents who modify their features to obtain a better
outcome~\cite{hardt2016strategic}. Kleinberg and Raghavan model how agents
allocate effort between actions that improve the attributes the evaluator cares
about and actions that merely game the evaluation, and characterize when an
evaluation rule can induce the desired effort~\cite{kleinberg2020classifiers};
Efthymiou et al.\ take into account causal relations between
features~\cite{efthymiou2025incentivizing}. Biological analogies are established
in computer security, from computer immunology~\cite{forrest1997computer} to the
use of diversity against software monocultures~\cite{forrest1997building}.
Evolutionary medicine has developed a detailed understanding of how combination
therapies prevent, or fail to prevent, the evolution of
resistance~\cite{coffin1995hiv,pennings2013hiv,feder2021understanding,morenogamez2015imperfect},
and of collateral sensitivity, where resistance to one drug increases
sensitivity to another~\cite{pal2015collateral,imamovic2013use}.

\subsection{Closest works and positioning}
\label{subsec:positioning}

Table~\ref{tab:comparison} compares MagMell with the closest systems and
mechanisms along the dimensions that matter for our research question: how
identities are made costly, how many signals are used and how they are
combined, whether reputation is put at risk by assessments, and how time is
handled.

\begin{table}[!htbp]
\centering
\caption{Comparison with the closest works. ``--'': not a design focus of the cited
work, to our reading; ``partly'': application to peer review discussed.}
\label{tab:comparison}
\scriptsize
\renewcommand{\arraystretch}{1.15}
\setlength{\tabcolsep}{3pt}
\begin{tabularx}{\textwidth}{>{\raggedright\arraybackslash\hspace{0pt}}p{2.1cm}LLLLL>{\centering\arraybackslash}p{1.05cm}}
\toprule
System & Identity / Sybil defense & Signals & Aggregation & Reputation at risk & Time & Science \\
\midrule
EigenTrust~\cite{kamvar2003} & pre-trusted peers & one (local trust) & eigenvector, global & -- & -- & -- \\
PeerTrust, TrustGuard~\cite{xiong2004peertrust,srivatsa2005trustguard} & -- & several factors & weighted combination & -- & oscillation-aware (TrustGuard) & -- \\
SybilRank~\cite{cao2012aiding} & few attack edges, seeds & topology only & short random walks & -- & -- & -- \\
Peer prediction, Bayesian truth serum~\cite{miller2005eliciting,prelec2004bayesian} & assumes non-colluding raters & one (reports) & scoring rule & scoring-rule payments & -- & partly \\
Stack Overflow~\cite{stackoverflow-downvote} & accounts, vote limits & votes & sum & down-votes cost the voter & -- & -- \\
Community Notes~\cite{wojcik2022birdwatch} & -- & ratings & matrix factorization (bridging) & rater filtering & -- & -- \\
Steemit~\cite{li2019incentivized,sun2023exploring} & -- & votes weighted by stake & stake-weighted sum & purchasable influence & -- & -- \\
DecSci~\cite{tenorio2019towards,tenorio2021decentralizing} & -- & reviewer reputation & -- & -- & -- & \checkmark \\
\midrule
MagMell (this paper) & plural entry, vouching stakes, probation & ten sub-indicators in five families; merit ranked, integrity screened & conjunctive (min), screen alerts & escrowed, non-transferable stakes; losses burned & decay, confidence, escrow & \checkmark \\
\bottomrule
\end{tabularx}
\end{table}

Table~\ref{tab:comparison-dist} compares the same works from the standpoint of
distributed systems, with DHT for distributed hash table and BFT for Byzantine
fault-tolerant. Many of these dimensions are not reported in the cited works; we mark
them as such rather than guess.

\begin{table}[!htbp]
\centering
\caption{Comparison from the standpoint of distributed systems. ``n.r.'': not
reported in the cited work, to our reading.}
\label{tab:comparison-dist}
\scriptsize
\renewcommand{\arraystretch}{1.15}
\setlength{\tabcolsep}{2pt}
\begin{tabularx}{\textwidth}{>{\raggedright\arraybackslash\hspace{0pt}}p{1.95cm}LLLLL}
\toprule
System & Architecture & Trust and node assumptions & Sybil tolerance & Computation and propagation & Identities and events \\
\midrule
EigenTrust~\cite{kamvar2003} & decentralized; score managers on a DHT & pre-trusted peers are honest & bounded by pre-trusted peers; not Sybil-proof~\cite{cheng2005sybilproof} & distributed power iteration over local trust values & DHT identifiers; local transaction ratings \\
PeerTrust~\cite{xiong2004peertrust} & decentralized; feedback stored in a DHT & credibility of feedback estimated & n.r. & trust computed on demand from retrieved feedback & peer identifiers; transaction feedback \\
TrustGuard~\cite{srivatsa2005trustguard} & decentralized, on top of a reputation system & dishonest feedback and oscillating peers & fake transactions require proofs & periodic, with summarized history & transaction proofs \\
SybilGuard~\cite{yu2006sybilguard} & decentralized & few attack edges between honest and Sybil regions & bounded per attack edge & random routes on the social graph & social-graph nodes \\
SybilRank~\cite{cao2012aiding} & centralized (operator) & few attack edges; trusted seeds & bounded per attack edge & early-terminated power iteration & accounts of a social network \\
Steemit~\cite{li2019incentivized,sun2023exploring} & blockchain & stake-weighted voting & stake required for influence & on-chain rewards & blockchain accounts; votes on chain \\
Community Notes~\cite{wojcik2022birdwatch} & centralized (platform) & platform controls accounts & platform account controls & matrix factorization & platform accounts; ratings \\
DecSci~\cite{tenorio2019towards} & blockchain and IPFS & n.r. & n.r. & n.r. & blockchain accounts; reviews \\
\midrule
MagMell (this paper) & ordered log (BFT committee in the analyzed instantiation), full nodes, content DHT (Sect.~\ref{subsec:sysmodel}) & at most $f$ of $3f+1$ validators Byzantine; any participant Byzantine & costly identities; screens (Sect.~\ref{subsec:identities}) & local re\-computation from the log; gossip; costs in Sect.~\ref{subsec:costs} & keys; signed, totally ordered events \\
\bottomrule
\end{tabularx}
\end{table}

\paragraph{Decentralized science.} DecSci~\cite{tenorio2019towards,tenorio2021decentralizing}
is the work closest to our application: it relies on a ledger and IPFS and
supports reviews that are both anonymous and accountable, a requirement we share.
Its reputation system concerns reviewers within a peer-review workflow.
MagMell targets the post-publication ranking of authors, publications and
assessments, and its contribution lies in the design of the signals
(Section~\ref{sec:theory}) and in the incentive rules for assessments
(Section~\ref{subsec:incentives}).

\paragraph{Blockchain social media.} Steemit already combines a ledger, a
reputation score, a cryptocurrency and down-votes~\cite{li2019incentivized}, but
the influence of a vote grows with the stake of the voter, so that influence can
be bought, and its down-voting is largely performed by bots~\cite{sun2023exploring}.
MagMell separates the currency from influence: reputation is non-transferable,
it can only be put at risk, and what is lost is burned rather than transferred.

\paragraph{Platform reputation.} Stack Overflow shows that making negative
assessments costly is practical~\cite{stackoverflow-downvote}, but votes are
settled immediately and never corroborated. Community Notes shows that bridging
across communities can be computed at scale~\cite{wojcik2022birdwatch}; MagMell
uses bridging as one family of signals among five rather than as the sole
aggregation.

\paragraph{Trust and Sybil-detection algorithms.} EigenTrust, PeerTrust,
TrustGuard and SybilRank provide components that MagMell reuses: seed-based trust
propagation, credibility weighting, oscillation detection. Each, however, relies
on one signal or combines several in a compensatory way. Peer prediction and the
Bayesian truth serum provide scoring rules for assessments but assume
non-colluding raters; MagMell uses them as one component of settlement, together
with long-term corroboration by independent identities.

\paragraph{Gap.} To our knowledge, no previous work treats the \emph{choice and
combination} of reputation signals as the central design problem, with an
explicit criterion against combinations of gaming strategies. We are not aware of
previous work using collateral sensitivity as such a criterion; this is the gap
addressed by Section~\ref{sec:theory}.

\section{Design constraints for digital reputation}
\label{sec:constraints}

This section answers RQ1. We first fix the vocabulary and make the adversary
explicit, then state six design constraints. The first five are necessary but not
sufficient conditions; the sixth is constructive and is developed in
Section~\ref{sec:theory}. Table~\ref{tab:constraints} summarizes them, together
with the threats they address and the mechanisms that implement them in MagMell.

\subsection{Vocabulary: trust, reputation, and what they apply to}
\label{subsec:vocabulary}

Following~\cite{josang2007survey}, we distinguish \emph{trust}, a subjective
and directed relation from one actor towards another, which may rely on private
experience, from \emph{reputation}, the collective and publicly computable
opinion about an actor. Local metrics compute trust while global metrics compute
reputation~\cite{massa2007trust}.

In a scientific network, reputation attaches to at least three kinds of
objects: \emph{identities} (authors), \emph{publications}, and
\emph{assessments} (a review, a vote, a stake). The three are coupled: the
reputation of an author aggregates those of their publications and assessments,
while the weight of an assessment depends on the reputation of its author. Most
of the difficulties discussed below come from this coupling.

Finally, the incentive to manipulate a reputation metric once it is in use is an
instance of Goodhart's law, in Strathern's formulation: when a measure becomes a
target, it ceases to be a good measure~\cite{strathern1997improving};
see~\cite{manheim2018categorizing} for a taxonomy of its variants.

\subsection{Threat model}
\label{subsec:threat}

Requiring a reputation system to be ``fair and not gamed'' states an objective,
not a property; turning it into a design criterion requires an explicit
adversary. We assume, following Kerckhoffs' principle~\cite{kerckhoffs1883},
that the adversary knows the rules of the system, and adapt the taxonomy
of~\cite{hoffman2009survey} to a scientific P2P network:

\begin{description}[leftmargin=1em,style=nextline]
  \item[Sybil attack.] A single actor controls many identities~\cite{douceur2002sybil}.
        In an open P2P network where an identity is a public key, identities
        are essentially free.
  \item[Whitewashing.] An actor with a bad record abandons the identity and
        starts afresh. When pseudonyms are cheap, newcomers must be distrusted
        and made to ``pay their dues'', which is an inherent social
        cost~\cite{friedman2001social}.
  \item[Collusion.] A group mutually inflates its reputation (ballot stuffing);
        in science, citation cartels~\cite{fister2016toward}.
  \item[Bad-mouthing.] Unfair negative assessments aimed at competitors~\cite{dellarocas2000immunizing}.
  \item[Oscillation.] An actor builds a reputation and then exploits it, possibly
        repeatedly~\cite{srivatsa2005trustguard}.
  \item[Reputation trading.] Buying an established identity or its
        endorsements; reputation behaves as a tradeable
        asset as soon as names can change hands~\cite{tadelis1999name}.
  \item[Herding.] Conforming to the expected majority instead of reporting one's
        own assessment, which is cheap and safe but uninformative~\cite{miller2005eliciting,prelec2004bayesian}.
  \item[Automation.] Bots acting at machine speed. With generative models, the
        production of plausible scientific text has itself become almost free,
        which weakens every signal that relies on the mere existence of content.
\end{description}

Two results frame what can be achieved. First, no \emph{symmetric} reputation
function, that is, no function invariant under a renaming of the nodes, such as
PageRank with uniform teleportation, is sybilproof, whereas certain asymmetric functions,
computed relative to trusted nodes, can be~\cite{cheng2005sybilproof}. Second,
graph-based Sybil defenses rely on the assumption that Sybils can only create few
attack edges towards honest nodes~\cite{yu2006sybilguard,cao2012aiding}, an
assumption that fails against attacks at the periphery of the
network~\cite{schoenebeck2016sybil}. No single defense is sufficient.

\begin{table}[!htbp]
\centering
\caption{Design constraints, threats, and their implementation in MagMell.}
\label{tab:constraints}
\small
\renewcommand{\arraystretch}{1.15}
\begin{tabularx}{\textwidth}{p{0.5cm}>{\raggedright\arraybackslash}p{3.4cm}LL}
\toprule
& Constraint & Main threats addressed & Mechanisms in MagMell (Section~\ref{sec:multirep}) \\
\midrule
C1 & Actions are bounded by human resources & Sybil (with C5), automation & identity cost, vouching, reading/writing budgets \\
C2 & Assessing is an action with consequences & bad-mouthing, ballot stuffing, herding, retaliation & staked assessments, settlement, pseudonymous assessments (construction open) \\
C3 & Score and confidence evolve with time & oscillation, whitewashing & forgetting factor, uncertainty, escrow \\
C4 & Losses are easier than gains & bad-mouthing, oscillation & asymmetric settlement, probation \\
C5 & Reputation is both local and global & Sybil, collusion, echo chambers & plural rotating seed sets, bridging, user-chosen aggregation \\
C6 & Signals act as a combination therapy & combined and sequential gaming & five families, minimum aggregation, matched coverage \\
\bottomrule
\end{tabularx}
\end{table}

\subsection{C1: actions are bounded by human resources}
\label{subsec:soccons}

\constraintbox{C1.}{Reputation must be built from human actions, and the actions
that create or move reputation must be bounded by human resources: time,
attention, and identity.}

Reputation is a human perception of events, used to forecast future behavior from
past behavior. It is not a law of nature, since a well-reputed actor may still behave
inappropriately, and it is not a mathematical object; reputation systems nevertheless
have to reduce it to one number or a few numbers, which are useful insofar as they
support reasonable statistical forecasts.

The human origin of reputation can be leveraged against automated attacks. Limits on
the number of actions per unit of time constrain machines far more than humans, as
the daily caps on likes of some social platforms illustrate. In a scientific network,
such budgets can be expressed in \emph{human time} and made much more stringent.
Silent reading of English non-fiction proceeds on average at about 238 words per
minute~\cite{brysbaert2019many}, techniques that claim to increase this rate
substantially do so at the expense of comprehension~\cite{rayner2016so}, and typing
speed is similarly bounded and skill-dependent~\cite{salthouse1984effects}. An
assessment of a 7{,}000-word paper submitted five minutes after its first access
therefore cannot result from a reading. Such budgets are signals rather than proofs,
since humans skim and reviewers do not read every paper they cite.

\emph{Interaction.} Budgets are enforced \emph{per identity}. If identities are free,
an attacker multiplies budgets by multiplying identities, so that C1 is effective
only together with a cost on identities.

Derived signals do not escape this logic. On X, likes and reposts are the basic
quantities from which reputation is inferred, and users have built derived signals on
top of them, such as the ``ratio'' (a reply that receives more likes than the
original post). A derived signal remains harder to game only as long as it is not
used as a target; moreover, since the ratio is computed from likes, an attacker able
to forge likes also forges the ratio, and the two signals share the same target
(Section~\ref{sec:theory}).

\subsection{C2: assessing is an action with consequences}
\label{subsec:pervasive}

\constraintbox{C2.}{Every visible action affects the reputation of its author,
including the action of assessing others; assessments must therefore be
assessable, and their authors protected from retaliation.}

Reputation results from a multi-level game: it is earned through actions
acknowledged by others, and assessing others is itself a way of building one's own
reputation. An assessment, positive or negative, should increase the reputation of
its author when it proves accurate and decrease it when it proves unfounded. Social
media likes are free and do not engage the reputation of their author. Other systems
make assessments costly or assessable: Slashdot's meta-moderation lets users judge the
judgments of moderators~\cite{lampe2004slash}; on Stack Overflow, down-voting an
answer costs the voter one reputation point~\cite{stackoverflow-downvote}; PeerTrust
weights feedback by the credibility of its issuer~\cite{xiong2004peertrust}.

The central difficulty is to determine who assesses an assessment, and against which
ground truth, since the truth value of a scientific claim may remain unknown for
years. Peer-prediction mechanisms score a report against the reports of others, so
that honest reporting is an equilibrium~\cite{miller2005eliciting}, and their authors
explicitly discuss academic reviewing; the Bayesian truth serum rewards answers that
are more common than collectively predicted~\cite{prelec2004bayesian}, which
penalizes herding. These mechanisms are vulnerable to coordinated raters, but they
reward an assessment before the ground truth is known.

Accountability also has a cost. When assessments are public and attributable, fear
of retaliation biases them: on eBay, feedback is overwhelmingly
positive~\cite{resnick2002trust}, and silence itself carries
information~\cite{dellarocas2008sound}. In science, the asymmetry of power between
junior and established researchers is a main reason for anonymous reviewing. A system
in which every assessment is signed and permanent may therefore suppress precisely the
negative assessments that it needs.

C2 therefore requires mechanisms that create reputation, since reputation is not a
zero-sum quantity; mechanisms that destroy it without necessarily transferring what is
lost (C4); consequences for every action; the possibility of staking reputation on an
assessment and gaining from it once the assessment is corroborated, which differs from
the fees that pay for the operation of a cryptocurrency; and protection of assessors
against retaliation.

\subsection{C3: score and confidence evolve with time}
\label{subsec:time}

\constraintbox{C3.}{Reputation must be represented by at least two quantities, a
recency-weighted score and a confidence that grows with evidence, and rewards
must be timed according to the slowest relevant signals.}

Unlike a currency balance, reputation changes with the passage of time in two
opposite ways: recent actions should weigh more than past ones, and a longer track
record should yield a more accurate estimate. The two effects can be reconciled only
if reputation is not summarized by a single number. The beta reputation
system~\cite{josang2002beta} and subjective logic~\cite{josang2016subjective}
make this explicit: given (discounted) amounts $r$ of positive and $s$ of
negative evidence, an opinion is the triple
\[
  b = \frac{r}{r+s+2},\qquad d = \frac{s}{r+s+2},\qquad u = \frac{2}{r+s+2},
\]
of belief, disbelief and uncertainty, with $b+d+u=1$. A forgetting factor
$\lambda\in(0,1]$ applied to $r$ and $s$ at each period implements the first
effect, while $u$ captures the second: a newcomer and a researcher with a long,
mixed record may have the same expected value but very different uncertainties.

Time is also an attack surface. An actor may build a reputation and then exploit
it, possibly in cycles, and detecting such oscillations requires comparing short- and
long-term behaviors~\cite{srivatsa2005trustguard}. Time constants must also match the
dynamics of the network: the half-life of a post ranges from minutes on X to months
on Pinterest or blog platforms~\cite{Grafius24}, whereas the half-life of the
scientific literature, a classical bibliometric notion~\cite{burton1960half}, is
measured in years. Rewards attached to assessments should therefore be held in escrow
until slow signals, such as citations, become available. Finally, recording only the
current value of reputation is insufficient: its time series must be kept, in a
shared append-only ledger when one exists, or locally otherwise.

\subsection{C4: losses are easier than gains}
\label{subsec:anisotropy}

\constraintbox{C4.}{Reputation must be harder and slower to earn than to lose,
unfounded negative assessments must cost more than unfounded positive ones, and a
fresh identity must be worth less than a damaged one.}

Reputation does not evolve symmetrically, which further distinguishes it from
conventional assets and from zero-sum games. In social psychology, the asymmetry
principle states that trust is much easier to destroy than to
create~\cite{slovic1993perceived}, an instance of the general negativity
bias~\cite{baumeister2001bad}. The principle is not universal, since its strength
depends on prior attitudes and trust can be symmetric under some
conditions~\cite{poortinga2004trust}, but it is supported by field data: on eBay, a
seller's first negative feedback causes a sharp drop in sales, and subsequent negative
feedback arrives markedly faster than the first~\cite{cabral2010dynamics}.

A single event may therefore destroy a reputation, and the reputation lost may exceed
the reputation earned. Correctly identifying a poor publication should be rewarded,
because detecting unreliable work benefits the network, but the reward need not
compensate the loss of the assessed author. An honest error should also be
distinguished from persistent error or bad faith, and a series of poor outcomes is
more informative than an isolated one. Finally, unfounded criticism should be deterred
more strongly than unfounded praise.

\emph{Interaction.} The harsher the loss, the greater the incentive to abandon a
damaged identity: C4 is effective only if newcomers start with low reputation or
high uncertainty~\cite{zacharia2000collaborative,friedman2001social}, i.e.\ if
identities have a cost (C1).

\subsection{C5: reputation is both local and global}
\label{subsec:topology}

\constraintbox{C5.}{Assessments must be weighted by the relation between assessor
and assessed, reputation must be computed relative to trusted nodes rather than
symmetrically, and endorsements across communities must weigh more than
endorsements within them.}

An assessment by a newcomer should not have the same influence as one by a highly
reputed actor, and some systems allow only the buyers of a product to rate it:
reputation is recursive and depends on the topology of interactions. Locality is also
required for a formal reason. Since no symmetric reputation function is
sybilproof~\cite{cheng2005sybilproof}, robust systems compute reputation relative to
trusted nodes, as EigenTrust does with pre-trusted peers~\cite{kamvar2003} and
attack-resistant trust metrics do with seeds~\cite{levien2009attack}; moreover,
controversial users cannot be captured by any global score~\cite{massa2007trust}.

In science, researchers form communities that interact heavily around a theme and can
be identified from citation patterns. Interactions within a community are not
perceived like interactions with outsiders, which creates a trade-off: a small group
can game a naive implementation by exchanging citations, whereas ignoring in-group
judgments discards expertise and over-weighting them creates echo
chambers~\cite{massa2007trust}. Bridging-based ranking~\cite{wojcik2022birdwatch}
suggests a way out: an endorsement from outside the citation community of a paper
weighs more than one from inside it. This also favors novelty, since novel papers are
significantly more cited in other fields than in their home
field~\cite{wang2017bias}. One may also consider a zero-sum game within a community
and a positive-sum game with outsiders. How communities should be defined from
interactions alone, and how actors who belong to several communities should be
handled, remain open problems.

\subsection{C6: signals act as a combination therapy}
\label{subsec:c6}

\constraintbox{C6.}{Signals must rely on distinct resources of the attacker, cover
the same actions, be aggregated conjunctively, and be chosen so that gaming one of
them exposes the attacker to another.}

Each of the constraints C1--C5 calls for one or several signals, and each signal,
taken in isolation, can be circumvented. C6 states how signals should be assembled
and is the object of Section~\ref{sec:theory}.

\paragraph{Interactions between constraints.} The constraints are not
independent. C1 and C4 both require identities to have a
cost. C2 and C3 together call for escrow: assessments can only be rewarded when
slow signals have corroborated them. C5 is both a requirement (Sybil resistance)
and a source of side effects on small communities (Section~\ref{subsec:sideeffects}).

\section{A combination-therapy theory of reputation signals}
\label{sec:theory}

This section answers RQ2. We first recall what evolutionary medicine has learned
about combination therapies, including how they fail; we then set up a
correspondence with reputation systems, state a model with explicit assumptions,
and derive results that justify conjunctive aggregation and a set of design
rules.

\subsection{Lessons from evolutionary medicine}
\label{subsec:multitherapy}

\paragraph{Absence of cross-resistance.} When the first drugs against the human immunodeficiency virus were used
alone, resistance evolved in nearly all treated individuals within the first
months of treatment~\cite{feder2021understanding}: the enormous number of viral
replication cycles in an infected individual generates genetic variation at a
high rate~\cite{coffin1995hiv}, so that single resistance mutations are readily
available. Triple-drug therapies, introduced in the mid-1990s, rely on the fact
that mutations conferring resistance to one class of antiretrovirals do not
confer resistance to the other classes. A virus carrying a single resistance
mutation therefore remains suppressed by the other drugs, and several mutations
must be present in the same viral genome for the virus to replicate under
treatment~\cite{feder2021understanding,pennings2013hiv}. To a first
approximation, if these mutations arise independently, the probability that a
virus carries all of them is the product of the individual probabilities. What
matters is the absence of cross-resistance, not the independence of molecular
targets: standard regimens combine two nucleoside reverse transcriptase
inhibitors, which act on the same enzyme, with a third drug from another class.

\paragraph{Sanctuaries and sequential escape.} The multiplicative picture is too
optimistic. Triple therapies drastically reduced, but did not eliminate,
resistance evolution; clinical data show that resistance evolves even after years
of successful therapy, sequentially, often one mutation at a time, and in a
partially predictable order~\cite{feder2021understanding}. Two mechanisms
explain this pattern. Temporally, imperfect adherence or different drug
half-lives create periods of \emph{effective monotherapy}. Spatially, imperfect
drug penetration creates compartments in which only one drug reaches an
effective concentration; such ``spatial monotherapy'' lets resistance mutations
accumulate stepwise and dramatically increases the risk of multidrug resistance,
even when the single-drug compartment is very small~\cite{morenogamez2015imperfect}.
The same authors find that drug combinations with matched penetration profiles
are often preferable, and recall that adding drugs one after the other, or
rotating antibiotics, has been observed to promote multidrug
resistance~\cite{morenogamez2015imperfect}.

\paragraph{Collateral sensitivity.} In some cases the trade-off is stronger:
mutations that confer resistance to one antibiotic increase the sensitivity to
another one~\cite{pal2015collateral,lazar2013bacterial}. Collateral sensitivity
networks can be mapped experimentally and used to design drug cycling protocols
that select against resistance~\cite{imamovic2013use}; alternating treatments
constrain the evolutionary paths to multidrug resistance~\cite{kim2014alternating}.
The trade-off is not always evolutionarily stable: hypersensitivity can sometimes
turn into multidrug resistance, and the order in which drugs are applied
matters~\cite{barbosa2019evolutionary}.

\subsection{A correspondence and its limits}
\label{subsec:dictionary}

Table~\ref{tab:dictionary} sets up the correspondence on which the rest of the
section relies. A pathogen population explores mutations at random; an attacker
explores gaming techniques, and a population of attackers (or one attacker
learning by trial and error) adapts to the rewards the system provides.

\begin{table}[!htbp]
\centering
\caption{Correspondence between evolutionary medicine and reputation systems.}
\label{tab:dictionary}
\small
\renewcommand{\arraystretch}{1.1}
\begin{tabularx}{\textwidth}{>{\raggedright\arraybackslash}p{5cm}L}
\toprule
Evolutionary medicine & Reputation system \\
\midrule
Pathogen population & Population of attacker strategies \\
Drug & Signal \\
Mechanism of action & Resource the attacker must forge (R1--R5, Section~\ref{subsec:rules}) \\
Resistance mutation & Gaming action, i.e.\ a row of the collateral matrix $C$ \\
Cross-resistance & A gaming action that raises several signals (shared resource) \\
Collateral sensitivity & A negative entry $C_{ij}<0$ \\
Fitness cost of resistance & Cost $c_i$ of a gaming action \\
Combination therapy & All signals active, conjunctive aggregation \\
A single mutant remains suppressed & No first-order gain from partial resistance (Proposition~\ref{prop:partial}) \\
Single-drug compartment & Action observed by a single signal \\
Temporal monotherapy & Reward computed before slow signals are available \\
Drug cycling & Variation of weights within the admissible set $W$ \\
Side effects & False positives on honest users \\
\bottomrule
\end{tabularx}
\end{table}

The correspondence has limits that must be kept in mind. An attacker is not
blind: it knows the rules (Kerckhoffs' principle), designs combined strategies
deliberately instead of waiting for random mutations, and coordinates with other
attackers. On the other hand, unlike a bacterial population, it has a budget.
Finally, the goal is not eradication but to make manipulation unprofitable. The
model below is therefore a worst-case, game-theoretic one: it does not assume
random exploration.

\subsection{Model and assumptions}
\label{subsec:model}

We consider one attacker controlling one account, or a coalition of accounts (see
H6). The account is evaluated through $n$ signals $s=(s_1,\dots,s_n)\in\R^n$; how
signals are built from observable events is the object of
Section~\ref{subsec:pipeline}. The attacker can perform $k$ gaming actions with
intensities $g=(g_1,\dots,g_k)\in\R^k_{\geq 0}$. Genuine contribution is treated
separately (Remark~\ref{rem:honest}). We write $\Delta_n=\{w\in\R^n_{\geq0} :
\one^\top w = 1\}$ for the simplex of weight vectors. The first assumption comes in
two versions, which lead to two different kinds of results.

\begin{description}[leftmargin=2.6em,labelwidth=2.2em,labelsep=0.4em]
  \item[H1g] \emph{Global linear model.} For all $g\geq0$, $s(g)=s^0+C^\top g$, where
        $s^0$ is the profile obtained without gaming (the \emph{honest profile})
        and $C\in\R^{k\times n}$ is the \emph{collateral matrix}: $C_{ij}$ is the
        effect of a unit of gaming action $i$ on signal $j$.
  \item[H1$\ell$] \emph{Local model.} The map $g\mapsto s(g)$ is differentiable at
        $g=0$ on $\R^k_{\geq0}$, with $s(g)=s^0+C^\top g+o(\|g\|)$; $C$ is then the
        transposed Jacobian at the honest profile, and it may change from one
        profile to another.
  \item[H2] \emph{Commensurability.} Signals are normalized on a common scale,
        e.g.\ as percentiles within a community, so that comparing $s_j$ and
        $s_{j'}$ is meaningful.
  \item[H3] \emph{Linear costs.} Gaming action $i$ has a constant marginal cost
        $c_i\geq 0$, expressed in the unit of reputation. Identity creation,
        budgets and stakes contribute to $c$.
  \item[H4] \emph{Worst-case attacker.} The designer commits to an aggregation
        rule $A:\R^n\to\R$; the attacker knows $A$, $C$ and $c$, and maximizes its
        net payoff $U(g)=A(s(g))-c^\top g$.
  \item[H5] \emph{Monotone aggregation.} $A$ is non-decreasing in each signal. We
        compare the conjunctive rule $A_{\min}(s)=\min_j s_j$ with linear rules
        $A_v(s)=v^\top s$, $v\in\Delta_n$.
  \item[H6] \emph{Single decision-maker.} A coalition of accounts controlled by one
        attacker is treated as a single decision-maker whose gaming actions and
        costs are aggregated; creating a Sybil identity is one gaming action.
\end{description}

H1g is an idealization: real signals are bounded and saturate, so that an
unbounded gain in the global model means ``a gain that grows with the intensity of
gaming until the model breaks down'', not an unbounded gain in the real system.
H1$\ell$ is realistic but only supports statements about small deviations from
the honest profile, with a matrix $C$ that must be re-estimated around each
profile. We state the results separately for the two versions.

\begin{definition}
The \emph{admissible set} is $W(C,c)=\{w\in\Delta_n : Cw\leq c\}$: the weightings of
the signals under which no gaming action yields more than it costs. For a subset
$J$ of signals, $W_J(C,c)=\{w\in W(C,c) : \mathrm{supp}(w)\subseteq J\}$. A gaming mix
$g\geq 0$ is \emph{cost-adjusted multi-resistant} on $J$ if $(C^\top g)_j>c^\top g$
for every $j\in J$, i.e.\ if it raises every signal of $J$ by more than its total
cost; when $J$ contains all signals we omit it, and when $c=0$ we simply say
\emph{multi-resistant}.
\end{definition}

\subsection{Exact results in the global linear model}
\label{subsec:theorem}

\begin{theorem}[Global linear model]\label{thm:value}
Under H1g and H2--H6, with the conjunctive rule $A=A_{\min}$:
\begin{enumerate}[label=(\roman*)]
  \item If $W(C,c)=\emptyset$, there exists a cost-adjusted multi-resistant gaming
        mix, and the attacker's net payoff in the model grows without bound with
        the intensity of that mix.
  \item If $W(C,c)\neq\emptyset$, the attacker's optimal net payoff is
        \[
          V^* \;=\; \min_{w\in W(C,c)} w^\top s^0 ,
        \]
        and the net gain of gaming $G^*=V^*-\min_j s^0_j$ satisfies
        \[
          0\;\leq\; G^* \;\leq\; \max_j s^0_j-\min_j s^0_j .
        \]
        Moreover, $G^*=0$ if and only if $W_{J_0}(C,c)\neq\emptyset$, where
        $J_0=\arg\min_j s^0_j$.
\end{enumerate}
\end{theorem}

\begin{proof}
The attacker's problem $\max_{g\geq 0}\ \min_j s_j(g)-c^\top g$ is the linear
program
\[
  (P)\qquad \max_{t\in\R,\ g\geq 0}\ t-c^\top g
  \quad\text{s.t.}\quad t\,\one - C^\top g\leq s^0 .
\]
Its Lagrangian with multipliers $w\geq 0$ is
$w^\top s^0 + t\,(1-\one^\top w) + g^\top (Cw-c)$, whose supremum over $t\in\R$ and
$g\geq 0$ is finite if and only if $\one^\top w=1$ and $Cw\leq c$, i.e.\ $w\in W(C,c)$;
the dual program is therefore
\[
  (D)\qquad \min_{w}\ w^\top s^0 \quad\text{s.t.}\quad w\in W(C,c).
\]
$(P)$ is feasible ($g=0$, $t=\min_j s^0_j$). By linear programming
duality~\cite{schrijver1986theory}, if $(D)$ is feasible both programs have the
same optimal value, which gives $V^*$ in (ii); if $(D)$ is infeasible, $(P)$ is
unbounded, which gives the second part of (i).

For the first part of (i), write $W(C,c)=\emptyset$ as: there is no $w\geq0$,
$w\neq 0$ with $(C-c\,\one^\top)\,w\leq 0$ (normalize $w$ to sum to $1$). By Ville's
theorem of the alternative~\cite{mangasarian1969nonlinear}, for any real matrix
$M$ exactly one of the systems $\{M^\top g>0,\ g\geq 0\}$ and
$\{Mw\leq 0,\ w\geq0,\ w\neq 0\}$ has a solution. With $M=C-c\,\one^\top$, the
first system reads $C^\top g-(c^\top g)\one>0$, $g\geq0$.

For the bounds in (ii), $g=0$ is feasible, so $V^*\geq\min_j s^0_j$; and any
$w\in\Delta_n$ satisfies $w^\top s^0\leq\max_j s^0_j$. Finally, $W(C,c)$ is compact,
so the minimum is attained, and for $w\in\Delta_n$ we have $w^\top s^0=\min_j s^0_j$
if and only if $\mathrm{supp}(w)\subseteq J_0$.
\end{proof}

\begin{remark}[Bounded intensities]\label{rem:budget}
If gaming intensities are bounded, e.g.\ by a budget $\ell^\top g\leq B$ with
$\ell>0$, the feasible set of $(P)$ shrinks: in case (ii) the bound on $G^*$ still
holds, and in case (i) the gain is finite but grows linearly with $B$. Budgets
(C1) therefore limit the damage of a multi-resistant mix; they do not remove it.
\end{remark}

\begin{corollary}[No costs]\label{cor:ville}
When $c=0$, exactly one of the following holds: there is a multi-resistant mix
$g\geq 0$ with $C^\top g>0$; or there is a weight vector $w\in\Delta_n$ with
$Cw\leq 0$, i.e.\ a weighting of the signals that no gaming action can increase.
\end{corollary}

\begin{corollary}[Conjunctive versus linear aggregation]\label{cor:linear}
Under H1g and H2--H6, with a linear rule $A=A_v$, the net gain of gaming is $0$
if $v\in W(C,c)$ and grows without bound otherwise. Consequently:
\begin{enumerate}[label=(\alph*)]
  \item if some linear rule makes gaming unprofitable, then $W(C,c)\neq\emptyset$
        and the conjunctive rule bounds the gain of gaming;
  \item the conjunctive rule does not require knowing $C$: it attains the bound
        $\min_{w\in W}w^\top s^0$ whatever the admissible set, whereas a linear rule
        is safe only if its weights were chosen inside $W(C,c)$, which requires
        knowing $C$ and $c$;
  \item the price of this robustness is a residual gain at most equal to the
        spread $\max_j s^0_j-\min_j s^0_j$ of the honest profile, which vanishes for
        balanced profiles.
\end{enumerate}
\end{corollary}

\begin{proof}
With $A_v$, $U(g)=v^\top s^0+g^\top(Cv-c)$. If $Cv\leq c$, the maximum over $g\geq
0$ is reached at $g=0$; otherwise some coordinate $i$ has $(Cv-c)_i>0$ and $U$ is
unbounded along $g=t e_i$. Points (a)--(c) follow from Theorem~\ref{thm:value}.
\end{proof}

\subsection{Local results around a profile}
\label{subsec:local}

\begin{theorem}[Local model, first order]\label{thm:local}
Under H1$\ell$ and H2--H6, with $A=A_{\min}$, let $J_0=\arg\min_j s^0_j$ be the set of
the attacker's weakest signals.
\begin{enumerate}[label=(\roman*)]
  \item If $W_{J_0}(C,c)=\emptyset$, there is a direction $d\geq0$ such that
        $U(\varepsilon d)>U(0)$ for every sufficiently small $\varepsilon>0$:
        gaming is locally profitable.
  \item If $W_{J_0}(C,c)\neq\emptyset$, then for every $d\geq 0$ the right derivative
        of $\varepsilon\mapsto U(\varepsilon d)$ at $0$ is non-positive: gaming
        brings no first-order gain.
  \item If some $w\in W_{J_0}(C,c)$ satisfies $Cw<c$ componentwise, then $g=0$ is a
        strict local maximizer of $U$ on $\R^k_{\geq0}$.
\end{enumerate}
\end{theorem}

\begin{proof}
For $d\geq0$ and $\varepsilon>0$ small enough, the signals outside $J_0$ keep a
positive slack, so $\min_j s_j(\varepsilon d)=\min_{j\in J_0}\big(s^0_j+\varepsilon
(C^\top d)_j\big)+o(\varepsilon)$, and the right derivative of $U(\varepsilon d)$ at
$0$ is $\min_{j\in J_0}(C^\top d)_j-c^\top d$.

(i) Applying Ville's theorem to the columns of $C$ indexed by $J_0$, as in the
proof of Theorem~\ref{thm:value}, $W_{J_0}(C,c)=\emptyset$ yields $d\geq 0$ with
$(C^\top d)_j>c^\top d$ for all $j\in J_0$, hence a positive right derivative.

(ii) If $w\in W_{J_0}(C,c)$, then for every $d\geq 0$,
\[
  \min_{j\in J_0}(C^\top d)_j\;\leq\;\sum_{j\in J_0}w_j(C^\top d)_j\;=\;d^\top Cw\;\leq\; d^\top c .
\]

(iii) For every $g\geq 0$, $\min_j s_j(g)\leq w^\top s(g)=\min_j s^0_j+g^\top Cw+o(\|g\|)$,
since $w$ is supported on $J_0$. Hence $U(g)-U(0)\leq g^\top(Cw-c)+o(\|g\|)\leq
-m\,\|g\|_1+o(\|g\|)$ with $m=\min_i(c-Cw)_i>0$, which is negative for $g\neq0$
small enough.
\end{proof}

The two theorems answer different questions. Theorem~\ref{thm:local} only
involves the attacker's weakest signals: near the honest profile, an account is
safe from gaming exactly when some admissible weighting lives on its weakest
signals, which is the condition $G^*=0$ of Theorem~\ref{thm:value}.
Theorem~\ref{thm:value} describes what happens beyond first order, but only in the
idealized global model. In a nonlinear system, a global guarantee would require
the local condition to hold at every profile reachable by gaming, with the matrix
$C$ of that profile; an attacker can be seen as following a path along which the
set of weakest signals changes.

\begin{proposition}[No reward for partial resistance]\label{prop:partial}
Under H1$\ell$, for every $g\geq 0$, the right derivative of
$\varepsilon\mapsto A_{\min}(s(\varepsilon g))$ at $0$ is
$\min_{j\in J_0}(C^\top g)_j$. In particular, at a balanced profile
($J_0=\{1,\dots,n\}$), a gaming mix that raises some signals but leaves at least one
unchanged or lowered yields no first-order gain, whereas under a linear rule $A_v$
with $v>0$ any mix with $g^\top Cv>0$ does.
\end{proposition}

\begin{proof}
This is the first step of the proof of Theorem~\ref{thm:local}, without costs.
\end{proof}

Proposition~\ref{prop:partial} is the formal counterpart of the observation that a
virus carrying a single resistance mutation remains suppressed by the other drugs.
An attacker who learns by trial and reward --a bot tuned by reinforcement, a
community imitating successful tricks-- receives no reward for defeating one
signal at a time and must find a multi-resistant strategy at once. Under a
weighted sum, every partial success is rewarded and the attacker can climb step by
step: the reputation analogue of sequential escape.

\subsection{What aggregation can and cannot do: non-emptiness of the admissible set}
\label{subsec:nonempty}

All the guarantees above are conditional on the admissible set being non-empty
(globally, or on the weakest signals locally). Conjunctive aggregation does not
create admissible weightings: it exploits them without knowing them. Whether
$W(C,c)$ is empty is a property of the set of signals and of the costs, not of the
aggregation rule; it must be established by design and estimated.

\emph{By design}, non-emptiness requires that every combination of gaming actions
be penalized on some signal that carries weight, or be costly (Rule~3 below). The
sign pattern of $C$ only gives necessary conditions (Remark~\ref{rem:coverage});
the magnitudes of the entries decide.

\emph{By estimation}, $C$ can be estimated by simulation or from data, with
uncertainty. Suppose entrywise intervals $C\in[\underline C,\overline C]$ are
available. Since $w\geq0$, each coordinate of $Cw$ is non-decreasing in every entry
of $C$, so the worst case over the box is attained at $\overline C$:
\[
  W(\overline C,c)\subseteq W(C,c)\quad\text{for every } C\in[\underline C,\overline C].
\]
Robust non-emptiness, globally or on a set $J$ of signals, is thus the feasibility
of a single linear program, $W(\overline C,c)\neq\emptyset$ or
$W_J(\overline C,c)\neq\emptyset$, which can be checked for each family of profiles
in simulation.

\emph{Are the necessary costs realistic?} This is the practical question, and it
can be asked action by action.

\begin{definition}[Required cost]\label{def:reqcost}
For a gaming action $i$ and a set $J$ of signals, the \emph{required cost} is
\[
  c_i^{\mathrm{req}}(J)=\min\big\{(Cw)_i \;:\; w\in\Delta_n,\ \mathrm{supp}(w)\subseteq J,\
  (Cw)_{i'}\leq c_{i'}\ \text{for all } i'\neq i\big\},
\]
with $c_i^{\mathrm{req}}(J)=+\infty$ when the set is empty.
\end{definition}

By definition, $W_J(C,c)\neq\emptyset$ if and only if $c_i\geq c_i^{\mathrm{req}}(J)$.
The required cost is the value of a linear program; computed with the upper bound
$\overline C$, it can be compared with the costs the system can actually impose
(fees, budgets, vouching stakes). Its simplest case is instructive. For an account
whose weakest signal is $j$ alone ($J=\{j\}$), and provided the other actions satisfy
$C_{i'j}\leq c_{i'}$, $c_i^{\mathrm{req}}(\{j\})=C_{ij}$: every gaming action must cost
at least its effect on $j$. By Theorem~\ref{thm:local}, incentives to game are thus
concentrated on accounts bottlenecked on a gameable signal, and these are not
necessarily dishonest accounts. Section~\ref{subsec:num-gaming} shows that, in
MagMell, the required costs are highest for newcomers and, to a lesser extent, for
researchers in narrow fields.

\begin{remark}[Geometric mean and soft minimum]\label{rem:geo}
The geometric mean is often presented as a non-compensatory aggregation, but its
logarithm is the uniform average of the log-signals. It therefore behaves as a
linear rule with fixed weights in log-space, and is safe only if uniform weights
happen to be admissible for the log-linearized collateral matrix. A soft minimum
$-\theta\log\sum_j e^{-s_j/\theta}$ interpolates between the average ($\theta\to\infty$)
and the minimum ($\theta\to0$), and can be used when a differentiable rule is needed.
\end{remark}

\begin{remark}[A zero-sum game]\label{rem:game}
When $c=0$, Corollary~\ref{cor:ville} has a game-theoretic reading: in the zero-sum
game where the attacker chooses a mixed strategy $g\in\Delta_k$ over gaming actions,
the designer a mixed strategy $w\in\Delta_n$ over signals, and the attacker receives
$g^\top Cw$, the second alternative holds if and only if the value of the game is
non-positive~\cite{vonneumann1928}. The ``evolutionary vise'' exists exactly when
the designer wins this game.
\end{remark}

\begin{remark}[Costs, coverage and timing]\label{rem:coverage}
$W(C,c)$ is monotone in $c$: raising the cost of a gaming action (identity cost,
budgets, stakes) can only enlarge the admissible set and decrease $V^*$. If a
gaming action $i$ is free ($c_i=0$) and raises some signals without lowering any,
the constraint $(Cw)_i\leq 0$ forces $w_j=0$ on every signal $j$ it raises: a signal
that can be raised for free and without collateral effect is useless, like a drug
against which resistance is free. Similarly, if rewards are computed from a subset
$S$ of the signals (for instance before slow signals are available), the relevant
set is $W_S(C,c)\subseteq W(C,c)$, which may be empty even when $W(C,c)$ is not: this
is temporal monotherapy.
\end{remark}

\begin{remark}[Genuine contribution]\label{rem:honest}
Let genuine contribution be an action with effect $\eta\in\R^n$ on the signals and
cost $c_h$. At a balanced profile, by Proposition~\ref{prop:partial}, it pays under
$A_{\min}$ if and only if $\min_j\eta_j>c_h$: genuine work must improve every
signal. The design goal can thus be stated compactly: genuine contribution should
be the only cost-adjusted multi-resistant direction. When the designer does know
$C$, it can instead use a linear rule with weights solving
$\max\{\eta^\top w : w\in W(C,c)\}$; a positive optimum rewards genuine work while
making gaming unprofitable, consistently with the result of Kleinberg and Raghavan
that, in their model, whenever some reasonable evaluation rule induces the
desired effort, a linear one does~\cite{kleinberg2020classifiers}.
\end{remark}

\subsection{From observable events to signals}
\label{subsec:pipeline}

A row of $C$ is only well defined once it is clear what a ``signal'' is. We
distinguish three levels.
\begin{enumerate}[label=\textbf{L\arabic*.},leftmargin=2.2em]
  \item \textbf{Observables} $o$: the events recorded in the ledger (publications,
        citations, accesses, assessments with their polarity, stake and prediction,
        settlements, attestations, timestamps).
  \item \textbf{Sub-indicators} $z_\ell=\zeta_\ell(o)\in[0,1]$, $\ell=1,\dots,m$: functions
        of the observables, each capturing one failure mode and normalized as a
        percentile within a community (H2). Each sub-indicator belongs to one
        family, i.e.\ relies mainly on one resource (Rule~1).
  \item \textbf{Family signals} $s_j=\varphi_j(z_{F_j})$, where $F_j$ is the set of
        sub-indicators of family $j$ and $\varphi_j$ a fixed within-family
        aggregation.
\end{enumerate}
The gaming matrix is first hypothesized or estimated at the level of
sub-indicators: $B\in\R^{k\times m}$, with $B_{i\ell}$ the effect of gaming action $i$
on $z_\ell$. The matrix $C$ at the level of family signals then depends on the
functions $\varphi_j$.

\begin{proposition}[Nesting]\label{prop:nesting}
If every $\varphi_j$ is the minimum of its sub-indicators and $A=A_{\min}$, then
$A(s)=\min_{\ell}z_\ell$, and Theorems~\ref{thm:value} and~\ref{thm:local} apply with
the sub-indicators as signals and $C=B$. If every $\varphi_j$ is linear,
$s_j=v_j^\top z_{F_j}$, then the family signals are linear in $z$ and $C=BV$, where
$V\in\R^{m\times n}$ is the block matrix of the within-family weights.
\end{proposition}

\begin{proof}
A minimum of minima over a partition is the minimum over the union. In the linear
case, $s=V^\top z$ and $z=z^0+B^\top g$, so $s=s^0+(BV)^\top g$.
\end{proof}

Proposition~\ref{prop:nesting} suggests a principle for within-family
aggregation: \emph{compensate within a resource, conjoin across failure modes}.
Components that measure the same quantity with independent noise can be averaged,
which reduces noise; components that capture distinct failure modes must be
combined by a minimum. For example, the judgement family of MagMell has two
sub-indicators: the accuracy of settled assessments and their information content.
Herding leaves the first unchanged and lowers the second; with a minimum inside
the family, its effect on the family signal is well defined, whereas an
unspecified combination would leave it ambiguous.

\subsection{Merit, screens, missing values and evidence}
\label{subsec:screens}

Applying the theory to an actual pipeline raises four further design questions.
The numerical analyses of Section~\ref{sec:numerics} show that each of them
matters: a first version of MagMell's pipeline, which ignored them, ranked honest
accounts by noise and made whitewashing profitable for most of them.

\paragraph{Merit and integrity.} Some sub-indicators measure \emph{merit}: their
variation among honest accounts is meaningful (recognition, accuracy of
assessments). Others measure \emph{integrity}: honest accounts are all ``normal''
on them, and their variation among honest accounts is noise (reading margin,
reciprocity, stability). Normalizing an integrity sub-indicator as a percentile
spreads this noise uniformly over $[0,1]$, and a minimum then ranks honest accounts
by their worst noise. Integrity sub-indicators should be \emph{screened} rather
than ranked: $\sigma_\ell=\min(1,p_\ell/\tau)$, where $p_\ell$ is the percentile of
the sub-indicator and $\tau$ a tolerance.

\begin{remark}[Screens fit the theory]\label{rem:screens}
A screen is the minimum of two affine functions of its percentile, $1$ and
$p_\ell/\tau$. An aggregate $R=\min(\text{merit},\text{screens})$ is therefore the
minimum of the merit sub-indicators, of the rescaled percentiles $p_\ell/\tau$ and
of the constant $1$, and Theorems~\ref{thm:value} and~\ref{thm:local} apply to this
extended set of signals, the constant being unaffected by gaming. Rescaling
multiplies the collateral entries of screen $\ell$ by $1/\tau$, which makes
admissible weightings easier to find. It also raises the honest value $p^0_\ell/\tau$
far above the merit level, so that the screen does not bind near honest profiles:
by Theorem~\ref{thm:local}, the vise only operates once gaming has pushed the screen
down to the merit level. The tolerance $\tau$ thus trades false positives against
the slack left to gaming. With $m$ independent screens, a fraction of about
$1-(1-\tau)^m$ of honest accounts has at least one screen below $1$.
\end{remark}

\paragraph{Percentile and absolute screens.} A percentile screen with tolerance
$\tau$ can only flag a fraction about $\tau$ of the reference population. A coalition
that represents a larger fraction dilutes itself: most of its members are above the
$\tau$-quantile, however abnormal they are. When honest behavior is bounded by human
constraints --reading speed, reciprocity of citations-- an absolute threshold should
be combined with the percentile:
$\sigma_\ell=\min(1,p_\ell/\tau,f_\ell/\theta_\ell)$. The absolute part catches groups
of any size, at the cost of false positives on honest accounts whose legitimate
behavior crosses the threshold (Section~\ref{subsec:num-dynamics}).

\paragraph{Missing values.} A sub-indicator is missing when the account has
produced no observation of it (no assessment, no publication). It can be imputed or
dropped from the minimum, but dropping is only safe under a condition that follows
from Remark~\ref{rem:coverage}: \emph{a sub-indicator may be dropped when missing
only if no gaming action can make it missing}. Otherwise the attacker escapes it by
resetting it, which amounts to being evaluated on a subset of the signals.
Whitewashing resets every history-dependent sub-indicator, so those that penalize
whitewashing must be imputed, not dropped. Sub-indicators that are triggered by the
activity they penalize can be dropped: an assessment produces a reading margin, and
a received citation produces a reciprocity value.

\paragraph{Scale of imputation.} An imputed value must be calibrated on the scale of
the aggregate, not of the sub-indicator. The minimum of several percentiles is
typically much lower than each of them, so a value that is low for a sub-indicator
(its first quartile, say) can be high for the aggregate. The imputed level should be
derived from the rank of the aggregate at which a fresh identity is to be placed.
This makes the trade-off of C4 explicit: identities ranked below that level gain by
whitewashing, unless the cost of a new identity exceeds the gain.

\paragraph{Evidence.} Shrinkage weighs the observed value of a sub-indicator by its
evidence. If evidence is measured by a quantity that gaming can inflate cheaply, such
as the number of one's own publications, then gaming weakens shrinkage and opens an
additional channel of gain, even when the gaming action has no direct effect on the
raw statistic. Evidence should be measured by quantities that the gaming actions
raising the sub-indicator cannot inflate, such as exposure time.

\subsection{Design rules}
\label{subsec:rules}

The results above translate into four design rules, which refine C6.

\paragraph{Rule 1: orthogonality of resources.} What matters for robustness is not
the formula of a signal but the \emph{resource} an attacker must forge to inflate
it. We distinguish five resources: (R1)~identities; (R2)~human time and attention;
(R3)~capital at risk, i.e.\ staked reputation; (R4)~persistence, i.e.\ consistent
behavior over long periods; (R5)~diversity of endorsers, i.e.\ support from
distinct communities. Two signals relying on the same resource are raised by the
same gaming action: they share a row of positive entries in $C$, like two drugs
with cross-resistance. The ratio on X is a case in point: it is computed from
likes, so a single farm of fake accounts raises both.

\paragraph{Rule 2: conjunctive aggregation.} By Corollary~\ref{cor:linear} and
Proposition~\ref{prop:partial}, signals should be aggregated with the minimum (or a
soft minimum), not with a weighted sum nor a geometric mean, and within a family
the principle of Section~\ref{subsec:pipeline} applies. In the global model, the
residual gain of gaming is bounded by the spread of the honest profile
(Theorem~\ref{thm:value}); the \emph{discordance} $\max_j s_j-\min_j s_j$ is
therefore both an indicator of what gaming can buy and a natural alert. Honest profiles are not uniform either (a theorist may review
little), so the alert threshold has to be calibrated per community.

\paragraph{Rule 3: collateral sensitivity.} The admissible set is non-empty only
if each gaming action is penalized on some signal that carries weight, or is
costly. Signals should therefore be chosen in pairs such that the gaming strategy
that inflates one of them degrades the other, and non-emptiness should be checked
on estimated matrices with the robust test of Section~\ref{subsec:nonempty}.
Table~\ref{tab:pairs} lists candidate pairs.

\paragraph{Rule 4: matched coverage and timing.} Every type of action
(publishing, citing, assessing, staking) should be observed by at least two
signals relying on different resources; an action observed by a single signal is
a single-drug compartment~\cite{morenogamez2015imperfect}. Rewards should only be
computed once the slow signals are available, otherwise the evaluation runs on a
subset of signals (Remark~\ref{rem:coverage}); this is the rationale for escrow.
Varying the weights over time is acceptable \emph{within} the admissible set and
with every signal active; switching from one signal to another would reproduce
the sequential monotherapies that the medical literature warns
against~\cite{morenogamez2015imperfect,imamovic2013use,kim2014alternating}.

\begin{table}[!htbp]
\centering
\caption{Candidate pairs of signals exhibiting collateral sensitivity.}
\label{tab:pairs}
\small
\begin{tabularx}{\textwidth}{>{\raggedright\arraybackslash}p{2.6cm}>{\raggedright\arraybackslash}X>{\raggedright\arraybackslash}X>{\raggedright\arraybackslash}p{2.4cm}}
\toprule
Gaming strategy & Inflates & Degrades & References \\
\midrule
Sybil farm & Volume of endorsements & Trust propagated from seeds (few attack edges) &
  \cite{yu2006sybilguard,cao2012aiding}; limits:~\cite{schoenebeck2016sybil} \\
Echo chamber & In-group endorsement & Bridging score across communities &
  \cite{massa2007trust,wojcik2022birdwatch} \\
Herding & Accuracy of stakes w.r.t.\ consensus & Information score (``surprisingly common'') &
  \cite{prelec2004bayesian,miller2005eliciting} \\
Bot automation & Volume of actions & Plausibility w.r.t.\ human time budget &
  \cite{brysbaert2019many,rayner2016so} \\
Citation ring & Citation count & Reciprocity and out-group share; diluting the ring requires more members and more attack edges &
  \cite{fister2016toward} \\
Oscillation, hype & Short-term score & Long-term corroboration by independent signals &
  \cite{srivatsa2005trustguard,wang2017bias} \\
Whitewashing & Escape from a negative record & Confidence (a fresh identity has high uncertainty) &
  \cite{friedman2001social,josang2002beta} \\
\bottomrule
\end{tabularx}
\end{table}

\subsection{Limits of the theory}
\label{subsec:limits}

The results rely on explicit assumptions, which delimit their scope.
\begin{itemize}
  \item \emph{Global versus local.} Theorem~\ref{thm:value} is exact in the global
        linear model only; in a real system, its unbounded case means a gain that
        grows until the model breaks down, and its bound relies on the linearity
        of the effects. Theorem~\ref{thm:local} is valid under the realistic
        assumption H1$\ell$ but only near a given profile.
  \item \emph{Knowledge of $C$.} The matrix $C$ is unknown, must be estimated by
        simulation or from data, and drifts as attackers innovate, which adds rows
        (new gaming actions, the analogue of new mutations) and changes entries (the
        analogue of epistasis~\cite{barbosa2019evolutionary}). The conjunctive rule
        does not need $C$ to be robust, but non-emptiness of the admissible set does
        need to be checked (Section~\ref{subsec:nonempty}).
  \item \emph{Normalization.} If signals are percentiles (H2), the profile of an
        account depends on the others: lowering a competitor raises one's own
        percentile. This strategic interaction is not captured by the model; it
        appears in the attack matrix of MagMell as an indirect effect of
        bad-mouthing.
  \item \emph{Coalitions.} H6 treats a coalition as one decision-maker, which is
        conservative for coordination but ignores the costs of coordinating.
  \item \emph{Noise.} The minimum of $m$ noisy sub-indicators is biased downwards,
        and more so when $m$ is large and data are scarce. Conjunction therefore
        trades robustness against false positives on accounts with little data
        (newcomers, small fields). Screening integrity sub-indicators instead of
        ranking them, shrinkage, and reporting uncertainty separately reduce this
        effect (Section~\ref{subsec:screens}); the analysis of
        Section~\ref{subsec:num-profiles} shows that it is large when these
        precautions are not taken.
\end{itemize}

\subsection{Side effects}
\label{subsec:sideeffects}

Every therapy has side effects, and the relevant ones here are false positives on
honest users. Researchers in small or niche fields receive few endorsements from
other communities; newcomers have little topological trust and high uncertainty;
novel work is riskier and its recognition is delayed~\cite{wang2017bias}, so it is
penalized by short time windows; and any reputation-weighted ranking reinforces
cumulative advantage, the Matthew effect~\cite{merton1968matthew}. Minimum
aggregation adds its own: an honest researcher whose profile is unbalanced is
ranked by her weakest signal. The hard case for a scientific reputation system is
not the flat-earther, it is the heterodox claim that turns out to be right. The
``toxicity'' of a set of signals on the honest population must therefore be
measured as carefully as its efficacy against attackers, and the mechanisms must be
designed to limit it: MagMell's rules G3--G7 (Section~\ref{subsec:governance})
address small fields, multidisciplinary authors, minority critiques, recent work and
the selection of assessors by capital. Section~\ref{subsec:num-newcomers} measures
how long a heterodox researcher recognized late takes to reach the median rank in a
simulation. Two further side effects
concern the permanence of records: the chilling effect on criticism
(Section~\ref{subsec:pervasive}), and the tension between immutable ledgers and
the right to erasure~\cite{finck2018blockchains}.

\section{MagMell: a two-tier P2P reputation system for scientific publication}
\label{sec:multirep}

This section answers RQ3. It specifies the mechanisms and incentive rules of
MagMell, which has not been implemented; numerical values are indicative and meant to
be calibrated by simulation (Section~\ref{subsec:evaluation}).

\subsection{Goals and two-tier architecture}
\label{sec:case_stdy}

MagMell is a P2P network on which researchers publish scientific material
(research articles, or discussions of topics and of other published material), with
a built-in reputation system. Its aim is not to abolish reviewing but to move it after
publication: the network is open, anyone can publish, and the reputation system
separates signal from noise. Work that is consistently judged unsound by independent
readers is ranked low and receives little attention. As noted in
Section~\ref{subsec:sideeffects}, this is the easy case; the design must also avoid
burying heterodox but correct work.

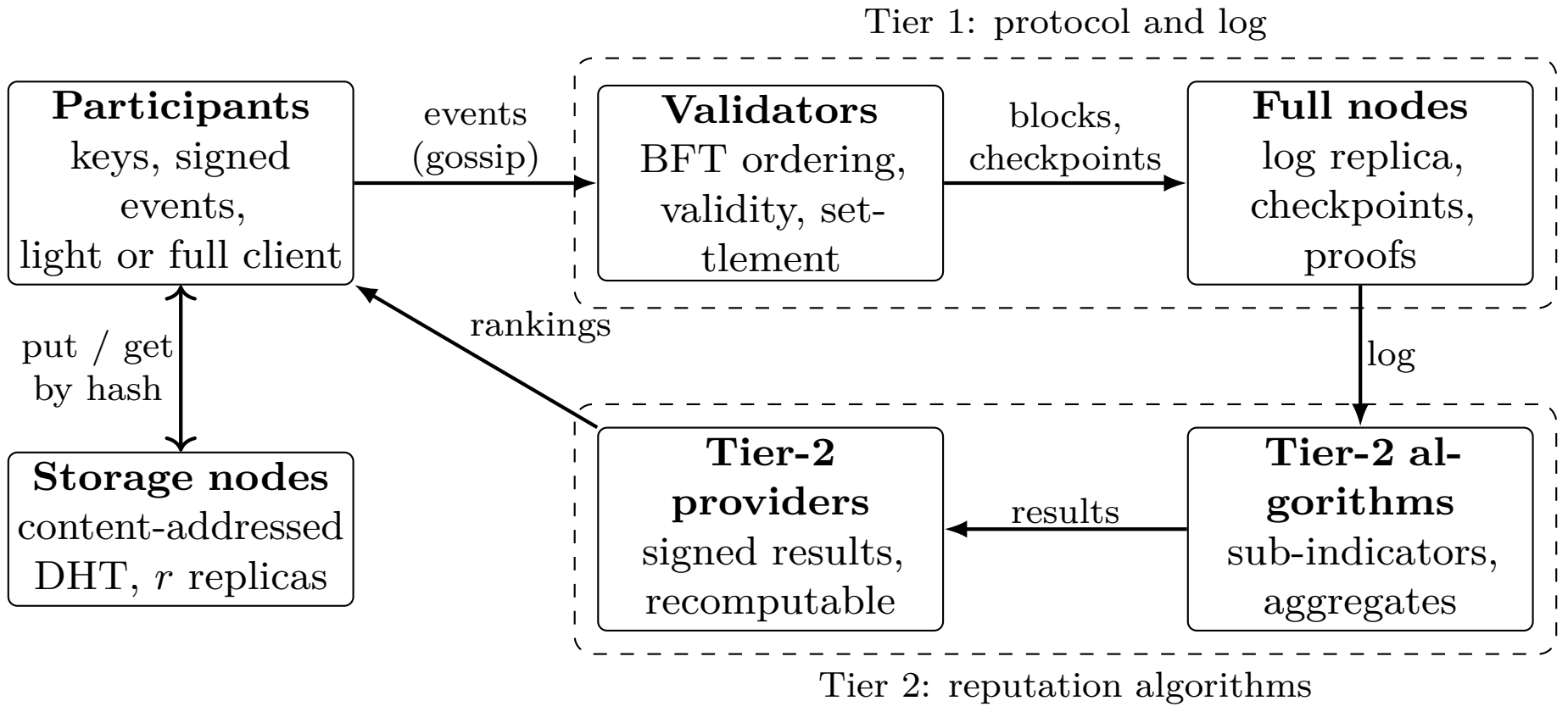
\begin{figure}[!htbp]
\centering
\begin{tikzpicture}[
  font=\footnotesize,
  box/.style={draw, rounded corners=2pt, align=center, text width=2.55cm,
              minimum height=10mm, inner sep=2pt},
  tier/.style={draw, dashed, rounded corners=4pt, inner sep=5pt},
  arr/.style={-{Latex[length=1.8mm]}, thick},
  lab/.style={font=\scriptsize, align=center, inner sep=1pt}]
  \node[box] (part)  at (0,0)     {\textbf{Participants}\\keys, signed events,\\light or full client};
  \node[box] (val)   at (4.6,0)   {\textbf{Validators}\\BFT ordering,\\validity, settlement};
  \node[box] (full)  at (9.2,0)   {\textbf{Full nodes}\\log replica,\\checkpoints, proofs};
  \node[box] (store) at (0,-2.7)  {\textbf{Storage nodes}\\content-addressed\\DHT, $r$ replicas};
  \node[box] (prov)  at (4.6,-2.7){\textbf{Tier-2 providers}\\signed results,\\recomputable};
  \node[box] (t2)    at (9.2,-2.7){\textbf{Tier-2 algorithms}\\sub-indicators,\\aggregates};
  \draw[arr] (part) -- node[lab, above]{events\\(gossip)} (val);
  \draw[arr] (val) -- node[lab, above]{blocks,\\checkpoints} (full);
  \draw[arr, <->] (part) -- node[lab, left]{put / get\\by hash} (store);
  \draw[arr] (full) -- node[lab, right]{log} (t2);
  \draw[arr] (t2) -- node[lab, above]{results} (prov);
  \draw[arr] (prov) -- node[lab, above right, pos=0.55]{rankings} (part);
  \begin{scope}[on background layer]
    \node[tier, fit=(val) (full), label={[font=\scriptsize]above:Tier 1: protocol and log}] {};
    \node[tier, fit=(prov) (t2), label={[font=\scriptsize]below:Tier 2: reputation algorithms}] {};
  \end{scope}
\end{tikzpicture}
\caption{Architecture of MagMell. Tier~1 ($n_v=3f+1$ validators and any number of
full nodes) orders,
validates and replicates signed events and executes the reference settlement;
contents are stored off-log in a content-addressed DHT; tier-2 algorithms run on any
replica, and their results can be recomputed from the log by anyone.}
\label{fig:architecture}
\end{figure}

MagMell separates a \emph{protocol} from the \emph{reputation algorithms} that
interpret it (Fig.~\ref{fig:architecture}). \emph{Tier~1}, the protocol, is a
replicated state machine. Its state consists of identities, balances, budgets,
stakes and escrows; its transitions are signed events, recorded in a totally ordered
log. Tier~1 checks the validity of events, enforces budgets and rate limits, and
executes the deterministic settlement and minting rules. From the layer that orders
events, tier~1 only requires an abstract interface: append a valid event, read the
ordered log, and obtain checkpoints of the state signed by the ordering layer. Any
consensus protocol that provides a totally ordered and available log can implement
this interface; Section~\ref{subsec:sysmodel} describes the instantiation that we
analyze. \emph{Tier~2} consists of
reputation algorithms that compute sub-indicators and aggregates from the log. Tier~2
is open and plural: any node can run any algorithm on its replica, and each user can
choose or tune the algorithm that ranks what she sees, because reputation rests on
local sensibilities (C5). One tier-2 algorithm, versioned and public, is designated
by governance as the \emph{reference algorithm}: it is the only one whose outputs
change tier-1 state, through settlement (A4--A8) and minting
(Section~\ref{subsec:minting}), and since it is deterministic every replica computes
the same result. The protocol only relies on \emph{objective data}, the events
recorded in the log; Section~\ref{subsec:guarantees} states what it guarantees
whatever tier-2 algorithm is used.

\subsection{System model: nodes, trust assumptions and failures}
\label{subsec:sysmodel}

\paragraph{Node roles.} MagMell distinguishes five roles, which a physical machine
may combine.
\begin{itemize}
  \item \emph{Participants} hold the keys of their identities (public keys of a
        signature scheme such as Ed25519~\cite{bernstein2012high}), sign events, and
        run either a full node or a light client.
  \item \emph{Validators} implement the ordered log. In the instantiation that we
        analyze, they form a committee of $n_v=3f+1$ nodes operated by independent and
        identifiable institutions (libraries, universities, learned societies)
        renewed by governance, and they run a BFT
        state-machine-replication protocol for the partially synchronous model, such
        as Practical Byzantine Fault Tolerance (PBFT)~\cite{castro1999practical} or HotStuff~\cite{yin2019hotstuff}. This
        is a design choice suited to the domain rather than a technical concession:
        scientific communication already relies on such institutions to operate
        shared infrastructure, such as repositories and archives, and identifiable
        operators are accountable, legally and reputationally, for misbehavior that
        the log makes provable (see the trust model of the committee below). The
        reputation layer is agnostic to this choice: a permissionless protocol such
        as proof of stake~\cite{gilad2017algorand} could implement the same
        interface, but we do not claim that it would have the same properties, and
        we analyze only the committee case.
  \item \emph{Full nodes} store the whole log and the tier-1 state, verify the
        validators' signatures on blocks and checkpoints, serve light clients with
        Merkle proofs, and run tier-2 algorithms. Any participant or institution can
        run one.
  \item \emph{Storage nodes} store contents in a content-addressed
        DHT~\cite{maymounkov2002kademlia}, as in IPFS~\cite{benet2014ipfs},
        with $r$ replicas per item; the institutions of the authors pin their
        contents.
  \item \emph{Tier-2 providers} are optional full nodes that publish signed tier-2
        results, with the log position at which they were computed, for light
        clients. They are not trusted: anyone holding the log can recompute their
        results.
\end{itemize}

\paragraph{Trust assumptions.} Signatures are unforgeable and hash functions
collision-resistant. At most $f$ of the $3f+1$ validators are Byzantine; the network
is partially synchronous~\cite{dwork1988consensus}: messages between correct nodes
are delivered within an unknown bound after an unknown global stabilization time.
Any number of participants, full nodes, storage nodes and tier-2 providers may be
Byzantine: they may withhold, delay or forge content, or equivocate, but cannot forge
signatures. A client that connects to at least one correct full node, or verifies
checkpoints signed by $2f+1$ validators, obtains the correct tier-1 state.

\paragraph{Failure model.} Nodes may crash, leave and rejoin. Storage nodes are
unavailable independently with probability $p$, an assumption. If more than $f$
validators are unavailable, the log stops growing but its safety is preserved:
assessments and settlements are delayed, not lost, and escrow horizons are counted in
ordered epochs.

\paragraph{Trust model of the committee.} MagMell is open at the level of
participants but not at the level of consensus: in the model analyzed, decentralization
is institutional. This has four consequences, which we state rather than solve.
\emph{Capture}: if $f+1$ validators collude, they can refuse to vote for blocks
that contain given events, and thus censor them or halt the log; if $2f+1$ collude,
they can commit blocks that violate the validity rules. Full nodes re-execute every
block, so that invalid transitions are detected and publicly provable, but not
prevented. \emph{Governance}: the committee is renewed by governance, so that its
honesty rests on the honesty of the renewal process, itself subject to capture.
\emph{Concentration}: institutions that run validators are few and unevenly
distributed among countries and fields; diversity of operators is a requirement on
governance, not a property of the protocol. \emph{Beyond $f$}: with more than $f$
Byzantine validators, neither safety (two conflicting blocks may be committed), nor
liveness, nor freedom from censorship is guaranteed.

\paragraph{Replication cost among validators.} Each block costs $O(n_v)$ messages
per phase with HotStuff, whose communication is linear in the number of
replicas~\cite{yin2019hotstuff}, and $O(n_v^2)$ with PBFT~\cite{castro1999practical}.
The consensus layer is specified but neither implemented nor benchmarked: confirmation
latency, achievable throughput, replication cost, behavior with faulty or malicious
validators and recovery after an interruption are not measured in this paper.

\paragraph{Out of scope.} Network-level attacks (eclipse of a client, routing
attacks), traffic analysis and the de-anonymization of pseudonyms are not addressed.
The Sybil resistance of the consensus layer is that of the committee; it is distinct
from the Sybil resistance of reputation, which onboarding
(Section~\ref{subsec:identities}) and the signals of Section~\ref{sec:theory}
address.

\subsection{Protocol: events, operations and validity}
\label{subsec:ledger}

\paragraph{Events.} An event is a tuple (type, identity, sequence number, epoch,
payload, signature). Per-identity sequence numbers make equivocation (two different
events signed with the same sequence number) detectable. Contents are stored
off-log and only their hashes are recorded: posts cannot be silently modified, but
contents can be removed when the law requires it while integrity proofs
remain~\cite{finck2018blockchains}. Each account has a currency balance (Curr) and a
reputation balance (Rep). Curr is transferable; it is created as in standard
cryptosystems, e.g.\ by participation rewards~\cite{gilad2017algorand} rather than
mining~\cite{nakamoto2008bitcoin}, and publishing costs Curr in proportion to the
size of the message, an anti-spam fee that bounds volume, not quality.

\paragraph{Budgets.} Per-identity budgets implement C1, with a time unit (epoch) of
length $\tau$:
\begin{enumerate}[label=\textbf{B\arabic*.},leftmargin=2.2em]
  \item \textbf{Reading budget} $B_r$ words per epoch, not accumulable. Assessing a
        document $D$ consumes $|D|$ words of the budget, and an assessment is
        admissible only if the delay between the first access to $D$ and the
        assessment is at least $|D|/\rho_{\max}$, where $\rho_{\max}$ is a generous
        upper bound on human reading rate~\cite{brysbaert2019many,rayner2016so}.
  \item \textbf{Writing budget} $B_w$ words per epoch, accumulable up to a cap,
        since drafts are written over time~\cite{salthouse1984effects}. As text
        generation is nearly free, this budget bounds volume only; quality must come
        from the other signals.
\end{enumerate}

\paragraph{Operations.} The protocol has six operations.
\begin{enumerate}[label=\textbf{O\arabic*.},leftmargin=2.2em]
  \item \textbf{Attest.} A full member attests a newcomer and locks a vouching stake
        (I1--I3, Section~\ref{subsec:identities}). Validity: the voucher is a full
        member, has a vouching right left (quota $m_v$) and enough free Rep.
  \item \textbf{Publish.} An event carries the hash and size of the content, the keys
        and signatures of all co-authors, and the list of referenced items, which
        are the citations counted as recognitions by tier~2. The authors' node puts
        the content to $r$ storage nodes. Validity: signatures, writing budget (B2),
        Curr fee. Validators do not fetch contents.
  \item \textbf{Access.} Once per epoch, a reader's node submits a signed batch of
        the items first accessed during the epoch, with their times. Validity:
        reading budget (B1). Accesses are the evidence used by the admissibility rule
        (A1) and by the independence and resolution rules (A5, A6).
  \item \textbf{Assess.} An assessment names its target, its polarity, its stake and
        a commitment to the reference forecast (A1--A4). Validity: admissibility
        (A1), stake limits (A2) and free Rep; the stake is locked in escrow (A3).
  \item \textbf{Settle and mint.} No transaction is needed: at each epoch boundary,
        every validator applies the reference algorithm to the assessments whose
        escrow expires (A4--A9) and to the publications that reach their minting
        horizon (Section~\ref{subsec:minting}), using only events ordered before the
        boundary. The resulting burns and mints are part of the state transition.
  \item \textbf{Checkpoint.} Every $K$ epochs, validators sign the Merkle root of
        the tier-1 state with $2f+1$ signatures.
\end{enumerate}

\paragraph{Validity versus quality.} Validators check what is decidable from the
log: syntax, signatures, sequence numbers, budgets, stakes, quotas and timing rules.
They do not check whether a publication is scientific, a citation relevant or an
assessment honest. Quality is the domain of tier~2, through the signals of
Section~\ref{subsec:profile}.

\paragraph{Distributed execution.} Table~\ref{tab:distributed} summarizes how the
operations are executed by the nodes. Participants submit events to several
validators, directly or through push gossip among full
nodes~\cite{demers1987epidemic,karp2000randomized}; validators order them in blocks,
which are disseminated to full nodes by gossip; a node that missed a block retrieves
it from any peer and verifies it against the validators' signatures (anti-entropy).
Contents are fetched lazily by hash from storage nodes.

\begin{table}[!htbp]
\centering
\caption{Distributed execution of MagMell.}
\label{tab:distributed}
\footnotesize
\renewcommand{\arraystretch}{1.15}
\begin{tabularx}{\textwidth}{>{\raggedright\arraybackslash}p{2.2cm}L}
\toprule
Question & MagMell \\
\midrule
Consensus & BFT state-machine replication by $n_v=3f+1$ validators (partial synchrony); specified, not implemented; a permissionless alternative is possible but not analyzed \\
Validation & validators check syntax, signatures, sequence numbers, budgets, stakes, quotas and timing; quality is left to tier~2 \\
Replication & the log and tier-1 state on every validator and full node; contents on $r$ storage nodes (content-addressed DHT) \\
Availability & writes need $2f+1$ responsive validators; reading the log needs one correct replica; a content item is available with probability $1-p^r$ \\
Synchronization & a new node loads the latest checkpoint signed by $2f+1$ validators and replays the blocks after it; light clients verify Merkle proofs \\
Conflicts & total order: of two events that together exceed a budget or a stake, the first is valid; equivocation is a proof of misbehavior that burns the vouching stake (I2) \\
Computation & tier-2 algorithms run locally on full nodes and providers; the reference settlement is recomputed by every validator at each epoch boundary \\
Cost & Section~\ref{subsec:costs} (analytic estimates) and Section~\ref{subsec:num-distributed} (simulated dissemination, measured computation) \\
\bottomrule
\end{tabularx}
\end{table}

\subsection{What the protocol guarantees}
\label{subsec:guarantees}

Table~\ref{tab:guarantees} separates the properties that tier~1 guarantees under the
assumptions of Section~\ref{subsec:sysmodel}, whatever the tier-2 algorithm, from
those that it does not guarantee. The latter are the object of the signals, of the
incentives and of the governance rules, and remain open problems where stated.

\begin{table}[!htbp]
\centering
\caption{Properties guaranteed by the protocol (tier~1) whatever the tier-2
algorithm, and properties it does not guarantee.}
\label{tab:guarantees}
\footnotesize
\renewcommand{\arraystretch}{1.15}
\begin{tabularx}{\textwidth}{LL}
\toprule
Guaranteed by tier~1 (assumption) & Not guaranteed by tier~1 \\
\midrule
Integrity and non-repudiation of events (cryptography) &
  Quality or truth of contents; relevance of citations \\
Agreement on a total order of events and on tier-1 state (at most $f$ Byzantine
  validators; liveness after stabilization) &
  Honesty or informativeness of assessments (incentives, Section~\ref{subsec:incentives}) \\
Enforcement of budgets, rate limits, quotas, stakes and escrows (validity rules) &
  Manipulation resistance of any tier-2 algorithm (Sections~\ref{sec:theory} and~\ref{sec:numerics}) \\
Conservation: Curr and Rep change only through fees, minting and burning (deterministic transitions) &
  Fairness of rankings and of governance (G1--G7) \\
Deterministic, verifiable settlement and minting under the reference algorithm &
  Privacy of reading behavior: access events are public in this design \\
Auditability: every tier-2 result can be recomputed from the log &
  Protection of assessors against retaliation: pseudonymous assessments (A10) are not yet constructed \\
Availability of the log (one correct replica) and of contents (probability $1-p^r$) &
  Safety, liveness and absence of censorship if more than $f$ validators are Byzantine \\
Detectability of equivocation &
  Availability of the \emph{evidence} needed to settle assessments (Section~\ref{subsec:num-dynamics}) \\
\bottomrule
\end{tabularx}
\end{table}

\subsection{Communication, storage and computation costs}
\label{subsec:costs}

We estimate the costs analytically from explicit event sizes and activity rates,
which are assumptions; only the computation time is measured, on synthetic event
tables (Section~\ref{subsec:num-distributed}). None of these figures is a
measurement of a deployed network. With 64-byte signatures and 32-byte keys and hashes, a publication with
three authors and 30 references takes about 1.3~kB, an assessment 178~bytes plus
44~bytes for its settlement record, and an access batch 107~bytes plus 36~bytes per
item read. With 1.6 publications per author and year shared among three authors, four
assessments per year and ten reads per week, the log grows by about 26~kB per
identity and year, 94\% of which are access batches. For $10^5$ identities, this is
2.6~GB per year, 0.19 transactions per second to order, and an ingest of
about 0.7~kbit/s per replica; push gossip with fanout $\lceil\ln N\rceil+3$ raises the
bandwidth of a node to about 10~kbit/s. For $10^6$ identities, these figures are
26~GB per year, 1.9 transactions per second and about 112~kbit/s, a load that does not
require high-throughput consensus. Contents dominate storage: with 2~MB per paper and
$r=5$ replicas, $10^5$ identities produce about 0.5~TB per year across storage nodes.
Computation is dominated by tier~2: a full recomputation of the aggregate of
Section~\ref{subsec:profile}, including seed-trust propagation, takes 1.0 to 1.6~s for $10^5$
identities and 15 to 24~s for $10^6$ on one processor core, measured on synthetic event
tables on two machines (Section~\ref{subsec:num-distributed}); it is needed once per epoch for the
reference settlement and at will for other algorithms.

\subsection{Identities and onboarding}
\label{subsec:identities}

Users are identified by a public key of a signature scheme. A key alone gives
access to reading and to the currency, but not to reputation: a
\emph{reputation-bearing} identity must be activated, which implements the cost of
identities required by C1 and C4.

\begin{enumerate}[label=\textbf{I\arabic*.},leftmargin=2.2em]
  \item \textbf{Attestation.} Activation requires an attestation by at least one
        full member, or by an institution, recorded as a non-transferable
        credential in the spirit of soulbound tokens~\cite{ohlhaver2022decentralized};
        a proof of personhood~\cite{borge2017proof} can be accepted as an
        alternative, and activation can also be earned through independent
        recognition (G2).
  \item \textbf{Vouching stake.} An attesting member locks a vouching stake
        $\sigma_v$ of her own reputation for $P_v$ epochs; if the newcomer is
        sanctioned for Sybil or collusive behavior during that period, the stake is
        burned. Each member can vouch for at most $m_v$ newcomers per year. The
        cost of a Sybil identity thus includes the expected loss of a vouching stake
        and the scarcity of vouching rights, which raises $c$ for Sybil-based
        strategies (Remark~\ref{rem:coverage}).
  \item \textbf{Probation.} During $P$ epochs, a new identity has maximal
        uncertainty, stakes capped at $\sigma_{\mathrm{prob}}$, and cannot vouch.
        A fresh identity is thus worth less than a damaged one (C4).
  \item \textbf{No monetary entry fee.} Entry fees exclude researchers with few
        resources~\cite{friedman2001social}; identity costs are social
        (attestation, vouching) rather than monetary.
\end{enumerate}

\subsection{Reputation balance, sub-indicators and profiles}
\label{subsec:profile}

We distinguish two objects. The \emph{Rep balance} is a resource (R3): it is
non-transferable, it can be put at stake, it is minted by recognized contributions
and burned by losses. The \emph{reputation profile} is an evaluation, built from
observable events in the three levels of Section~\ref{subsec:pipeline}.
Table~\ref{tab:subindicators} defines the ten sub-indicators of MagMell, grouped in
five families by the resource they mainly rely on, with the natural range of their
raw statistic, the measure of their evidence, their class (merit or screen,
Section~\ref{subsec:screens}) and their treatment when missing.

\begin{table}[!htbp]
\centering
\caption{Sub-indicators of MagMell. All raw statistics are oriented so that higher
is better; every sub-indicator is mapped to $[0,1]$ by the pipeline below.}
\label{tab:subindicators}
\footnotesize
\renewcommand{\arraystretch}{1.15}
\setlength{\tabcolsep}{3pt}
\begin{tabularx}{\textwidth}{>{\raggedright\arraybackslash}p{1.25cm}>{\raggedright\arraybackslash}p{1.75cm}L>{\raggedright\arraybackslash}p{1.65cm}>{\raggedright\arraybackslash}p{1.35cm}>{\raggedright\arraybackslash}p{1.05cm}}
\toprule
Family & Sub-indicator & Raw statistic $f_\ell$ (natural range) & Evidence $n_\ell$ & Class & If missing \\
\midrule
S1 (R2) & $z_1$ reading margin & median over the account's assessments of
  $\min(1,\bar\rho\,t/|D|)$, $t$ the delay between first access to $D$ and the
  assessment, $\bar\rho=238$ words/min~\cite{brysbaert2019many} ($[0,1]$) &
  assessments & screen & dropped \\
\addlinespace
S2 (R3) & $z_{21}$ settlement accuracy & stake-weighted mean of the agreement $a$ of
  settled assessments, A8 ($[-1,1]$) & settled assessments & merit ($R_{\mathrm{rev}}$), screen ($R_{\mathrm{auth}}$) & dropped \\
 & $z_{22}$ information & mean information score $\iota$, A7 ($[-1,1]$) &
  assessments with $\geq n_\iota$ co-assessments & merit ($R_{\mathrm{rev}}$), screen ($R_{\mathrm{auth}}$) & dropped \\
\addlinespace
S3 (R5) & $z_{31}$ independent recognition & the account's share of the minted
  recognition $m(P)$ of its publications (Section~\ref{subsec:minting}), per year of
  activity ($[0,\infty)$) & exposure time (quarters) & merit ($R_{\mathrm{auth}}$) & imputed at $\pi_0$ \\
 & $z_{32}$ out-group share & fraction of the trust-weighted recognition of each
  publication that comes from outside its communities ($[0,1]$) & recognitions & screen & dropped \\
 & $z_{33}$ yield & fraction of the account's publications older than $T_y$ that
  received at least one trust-weighted recognition ($[0,1]$) & publications older than $T_y$ & screen & dropped \\
\addlinespace
S4 (R1) & $z_{41}$ own seed trust & median over $M$ seed sets (G1) of the
  degree-normalized landing probability of short random walks from the
  seeds~\cite{cao2012aiding} ($[0,\infty)$) & degree & screen & never missing \\
 & $z_{42}$ endorser trust & median seed trust of the identities that cite or endorse
  the account, divided by the community median ($[0,\infty)$) & endorsers & screen & dropped \\
 & $z_{43}$ non-reciprocity & $1-{}$fraction of the trust-weighted recognition received
  that the account reciprocates within the window ($[0,1]$) & recognitions & screen & dropped \\
\addlinespace
S5 (R4) & $z_{51}$ stability & $1-{}$normalized divergence between short- and
  long-window values of $f_{21}$ and $f_{31}$~\cite{srivatsa2005trustguard} ($[0,1]$) &
  quarters of history ($\geq 2$) & screen & dropped \\
\bottomrule
\end{tabularx}
\end{table}

\paragraph{Pipeline.} For each account $a$ and sub-indicator $\ell$, the value
$z_\ell(a)\in[0,1]$ is computed by the following steps, in this order.
\begin{enumerate}[label=\textbf{P\arabic*.},leftmargin=2.4em]
  \item \textbf{Raw statistic and evidence.} $f_\ell(a)$, in its natural range, and
        $n_\ell(a)$ are computed over a sliding window with forgetting factor
        $\lambda$ per epoch. Evidence counts quantities that the gaming actions
        raising $\ell$ cannot cheaply inflate (Section~\ref{subsec:screens}); for
        $z_{31}$ it is exposure time, not the number of publications.
  \item \textbf{Reference population.} For a community $c$,
        $\mathcal P_\ell(c)$ is the set of activated identities of $c$ with
        $n_\ell\geq 1$: only identities on which $\ell$ is observed serve as reference.
        If $|\mathcal P_\ell(c)|<n_{\min}$, $c$ is replaced by its parent in a
        hierarchy of communities, obtained by multi-level community detection on
        the citation graph, until the threshold is met (G3).
  \item \textbf{Shrinkage.} $\tilde f_\ell(a)=\big(n_\ell f_\ell+n_0\,m_\ell(c)\big)/(n_\ell+n_0)$,
        where $n_0$ is expressed in units of the evidence of $\ell$, and the target
        $m_\ell(c)$ is the $q_0$-quantile of $f_\ell$ over $\mathcal P_\ell(c)$ for
        merit sub-indicators (a conservative prior, C4) and the median for screens
        (a neutral prior, so that scarce data do not trigger a screen).
  \item \textbf{Percentile.} $p_\ell(a;c)$ is the mid-rank percentile of
        $\tilde f_\ell(a)$ among the shrunk values $\{\tilde f_\ell(b):b\in\mathcal P_\ell(c)\}$.
  \item \textbf{Memberships.} $p_\ell(a)=\sum_c\pi_{a,c}\,p_\ell(a;c)$ over the
        communities of $a$ (G4), a linear combination of the same quantity.
  \item \textbf{Class transform.} For a merit sub-indicator, $z_\ell=p_\ell$; for a
        screen, $z_\ell=\min(1,p_\ell/\tau_s)$, and, for the screens whose honest
        values are bounded by human behavior, $z_\ell=\min(1,p_\ell/\tau_s,f_\ell/\theta_\ell)$
        with an absolute threshold $\theta_\ell$: $\theta_1$ on the reading margin
        (reading faster than twice the average rate) and $\theta_{43}$ on
        non-reciprocity (Section~\ref{subsec:screens}).
  \item \textbf{Missing values.} If $n_\ell(a)=0$: $z_{31}$ is imputed at $\pi_0$,
        defined as the $\pi_{\mathrm{rank}}$-quantile of $R_{\mathrm{auth}}$ over
        identities on which $z_{31}$ is observed, so that a fresh identity ranks at
        about $\pi_{\mathrm{rank}}$; screens are dropped, except $z_{41}$, which is
        never missing (an activated identity has at least its attestation edge).
        Whitewashing cannot escape $z_{31}$ and $z_{41}$, the two sub-indicators
        that penalize it, so the dropping rule of Section~\ref{subsec:screens} is
        respected.
\end{enumerate}

\paragraph{Role-specific aggregates.} Two aggregates are computed:
\begin{align*}
  R_{\mathrm{auth}}&=\min\big(\{z_{31}\}\cup\{\text{observed screens}\}\big),\\
  R_{\mathrm{rev}}&=\min\big(\{z_{21},z_{22}\}\cup\{\text{other observed screens}\}\big).
\end{align*}
$R_{\mathrm{auth}}$ ranks authors and their publications; $R_{\mathrm{rev}}$ weights
assessments and is undefined for identities that have not assessed. In
$R_{\mathrm{auth}}$, the accuracy and information of assessments act as screens:
unfounded criticism (bad-mouthing) is penalized, but an author is not ranked by the
quality of her assessments. Were $z_{21}$ and $z_{22}$ merit sub-indicators of
$R_{\mathrm{auth}}$, an excellent author with average assessments would lower her
rank by assessing and could avoid this by abstaining, since these sub-indicators are
dropped when missing; good assessment is rewarded instead through settlement gains
(A8) and $R_{\mathrm{rev}}$. Following Proposition~\ref{prop:nesting}, both aggregates
are minima over sub-indicators, and Theorems~\ref{thm:value} and~\ref{thm:local}
apply with screens rescaled as in Remark~\ref{rem:screens}.

The uncertainty $u$ of the profile, computed from the amount of discounted evidence
(C3), is reported alongside the aggregates and used for exploration (G6), so that
honest newcomers are shown as uncertain rather than bad. A screen below $1$ that
does not bind is only monitored; when a binding screen costs the account at least
$\delta_r$ rank points, the account is sent to human examination, which can lift the
screen. Section~\ref{subsec:num-robustness} distinguishes the corresponding shares of
honest accounts. Following Rule~4, every type of event is observed by at least two
families: assessments by S1, S2 and S5; publications and citations by S3, S4 and S5.

\subsection{Assessments, stakes and settlement}
\label{subsec:incentives}

An assessment is a tuple $(\text{target}, p, \sigma, \hat q, \text{report})$, where
the target is a publication or an account, $p\in\{+1,-1\}$ its polarity, $\sigma>0$
the staked Rep, $\hat q\in[0,1]$ the assessor's prediction of the fraction of
positive assessments the target will receive, and the report an optional text.
The following rules apply.

\begin{enumerate}[label=\textbf{A\arabic*.},leftmargin=2.4em]
  \item \textbf{Admissibility.} The assessor has read the target (B1); she may not
        assess her own work, the work of her co-authors, of identities she vouched
        for or that vouched for her.
  \item \textbf{Stake limits.} $\sigma\leq\alpha\cdot\mathrm{Rep}_{\mathrm{free}}$; a
        negative assessment requires $\sigma\geq\kappa\,\sigma_{\min}$ with
        $\kappa>1$, so that criticism requires reputational capital, which Sybils
        lack.
  \item \textbf{Escrow.} The stake is locked for at most $T_e$ epochs, a horizon long
        enough for slow signals (citations) to accumulate (C3, Rule~4). If the
        uncertainty of the outcome falls below $u_{\mathrm{rel}}$ earlier, the
        assessment is settled early.
  \item \textbf{Reference forecast.} At assessment time, the reference algorithm records
        $\hat O\in[-1,1]$, a forecast of the outcome $O$ of A5 \emph{on the scale of
        $O$}, calibrated on past settlements, from the merit sub-indicators of the
        authors (not from their screened aggregate) and the base rate of the
        community; for a new publication it has high uncertainty. Rewarding the
        difference with a reference forecast, as market scoring rules
        do~\cite{hanson2003combinatorial}, means that endorsing what is already well
        known earns nothing. A forecast on another scale, or one that includes the
        screens, biases settlement (Section~\ref{subsec:num-dynamics}).
  \item \textbf{Independent outcome.} At settlement, the reference algorithm computes an outcome
        $O\in[-1,1]$ for the target from the recognitions it received since the
        assessment from identities outside the assessor's community and not linked to
        her by vouching or co-authorship, and from replication reports when there are
        any. Recognitions are counted \emph{in excess of those expected from the
        independent exposure} of the target (its accesses by the same identities,
        times the mean recognition rate): raw counts would reward assessments that
        merely raise the visibility of their target, a self-fulfilling effect
        measured in Section~\ref{subsec:num-dynamics}. This independence rule
        prevents a cartel from corroborating its own assessments. The outcome never
        depends on the other assessments of the same target.
  \item \textbf{Unresolved outcomes.} If, at $T_e$, the target has had fewer than
        $d_{\min}$ independent accesses since the assessment, or the uncertainty of $O$
        is above $u_{\mathrm{res}}$, or $|O-\hat O|<\epsilon_O$, the stake is returned
        without gain or loss. The first condition matters most for criticism: criticized
        papers are little read, so that their outcome is often not measurable
        (Section~\ref{subsec:num-dynamics}). An assessment, critical or not, that the future has neither
        confirmed nor refuted is therefore not penalized.
  \item \textbf{Information score.} When the target has received at least $n_\iota$
        assessments, an information score $\iota\in[-1,1]$ is computed from their
        polarities and predictions $\hat q$ with a rule of the Bayesian-truth-serum
        type~\cite{prelec2004bayesian,miller2005eliciting}; otherwise $\iota=0$.
  \item \textbf{Asymmetric settlement.} Let $a=\clip_{[-1,1]}\big(p\,(O-\hat O)\big)$
        be the agreement between the assessment and the outcome. The assessor
        recovers
        \[
          \sigma\,\big(1+\beta\,a+\beta_\iota\,\iota\big)\ \text{ if } a\geq 0,
          \qquad
          \sigma\,\big(1+\gamma_p\,\beta\,a+\beta_\iota\,\iota\big)\ \text{ if } a<0,
        \]
        with $\gamma_{+1}=1$ and $\gamma_{-1}=\gamma>1$: unfounded criticism costs more
        than unfounded praise (C4). The condition $\gamma\beta+\beta_\iota\leq 1$
        ensures non-negative returns.
  \item \textbf{Burning and minting.} What is lost is burned, and what is gained is
        minted: no identity profits from the loss of another. This removes the
        incentive to provoke others' losses and the possibility of transferring
        reputation between colluding accounts through deliberately lost stakes.
  \item \textbf{Pseudonymous accountability.} An assessor may sign with a
        per-assessment pseudonym cryptographically linked to her identity, so that
        settlement is credited to her record without revealing who wrote the
        assessment, as DecSci also aims to allow~\cite{tenorio2019towards}. The
        cryptographic construction is left open: it must combine unlinkability with
        settlement, stake limits and sanctions, for which anonymous
        credentials~\cite{camenisch2001efficient} are a natural building block but
        not a complete solution. Until such a construction exists, assessments are
        signed with the assessor's key, and protection against retaliation is
        \emph{not} guaranteed.
\end{enumerate}

\begin{proposition}[Incentives of the settlement rule]\label{prop:settlement}
Consider a risk-neutral assessor, ignore the information score ($\beta_\iota=0$) and
rule A6, and let $\Delta=O-\hat O$, assumed in $[-1,1]$, with
$\Delta^+=\max(\Delta,0)$, $\Delta^-=\max(-\Delta,0)$ and expectations
$\mu^\pm=\mathbb{E}[\Delta^\pm]$ under her beliefs. Her expected net return is
$\sigma\beta(\mu^+-\mu^-)$ for a positive assessment and $\sigma\beta(\mu^-
-\gamma\mu^+)$ for a negative one. Hence she endorses only if she expects the
target to do better than its current reputation predicts ($\mathbb{E}[\Delta]>0$),
criticizes only if $\mu^->\gamma\mu^+$, and abstains otherwise. Since $\gamma>1$,
criticism requires stronger evidence than endorsement.
\end{proposition}

\begin{proof}
For $p=+1$, $a=\Delta$ and the net return is $\sigma\beta\Delta^+-\sigma\beta\Delta^-$.
For $p=-1$, $a=-\Delta$: the net return is $\sigma\beta\Delta^-$ when $\Delta\leq 0$
and $-\sigma\gamma\beta\Delta^+$ when $\Delta>0$. Take expectations.
\end{proof}

\paragraph{What Proposition~\ref{prop:settlement} establishes, and what it does
not.} The proposition is a decision-theoretic statement about one assessor facing
an exogenous outcome. It shows that the settlement rule points in the right
direction (informed endorsement and informed criticism pay, uninformed ones do
not, and criticism requires stronger evidence) and that the reference forecast
removes the gain from endorsing what is already known. It does \emph{not} show that
the whole mechanism makes honest assessment an equilibrium. We list the gaps.
\begin{itemize}
  \item \emph{Risk aversion.} With a concave utility, the certainty equivalent of
        the net return is below its expectation. The returns of criticism are more
        dispersed on the downside ($\gamma>1$), so risk aversion deters criticism
        more than endorsement and compounds the asymmetry; if criticism becomes too
        rare, $\gamma$ must be lowered or the variance reduced (stake caps,
        rule A6). Table~\ref{tab:settlement} quantifies this effect in a synthetic
        setting.
  \item \emph{Delays.} If the assessor discounts reputation by a factor $\delta_t$
        per epoch, a positive assessment is worth making only if
        $\delta_t^{T_e}(1+\beta\,\mathbb{E}[\Delta])\geq 1$, i.e.\
        $\mathbb{E}[\Delta]\geq(\delta_t^{-T_e}-1)/\beta$: the participation threshold
        grows with the escrow horizon. Early settlement (A3) mitigates this tension
        between Rule~4 and participation.
  \item \emph{Few assessments.} Scoring rules of the peer-prediction family need
        several reports on the same target; below $n_\iota$ the information score is
        switched off (A7) and settlement relies on the outcome only, which no longer
        penalizes herding before the outcome is known.
  \item \emph{Noisy or biased outcomes.} Mean-zero noise on $O$ leaves expected
        returns unchanged, since returns are piecewise linear, but increases their
        variance, which matters under risk aversion. A bias in $O$ shifts incentives
        in its direction. The independence rule (A5) removes direct links between
        assessors and the identities that produce $O$, but not correlations between
        communities, e.g.\ between schools of thought that systematically disagree.
        Rule A6 limits the damage for minority positions that remain unresolved.
  \item \emph{Retaliation.} The proposition does not model retaliation. Rule A10
        will reduce it only once constructed, and only insofar as pseudonyms cannot be
        linked to identities through style, timing or topic.
  \item \emph{Equilibrium.} In the real system, $O$ depends on the behavior of other
        identities, so incentives interact. Peer-prediction mechanisms make honest
        reporting \emph{an} equilibrium~\cite{miller2005eliciting}, not necessarily
        the only one; uninformative coordinated equilibria may coexist. The reference
        forecast makes uniform conformism unprofitable, and a coalition large enough
        to move $O$ is a gaming action in the sense of Section~\ref{sec:theory}, but a
        full equilibrium analysis of the mechanism is left for future work. The
        dynamic simulation of Section~\ref{subsec:num-dynamics}, in which outcomes are
        endogenous, is a first step in that direction.
\end{itemize}

\subsection{Minting reputation from publications}
\label{subsec:minting}

At each epoch, a publication $P$ mints Rep for its authors in proportion to
\[
  m(P)=\sum_{x\ \text{cites or uses}\ P} \rho(x)\, b(x,P),
\]
where $\rho(x)$ is the seed-propagated trust of the citing identity (S4) and
$b(x,P)$ a bridging coefficient. The communities of a publication are inferred
from its own references and citers, not from its authors (G4); $b(x,P)=1$ when $x$
is outside the communities of $P$, and $b(x,P)=b_{\mathrm{in}}(n_c)$ when $x$ belongs
to a community of size $n_c$ of $P$, with
\[
  b_{\mathrm{in}}(n_c)=b_{\min}+(1-b_{\min})\,\frac{n_{\min}}{n_c+n_{\min}},
\]
which tends to $b_{\min}$ for large communities and to $1$ for very small ones
(G3). Self-citations and citations along vouching or co-authorship links are
excluded. Downloads are recorded but not rewarded, since they are easy to automate.
Reputation is therefore created by recognition from trusted and, preferably,
distant identities, and destroyed by failed stakes and by the forgetting factor.

\subsection{Proposed governance and fairness rules}
\label{subsec:governance}

The side effects identified in Section~\ref{subsec:sideeffects} affect how the
mechanisms work, not only their adoption. We therefore propose the following rules.
They are proposed mechanisms, not established solutions, and none of them is a
guarantee of fairness: each introduces parameters and strategic opportunities of its
own, summarized in Table~\ref{tab:governance}, and their properties remain to be
evaluated.

\begin{enumerate}[label=\textbf{G\arabic*.},leftmargin=2.4em]
  \item \textbf{Plural, rotating seeds.} Trust is propagated from $M$ disjoint seed
        sets rather than one. Seeds are drawn by stratified sortition, across
        communities and types of institutions, for a limited term; a fraction $f_s$
        of each set is renewed every year, and the seed sets are public. Drawing
        seeds only among members with high reputation and low uncertainty would
        limit capture by a small group but could progressively reproduce the
        preferences of already dominant groups; a fraction $f_r$ of the seats is
        therefore drawn uniformly among all full members, irrespective of their
        reputation. The sub-indicator $z_{41}$ is the median over the $M$ sets, so
        that capturing fewer than half of the sets cannot move it outside the range
        of values given by the uncaptured sets. A seed whose propagated trust
        concentrates on clusters later flagged for collusion loses its status. Users
        remain free to choose their own seeds in tier~2 (C5).
  \item \textbf{Plural entry.} Activation does not depend on a single type of actor:
        it can result from vouching by any full member (within the quota $m_v$),
        from an institutional attestation, from a proof of
        personhood~\cite{borge2017proof}, or be \emph{earned}: an unactivated key that
        obtains, over $P_e$ epochs, trust-weighted recognition above $\theta_e$ from at
        least $q_e$ distinct communities is activated without attestation. The share
        of activations per institution is published, so that gatekeeping by
        established institutions can be detected.
  \item \textbf{Small fields.} Reference populations are taken in the smallest
        community with at least $n_{\min}$ observed identities, walking up a
        hierarchy of communities (P2); shrinkage uses the same population (P3); and
        in-group recognition is discounted less in small communities
        ($b_{\mathrm{in}}(n_c)$ above).
  \item \textbf{Multidisciplinary authors.} An author has a mixed membership
        $\pi_{a,c}\geq 0$, $\sum_c\pi_{a,c}=1$, inferred from her publications and
        citations rather than declared, and combined linearly across communities
        (P5). Since the classification of interdisciplinary researchers may be
        unstable, memberships are exponentially smoothed over time and memberships
        below a threshold $\pi_{\min}$ are ignored. Out-group recognition is defined
        relative to the communities of each publication, so that interdisciplinary
        work is not penalized for crossing its author's fields.
  \item \textbf{Minority critiques.} An assessment is never settled against the
        consensus of the other assessors, but against an independent, long-term
        outcome (A5); an unresolved assessment is returned without loss (A6); a
        critique that proves right earns more than an endorsement of consensus,
        because the reference forecast encoded the consensus (A4); and the
        information score rewards answers that are more common than predicted, which
        can favor minority positions~\cite{prelec2004bayesian}. Critiques are also
        publications: they can be cited and earn recognition.
  \item \textbf{Recent publications and exploration.} The reference forecast of a
        new publication is a prior with high uncertainty, so that informed early
        assessments carry the largest expected agreement (A4). The default search
        ranking reserves a share $\varepsilon_x$ of the positions for items with high
        uncertainty (recent publications, newcomers) selected by an
        upper-confidence-bound rule ($R_{\mathrm{auth}}+\kappa_x u$), in order to
        counter cumulative advantage~\cite{merton1968matthew} and to give new and
        heterodox work a chance to be seen. Exploration improves visibility; it does
        not by itself remove the structural position of newcomers, which is set by
        $\pi_0$ (Sections~\ref{subsec:num-profiles} and~\ref{subsec:num-newcomers}).
  \item \textbf{Capital and risk in assessments.} The settlement rule selects
        assessors by their tolerance to risk and by their reputational capital
        (Section~\ref{subsec:num-settlement}): informative criticism may become the
        privilege of those who can afford a loss, which biases the opinions actually
        expressed towards established actors. We propose three mitigations: stakes
        and returns are expressed as fractions of the assessor's free Rep, so that
        relative exposure is the same for all (A2); each full member in the lower half
        of the Rep distribution receives a small yearly allowance of $k_c$
        \emph{protected} criticisms, whose losses are capped at the stake floor and
        funded by minting; and the distribution of assessments and of criticisms by
        capital tier, seniority and community is published, with $\gamma$ and
        $\kappa$ revised if the share of criticisms from the lower tiers falls below a
        threshold.
\end{enumerate}

\begin{table}[!htbp]
\centering
\caption{Proposed governance rules: parameters, open risks, and how to evaluate
them.}
\label{tab:governance}
\footnotesize
\renewcommand{\arraystretch}{1.15}
\setlength{\tabcolsep}{3pt}
\begin{tabularx}{\textwidth}{>{\raggedright\arraybackslash}p{1.5cm}>{\raggedright\arraybackslash}p{1.9cm}LL}
\toprule
Rule & Parameters & Open risks & How to evaluate \\
\midrule
G1 seeds & $M$, $f_s$, $f_r$ & drift of the seeds towards the preferences of dominant
  groups; capture of a majority of seed sets & diversity of seed sets over time;
  sensitivity of rankings to the choice of seed sets \\
G2 entry & $m_v$, $P_e$, $\theta_e$, $q_e$ & Sybil entry through the earned path;
  slow activation of isolated researchers & time to activation by group; success rate
  of Sybil entry in simulation \\
G3 small fields & $n_{\min}$, $b_{\min}$, hierarchy & arbitrary hierarchy; pooled
  normalization that still disadvantages niche fields & distribution of ranks as a
  function of community size \\
G4 memberships & smoothing rate, $\pi_{\min}$ & unstable classification of
  interdisciplinary researchers & volatility of ranks under perturbations of
  community detection \\
G5 critiques & $u_{\mathrm{res}}$, $\epsilon_O$ & outcomes that remain unresolved, so
  that being right early is never rewarded & settlement of later-vindicated critiques,
  in replay on historical controversies \\
G6 exploration & $\varepsilon_x$, $\kappa_x$ & capture of exploration slots by fake
  newcomers & visibility of newcomers; share of exploration slots captured by Sybils
  in simulation \\
G7 capital and risk & $k_c$, capital tiers & protected criticisms exploited by Sybils
  or by coordinated critics & share of criticisms, and of informative criticisms, by
  capital tier and risk attitude, in simulation with heterogeneous assessors \\
\bottomrule
\end{tabularx}
\end{table}

\subsection{Parameters}
\label{subsec:parameters}

Table~\ref{tab:parameters} lists the parameters of MagMell. The values are
indicative orders of magnitude, informed for some of them by the
analyses of Section~\ref{sec:numerics}; their calibration is part of the
evaluation.

\begin{table}[!htbp]
\centering
\caption{Parameters of MagMell. The reading budget corresponds to about three
hours of reading per day, five days a week, at 238 words per
minute~\cite{brysbaert2019many}. The screen tolerance is written $\tau_s$ to
distinguish it from the epoch length.}
\label{tab:parameters}
\footnotesize
\renewcommand{\arraystretch}{1.05}
\begin{tabularx}{\textwidth}{>{$}l<{$}L>{\raggedright\arraybackslash}p{2.7cm}p{1.15cm}}
\toprule
\text{Symbol} & Meaning & Indicative value & Serves \\
\midrule
\tau & epoch length & 1 week & C3 \\
\lambda & forgetting factor per epoch & $0.99$ (half-life $\approx 69$ epochs) & C3 \\
T_e & maximal escrow horizon & 52--104 epochs & C3, Rule 4 \\
u_{\mathrm{rel}},\ u_{\mathrm{res}},\ \epsilon_O & early settlement, unresolved thresholds & $0.2$, $0.6$, $0.05$ & C2, G5 \\
P & probation period & 26 epochs & C1, C4 \\
\sigma_v,\ P_v,\ m_v & vouching stake, duration, quota per year & to calibrate; 52 epochs; 5 & C1 \\
P_e,\ \theta_e,\ q_e & earned entry: duration, threshold, communities & 26 epochs; to calibrate; 2 & G2 \\
B_r & reading budget & $\approx 2\times10^5$ words/epoch & C1 \\
\rho_{\max} & reading-rate bound & $2\times 238$ words/min & C1 \\
\alpha & maximal stake fraction & $0.1$ & C2 \\
\kappa & minimal stake multiplier for criticism & $2$ & C4 \\
\beta,\ \beta_\iota & settlement and information gains & $0.3$, $0.1$ & C2 \\
\gamma & loss multiplier for unfounded criticism & $1.5$ ($\gamma\beta+\beta_\iota=0.55\leq1$) & C4 \\
n_\iota & minimal number of co-assessments & 5 & C2 \\
n_0,\ q_0 & shrinkage weight; merit prior quantile & 5 (units of evidence); $0.25$ & P3 \\
\tau_s & screen tolerance & $0.02$ & C6 \\
\theta_1,\ \theta_{43} & absolute screen thresholds & $0.5$; $0.5$ & C6 \\
\delta_r & rank loss that triggers human examination & 10 points & C6 \\
d_{\min} & independent accesses needed to settle & 10 & C2 \\
k_c & protected criticisms per year (lower half of Rep) & 2 & G7 \\
\pi_{\mathrm{rank}} & rank of a fresh identity & $0.25$ & C4 \\
T_y & age of publications counted in the yield & 26 epochs & C6 \\
b_{\min},\ n_{\min} & in-group bridging floor; minimal reference population & $0.2$; 50 & C5, G3 \\
M,\ f_s,\ f_r & seed sets; yearly renewal; random seats & 5; 20\%; 20\% & C5, G1 \\
\pi_{\min} & minimal membership weight & $0.1$ & G4 \\
\varepsilon_x,\ \kappa_x & exploration share; uncertainty bonus & $0.1$; 1 & G6 \\
\bottomrule
\end{tabularx}
\end{table}

\subsection{Attack-by-sub-indicator matrix}
\label{subsec:matrix}

Table~\ref{tab:matrix} gives the hypothesized sign pattern of the gaming matrix $B$
at the level of sub-indicators, which by Proposition~\ref{prop:nesting} is the
matrix $C$ of the theory. Effects are those on the account that benefits from the
attack; whitewashing and identity purchase are compared with continuing with a
damaged identity and with creating a new one, respectively.

\begin{table}[!htbp]
\centering
\caption{Hypothesized sign pattern of the gaming matrix at the level of
sub-indicators, to be validated. ``$+$'': the attack raises the sub-indicator;
``$-$'': it lowers it (collateral sensitivity); ``$0$'': no first-order effect.}
\label{tab:matrix}
\small
\setlength{\tabcolsep}{3.2pt}
\begin{tabular}{l|c|cc|ccc|ccc|c}
\toprule
 & S1 & \multicolumn{2}{c|}{S2} & \multicolumn{3}{c|}{S3} & \multicolumn{3}{c|}{S4} & S5 \\
Attack & $z_1$ & $z_{21}$ & $z_{22}$ & $z_{31}$ & $z_{32}$ & $z_{33}$ & $z_{41}$ & $z_{42}$ & $z_{43}$ & $z_{51}$ \\
\midrule
Sybil farm              & $0$ & $0$ & $0$ & $+$ & $+$ & $+$ & $0$ & $-$ & $0$ & $-$ \\
Whitewashing            & $0$ & $+$ & $+$ & $-$ & $0$ & $0$ & $-$ & $0$ & $0$ & $0$ \\
Collusion ring          & $0$ & $0$ & $0$ & $+$ & $-$ & $+$ & $0$ & $0$ & $-$ & $0$ \\
Bad-mouthing            & $0$ & $-$ & $0$ & $+$ & $0$ & $0$ & $0$ & $0$ & $0$ & $0$ \\
Oscillation             & $0$ & $0$ & $0$ & $+$ & $0$ & $0$ & $0$ & $0$ & $0$ & $-$ \\
Identity purchase       & $0$ & $+$ & $+$ & $+$ & $0$ & $0$ & $+$ & $0$ & $0$ & $-$ \\
Herding                 & $0$ & $0$ & $-$ & $0$ & $0$ & $0$ & $0$ & $0$ & $0$ & $0$ \\
Automated assessments   & $-$ & $0$ & $-$ & $0$ & $0$ & $0$ & $0$ & $0$ & $0$ & $0$ \\
Generated paper flood   & $0$ & $0$ & $0$ & $+$ & $0$ & $-$ & $0$ & $0$ & $0$ & $0$ \\
\bottomrule
\end{tabular}
\end{table}

The table is a hypothesis, tested for three attacks in
Section~\ref{subsec:num-gaming}, and we do not claim that the admissible sets of MagMell
are non-empty: by Theorems~\ref{thm:value} and~\ref{thm:local}, this depends on the
magnitudes of the entries and on the costs, which must be estimated
(Section~\ref{subsec:nonempty}). What the table does provide is a list of
refutable predictions, and a diagnosis.
\begin{itemize}
  \item Rows without ``$+$'' (herding, automated assessments) are not profitable at
        all: the reference forecast (A4) and the reading margin remove their gains.
  \item Rows with both signs are candidates for a collateral vise. The ``$+$'' of
        bad-mouthing on $z_{31}$ is an indirect effect of percentile normalization:
        lowering competitors raises one's own percentile
        (Section~\ref{subsec:limits}). The ``$+$'' of Sybil farms on $z_{32}$ reflects
        that Sybils may pose as an out-group; weighting recognition by seed trust
        keeps this entry small.
  \item The flood of generated papers is the most problematic row. In the first
        version of the matrix, without the yield screen $z_{33}$, it had a ``$+$''
        and no ``$-$'': it was contained only by its cost and by the small magnitude
        of its effect. For an account whose weakest sub-indicator is $z_{31}$, the
        required cost of flooding is its effect on $z_{31}$
        (Section~\ref{subsec:nonempty}), so the incentive to flood is concentrated on
        accounts bottlenecked on recognition. The yield screen adds a ``$-$'', but its
        strength depends on the probability that a generated paper obtains
        recognition, which is the key empirical unknown. Section~\ref{subsec:num-gaming}
        estimates the matrix for three of these attacks.
\end{itemize}

\section{Evaluation on synthetic data}
\label{sec:numerics}

This section reports numerical analyses of MagMell on synthetic data. Their purpose
is to test the internal consistency of the design: whether its mechanisms do what
Sections~\ref{sec:theory} and~\ref{sec:multirep} claim under explicit assumptions,
where they fail, and what the orders of magnitude are. They do not estimate the
behavior of MagMell on real data: every distribution below is an assumption, and no
figure is an empirical estimate. All results reported here come from Studies~A to~D
and from a decision-theoretic analysis, which have been run; the evolutionary
simulation specified in Section~\ref{subsec:protocol} and Appendix~\ref{app:ga} has
not been run, and no result comes from it. Study~A measures the robustness of the static
pipeline and its false positives on honest populations
(Sections~\ref{subsec:num-profiles} and~\ref{subsec:num-robustness}); a
decision-theoretic analysis studies assessors with different attitudes to risk
(Section~\ref{subsec:num-settlement}); Study~B estimates the gaming matrix of three
attacks and compares aggregation rules on the same scenarios
(Section~\ref{subsec:num-gaming}); Study~C is a dynamic simulation in which authors,
readers and assessors interact and outcomes are endogenous
(Sections~\ref{subsec:num-dynmodel} to~\ref{subsec:num-dynamics}); Study~D evaluates
the distributed execution of the protocol: dissemination, equivocation, costs and
partial replicas (Section~\ref{subsec:num-distributed}). Several results
led to changes of the design, which are collected in
Section~\ref{subsec:num-corrections} and already included in
Section~\ref{sec:multirep}. Section~\ref{sec:discussion} states what the
analyses establish and what they cannot, and specifies the first full simulation.

\subsection{Protocol and reproducibility}
\label{subsec:num-protocol}

\paragraph{Honest population.} Studies~A and~B share one generator. An honest
identity has a latent quality $\theta\sim\mathcal N(0,1)$, a seniority uniform between
26 and 520 weekly epochs, a publication rate drawn from a gamma distribution (mean
$0.03$ per epoch) and an assessment rate drawn from a beta distribution (mean $0.075$
per epoch). Each publication receives a Poisson number of recognitions with mean
$e^{1+0.8\theta}$ multiplied by a gamma factor of shape $k=0.7$, so that many papers
are never recognized. A recognition has a trust weight uniform on $[0.5,1.5]$, comes
from outside the community with a probability drawn per identity from
$\mathrm{Beta}(3,3)$, and is reciprocated with a probability drawn from
$\mathrm{Beta}(2,8)$. Integrity statistics and seed trust do not depend on $\theta$.
All sub-indicators of Table~\ref{tab:subindicators} are computed from these events as
specified in Section~\ref{subsec:profile}.

\paragraph{Scenarios.} The baseline is varied one factor at a time: a weak (slope
$0.4$) or strong ($1.2$) quality signal in recognition; high ($k=0.3$) or low ($k=2$)
overdispersion of recognition; and a \emph{diverse} honest population in which 30\%
of the identities have one atypical but legitimate trait: fast reading, frequent
reciprocation, bursty activity, an in-group audience, or a low yield.

\paragraph{Rules compared.} On the same sub-indicators and the same draws: the
minimum of MagMell (pipeline P1--P7, merit and screens); the arithmetic mean with
equal weights; the geometric mean, with a floor of $0.005$; merit only ($z_{31}$);
and, as a reference, the \emph{naive} pipeline of the first version of this paper
(every sub-indicator a percentile, missing values imputed at the first quartile,
minimum).

\paragraph{Repetitions and uncertainty.} Study~A uses 10 independent populations of
3000 identities per scenario; Study~B uses 5 populations and 200 paired draws per
cell; Study~C uses 10 seeds per condition. Tables report means across seeds and, where
indicated, standard deviations across seeds. These measure the variability between
synthetic populations, not the uncertainty on the assumptions, which is explored by
the scenarios. Attacks are evaluated with common random numbers: the attacked and the
honest versions of a profile share every honest random draw, so that a difference is
due to the attack alone.

\paragraph{Reproducibility.} The scripts, parameters and seeds are provided as
ancillary files of this preprint (directory \texttt{anc/}), with the dependency versions (Python~3.13, NumPy~2.5,
SciPy~1.18), instructions, and a single script that reproduces every table of this
section in about twenty minutes on a single processor core. Each table names the script that
produces it. The ancillary files also contain numerical checks of
Theorems~\ref{thm:value} and~\ref{thm:local} and of
Proposition~\ref{prop:settlement} on random instances (3000 linear and 1500
nonlinear instances for the theorems); none produced a counterexample. The package
was also run on a second machine with older versions (Python~3.9, NumPy~2.0,
SciPy~1.13). Studies~A, B and~D and the numerical checks gave identical results, apart
from wall-clock times. The random streams of Study~C depend on the NumPy version:
Tables~\ref{tab:newcomers} and~\ref{tab:reflexivity} then differ from the reported
values by amounts within the variation across seeds, and the conclusions drawn from
them in Sections~\ref{subsec:num-newcomers} and~\ref{subsec:num-dynamics} are
unchanged.

\paragraph{Use of generative artificial intelligence.} The author used a large language model (Claude,
Anthropic) to assist with literature search, drafting and editing of the text, and
writing of the simulation code. The author reviewed and verified all content, checked
the results by rerunning every script, and takes full responsibility for the
manuscript.

\subsection{Typical profiles and pipeline corrections}
\label{subsec:num-profiles}

Table~\ref{tab:profiles} reports the median rank of $R_{\mathrm{auth}}$ for eight
typical profiles, T1 to T7 and a variant T5$'$, with the naive pipeline and with the
revised one at two screen tolerances.

\begin{table}[!htbp]
\centering
\caption{Median rank of $R_{\mathrm{auth}}$ (percent) of typical profiles over 100
draws, mean $\pm$ standard deviation across 10 populations, baseline scenario.
Script: \texttt{robustness.py}.}
\label{tab:profiles}
\footnotesize
\renewcommand{\arraystretch}{1.1}
\begin{tabularx}{\textwidth}{Lccc}
\toprule
Profile & Naive & Revised, $\tau_s=0.10$ & Revised, $\tau_s=0.02$ \\
\midrule
T1 established, strong ($\theta=1.5$)        & $75\pm 6$ & $90\pm 2$ & $89\pm 2$ \\
T2 established, median ($\theta=0$)          & $54\pm 5$ & $64\pm 3$ & $62\pm 2$ \\
T3 newcomer, vouched, inactive               & $89\pm 1$ & $27\pm 0$ & $27\pm 0$ \\
T4 newcomer, one strong paper                & $64\pm 3$ & $71\pm 3$ & $64\pm 6$ \\
T5 narrow field, $b_{\mathrm{in}}=b_{\min}$  & $44\pm 2$ & $49\pm 3$ & $41\pm 3$ \\
T5$'$ narrow field, $b_{\mathrm{in}}=0.9$    & $40\pm 3$ & $54\pm 4$ & $54\pm 3$ \\
T6 two papers, high impact                   & $83\pm 3$ & $87\pm 3$ & $93\pm 1$ \\
T7 prolific non-assessor                     & $80\pm 4$ & $87\pm 1$ & $87\pm 1$ \\
\bottomrule
\end{tabularx}
\end{table}

The naive pipeline exhibits the three flaws discussed in
Section~\ref{subsec:screens}. Because missing values are imputed on the wrong scale,
an inactive newcomer (T3) ranks at the 89th percentile, above an established strong
researcher; because integrity sub-indicators are percentiles of honest noise, the
strong researcher ranks only at the 75th percentile. In the revised pipeline with
$\tau_s=0.02$, merit binds for 93\% of honest accounts, the median aggregate is
$0.42$, and a fresh identity ranks at the 27th percentile, as set by
$\pi_{\mathrm{rank}}$. The bridging correction of G3 raises a researcher in a narrow
field by 12.5 points on average (T5 against T5$'$), and authors who publish little but with
high impact (T6), or who never assess (T7), are not penalized.

\paragraph{Whitewashing is not eliminated.} In every scenario, 23\% of honest
accounts would rank higher with a fresh identity than with their own. The design does
not remove this incentive. It makes its size explicit (it is set by
$\pi_{\mathrm{rank}}$) and moves its cost to the activation mechanisms of
Section~\ref{subsec:identities}: vouching with liability, probation and earned entry.
Lowering $\pi_{\mathrm{rank}}$ reduces the share of accounts tempted, but lowers the
starting rank of every honest newcomer (Section~\ref{subsec:num-newcomers}); a prior
at the median would make about half of the honest accounts better off with a new key.
Whether identity costs deter whitewashing is therefore a property of onboarding, not
of aggregation, and must be evaluated as such.

\subsection{Robustness and false positives}
\label{subsec:num-robustness}

We define a funnel of false positives on honest accounts: FP1, at least one screen
is below~$1$; FP2, a screen binds, that is, it lowers the aggregate below the merit
level; FP3, the binding screen costs at least $\delta_r=10$ rank points, and the
account is sent to human examination (Section~\ref{subsec:profile}); FP4, the account
is wrongly sanctioned, assuming that examination clears an honest account with
probability $0.9$, an assumption. Table~\ref{tab:robustness} reports the funnel and
the whitewashing statistics in the six scenarios, and Table~\ref{tab:rules} the rank
correlation between the aggregate and the latent quality of honest accounts under
each rule.

\begin{table}[!htbp]
\centering
\caption{False-positive funnel (percent of honest accounts), median rank loss of the
accounts with a binding screen (points), rank of a fresh identity (percent) and share
of honest accounts that would gain by whitewashing (percent). Means across 10
populations; standard deviations are at most $0.3$ points for the shares ($0.5$ for the
examination rate of atypical identities) and $2.8$ points for the loss. Script: \texttt{robustness.py}.}
\label{tab:robustness}
\footnotesize
\renewcommand{\arraystretch}{1.1}
\begin{tabularx}{\textwidth}{Lccccccc}
\toprule
Scenario & FP1 & FP2 & FP3 & FP4 & Loss & Fresh rank & Whitewash \\
\midrule
Baseline                      & 15.0 & 6.9 & 5.1 & 0.51 & 37 & 27.1 & 23.0 \\
Weak quality signal           & 15.1 & 7.2 & 5.4 & 0.54 & 35 & 27.1 & 23.0 \\
Strong quality signal         & 14.8 & 6.8 & 5.1 & 0.51 & 37 & 27.1 & 23.0 \\
High overdispersion           & 14.8 & 7.2 & 5.4 & 0.54 & 35 & 27.1 & 23.0 \\
Low overdispersion            & 15.0 & 7.0 & 5.2 & 0.52 & 35 & 27.2 & 22.9 \\
Diverse honest population     & 14.8 & 7.1 & 5.3 & 0.53 & 37 & 27.1 & 23.0 \\
\quad atypical identities     &      &     & 9.5 &      &    &      &      \\
\quad typical identities      &      &     & 3.5 &      &    &      &      \\
\bottomrule
\end{tabularx}
\end{table}

\begin{table}[!htbp]
\centering
\caption{Spearman correlation between the aggregate and the latent quality $\theta$
of honest accounts, mean across 10 populations (standard deviations at most $0.023$).
Script: \texttt{robustness.py}.}
\label{tab:rules}
\footnotesize
\renewcommand{\arraystretch}{1.1}
\begin{tabularx}{\textwidth}{Lccccc}
\toprule
Scenario & Minimum & Mean & Geometric mean & Merit only & Naive \\
\midrule
Baseline                  & 0.51 & 0.62 & 0.57 & 0.56 & 0.31 \\
Weak quality signal       & 0.31 & 0.49 & 0.46 & 0.33 & 0.26 \\
Strong quality signal     & 0.63 & 0.69 & 0.64 & 0.69 & 0.33 \\
High overdispersion       & 0.43 & 0.56 & 0.51 & 0.47 & 0.27 \\
Low overdispersion        & 0.56 & 0.65 & 0.60 & 0.61 & 0.32 \\
Diverse honest population & 0.50 & 0.61 & 0.56 & 0.55 & 0.29 \\
\bottomrule
\end{tabularx}
\end{table}

\paragraph{False-positive funnel.} The funnel hardly varies across scenarios. This is a
property of percentile screens: they flag a share of the reference population fixed
by $\tau_s$ and by the number of screens, whatever the population. The false-positive
rate is therefore predictable and tunable, but screens do not adapt to the number of
attackers (Section~\ref{subsec:screens}). Among honest accounts, 15\% have a screen
below~$1$, but only 7\% are affected, because for the others merit is lower still;
5\% lose at least 10 rank points and are sent to examination; with the assumed
accuracy of examination, 0.5\% are wrongly sanctioned. The accounts affected lose
substantially, with a median loss of 35 to 37 rank points. The burden is not evenly
shared: in
the diverse population, atypical honest identities are sent to examination 2.7 times
as often as typical ones (9.5\% against 3.5\%). Atypical but legitimate behavior is
what screens are built to detect; governance must therefore monitor the rate of
examination by type of behavior, not only its global rate (Table~\ref{tab:governance}).

\paragraph{The price of robustness.} In every scenario, the arithmetic mean ranks
honest accounts best and the minimum ranks them worst of the four rules, ahead only of
the naive pipeline; the geometric mean and merit only lie in between, in an order that
depends on the scenario. The minimum loses $0.06$ to $0.18$ of correlation against
the mean, the most when the quality signal is weak: the weaker the merit signal, the
more the minimum pays for discarding the information that the other sub-indicators
carry about quality. Section~\ref{subsec:num-gaming} shows what this price buys.

\subsection{Settlement under risk attitudes}
\label{subsec:num-settlement}

We simulated assessors facing targets whose true deviation from the reference
forecast is $\Delta\sim\mathcal N(0,0.3^2)$, so that targets are as often over- as
under-rated. Each assessor observes $\Delta$ with Gaussian noise of standard
deviation $0.3$, the settled outcome adds noise of standard deviation $0.15$, 20\% of
assessments remain unresolved (A6), as do those whose observed deviation is below
$\epsilon_O=0.05$, and the escrow costs 5\% of the stake
($\delta_t^{T_e}=0.95$). Assessors endorse, criticize or abstain so as to maximize
expected utility, with $\beta=0.3$ and $\beta_\iota=0$. We compared a risk-neutral
assessor, two assessors with constant absolute risk aversion (CARA) $A=2$ and $A=10$ per unit of
stake, and a loss-averse assessor (losses weighted $2.25$). With $A=\rho\alpha$, usual
relative risk aversion coefficients ($\rho$ between 1 and 5) and $\alpha=0.1$ give
$A\leq 0.5$, close to risk neutrality; the moderate and strong profiles represent
assessors who perceive their stake as a much larger share of what they value, for
instance because reputational losses have career consequences.
Table~\ref{tab:settlement} reports the results.

\begin{table}[!htbp]
\centering
\caption{Assessment behavior under the settlement rule, synthetic setting.
Participation is the share of encounters with a target that lead to an assessment;
``overrated targets criticized'' is the share of targets with $\Delta<-0.2$ that
receive a criticism. Script: \texttt{settlement.py}.}
\label{tab:settlement}
\footnotesize
\renewcommand{\arraystretch}{1.1}
\begin{tabularx}{\textwidth}{>{\raggedright\arraybackslash}X>{\centering\arraybackslash}p{0.6cm}>{\centering\arraybackslash}p{1.45cm}>{\centering\arraybackslash}p{1.45cm}>{\centering\arraybackslash}p{1.45cm}>{\centering\arraybackslash}p{1.7cm}}
\toprule
Assessor & $\gamma$ & Partici\-pation & Share of criticisms & Precision of criticisms & Overrated targets criticized \\
\midrule
Risk neutral        & 1   & 31\% & 49\% & 92\% & 42\% \\
                    & 1.5 & 29\% & 47\% & 93\% & 39\% \\
                    & 2   & 28\% & 44\% & 94\% & 36\% \\
CARA, $A=2$         & 1   & 25\% & 49\% & 94\% & 36\% \\
                    & 1.5 & 24\% & 47\% & 94\% & 33\% \\
                    & 2   & 23\% & 45\% & 95\% & 31\% \\
CARA, $A=10$        & 1   & 11\% & 49\% & 97\% & 19\% \\
                    & 1.5 & 10\% & 43\% & 98\% & 15\% \\
                    & 2   & 10\% & 39\% & 98\% & 13\% \\
Loss averse (2.25)  & 1   & 12\% & 49\% & 97\% & 19\% \\
                    & 1.5 & 11\% & 47\% & 97\% & 18\% \\
                    & 2   & 11\% & 44\% & 97\% & 16\% \\
\midrule
Risk neutral, no escrow cost & 1 & 100\% & 49\% & 75\% & 87\% \\
\bottomrule
\end{tabularx}
\end{table}

Three observations follow. First, the escrow cost, more than risk aversion, drives
participation: without it, a risk-neutral assessor assesses every target, with a
precision of 75\%; with it, only well-informed assessments are made (31\%), with a
precision above 90\%. The escrow thus acts as a filter, at the price of fewer
assessments, and early settlement (A3) matters. Second, risk aversion reduces
participation further, to about 11\% for strongly risk-averse or loss-averse
assessors, but with $\gamma=1$ it does not bias the mix: criticisms remain 49\% of
assessments. Third, $\gamma>1$ reduces the share of criticisms and the share of
overrated targets that get criticized, and the effect grows with risk aversion: for
strongly risk-averse assessors, $\gamma=2$ leaves only 13\% of overrated targets
criticized. Hence the indicative value $\gamma=1.5$ (Table~\ref{tab:parameters}).

\paragraph{A selection bias.} The settlement rule therefore selects who assesses:
assessors who tolerate risk and, since the weight of a given loss decreases with the
capital one holds, established actors. Informative criticism can become the
privilege of those who can afford to lose, and the opinions actually expressed are
then biased towards those of established actors. We treat this as a governance
problem rather than as a parameter choice: rule G7 expresses stakes as a share of
free Rep, gives a small allowance of protected criticisms to the lower half of the
distribution, and publishes the distribution of assessments and criticisms by capital
tier so that $\gamma$ and $\kappa$ can be revised. Its effect has not been tested.
The analysis also assumes that 20\% of assessments are unresolved; Study~C shows that
this share is much higher for criticisms (Section~\ref{subsec:num-dynamics}).

\subsection{Estimated gaming matrix and comparison of aggregation rules}
\label{subsec:num-gaming}

Study~B attacks three profiles: an established account of median quality (T2), a
newcomer with one strong paper (T4) and a researcher in a narrow field (T5). The
\emph{flood} adds $G$ generated papers, each receiving a Poisson number of
recognitions with mean $\lambda_g=0.1$. The \emph{Sybil farm} adds $x$ recognitions
from Sybil identities, with a low seed trust (uniform on $[0,0.2]$), presented as
coming from outside the community, recent and not reciprocated. The \emph{collusion
ring} adds $x$ recognitions from ring members with an ordinary trust (uniform on
$[0.5,1.5]$), from the same community, reciprocated, and mostly recent.
Table~\ref{tab:bmatrix} gives the estimated local gaming matrix: the mean change of
the percentile of each sub-indicator per unit of intensity, at small intensity (2
papers, 10 recognitions). Sub-indicators $z_1$, $z_{21}$ and $z_{22}$, which these
attacks do not touch, are omitted, and stability $z_{51}$ is missing for the
newcomer, whose history is too short.

\begin{table}[!htbp]
\centering
\caption{Estimated local gaming matrix $B$: change of the percentile of each
sub-indicator (points) per generated paper (flood) or per added recognition (Sybil,
ring), mean across 5 populations. Entries within two standard deviations of zero are
in parentheses. Script: \texttt{gaming\_matrix.py}.}
\label{tab:bmatrix}
\footnotesize
\setlength{\tabcolsep}{4pt}
\renewcommand{\arraystretch}{1.1}
\begin{tabular}{llrrrrrrr}
\toprule
Profile & Attack & $z_{31}$ & $z_{32}$ & $z_{33}$ & $z_{41}$ & $z_{42}$ & $z_{43}$ & $z_{51}$ \\
\midrule
T2 established & flood & $+0.15$ & $(+0.02)$ & $-9.14$ & $(0.00)$ & $(+0.08)$ & $(-0.05)$ & $(+0.51)$ \\
               & Sybil & $+0.25$ & $+0.36$ & $+3.55$ & $(0.00)$ & $-4.03$ & $+0.21$ & $(+0.37)$ \\
               & ring  & $+0.49$ & $-2.37$ & $+3.53$ & $(0.00)$ & $(+0.11)$ & $-4.15$ & $(-0.15)$ \\
T4 newcomer    & flood & $+0.53$ & $(-0.04)$ & $-18.76$ & $-0.20$ & $(-0.03)$ & $(-0.01)$ & -- \\
               & Sybil & $+0.91$ & $+0.93$ & $+0.74$ & $-0.72$ & $-5.06$ & $+0.79$ & -- \\
               & ring  & $+1.56$ & $-3.54$ & $+0.74$ & $-0.72$ & $-0.22$ & $-4.91$ & -- \\
T5 narrow field & flood & $+0.24$ & $(-0.04)$ & $-5.82$ & $(0.00)$ & $(+0.01)$ & $(-0.04)$ & $(+0.55)$ \\
               & Sybil & $+0.41$ & $+0.69$ & $+5.17$ & $(0.00)$ & $-4.84$ & $+0.43$ & $(-0.45)$ \\
               & ring  & $+0.78$ & $-3.48$ & $+5.05$ & $(0.00)$ & $(+0.04)$ & $-4.65$ & $-1.11$ \\
\bottomrule
\end{tabular}
\end{table}

\paragraph{Scope of the estimates.} These estimates are profile- and
intensity-dependent local effects, not a system-wide collateral matrix: by
Theorem~\ref{thm:local}, the matrix that matters is the one computed around the
honest profile of the attacker, and it changes with that profile. Their purpose is to
illustrate the estimation procedure and to identify vulnerable profiles, not to
establish the global non-emptiness of the admissible set.

\paragraph{Comparison with the hypothesis.} The estimated signs agree with
Table~\ref{tab:matrix} on every entry that the hypothesis marks ``$+$'' or ``$-$''
except one: the Sybil farm was expected to lower stability $z_{51}$, and the estimate
is not distinguishable from zero, plausibly because recent recognitions make a
profile less stable only if its honest short-term rate was not already above its
long-term rate. Conversely, several entries that the hypothesis marks ``$0$'' differ
from zero for at least one profile. Sybil recognitions slightly raise non-reciprocity
$z_{43}$, since Sybils ask for nothing in return. All three attacks lower the
newcomer's own seed trust $z_{41}$: additional endorsers provide more evidence on a
seed trust that is low for a newcomer, so that shrinkage no longer protects it. A
ring also lowers the newcomer's median endorser trust $z_{42}$ ($-0.22$) and the
stability $z_{51}$ of the narrow-field profile ($-1.11$). These second-order effects
all have the sign of collateral sensitivity.

\paragraph{Required costs.} For an account bound by merit, as are 93\% of honest
accounts, the binding set is $J=\{z_{31}\}$, the only admissible weighting is the
vertex $e_{z_{31}}$, and by Theorem~\ref{thm:local} gaming brings no first-order gain
exactly when each unit of attack costs more than its effect on $z_{31}$. The required
costs of Definition~\ref{def:reqcost} are therefore the entries of the first column:
from $0.15$ (T2) to $0.53$ (T4) rank points per generated paper, from $0.25$ to $0.91$
per Sybil recognition, and from $0.49$ to $1.56$ per ring recognition. They are
highest for the newcomer, whose merit evidence is thinnest: the incentive to game
concentrates on newcomers, as Section~\ref{subsec:nonempty} predicts. These figures
give the protocol a target (budgets, identity costs and burned stakes must cost an
attacker at least this much per unit, once converted into rank points) but the
conversion requires a model of what attackers value, which the first full simulation
must provide. The required costs are first-order quantities: beyond small
intensities, screens become binding and the change of rank itself becomes negative,
whatever the cost, as Table~\ref{tab:gains} shows.

\begin{table}[!htbp]
\centering
\caption{Change of the rank of the aggregate (percentage points) under each attack
and each aggregation rule, on the same draws; intensity is the number of generated
papers (flood) or of added recognitions (Sybil, ring). Means across 5 populations of
the mean over 200 paired draws; standard deviations across populations are at most
$3.2$ points ($5.2$ for one cell). Costs are not deducted. Script:
\texttt{gaming\_matrix.py}.}
\label{tab:gains}
\footnotesize
\setlength{\tabcolsep}{5pt}
\renewcommand{\arraystretch}{1.05}
\begin{tabular}{llrrrrr}
\toprule
Profile & Attack & Intensity & Minimum & Mean & Geometric & Merit only \\
\midrule
T2 established  & flood & 5   & $+0.8$  & $-6.0$  & $-6.6$  & $+0.7$ \\
                &       & 20  & $-14.7$ & $-10.8$ & $-23.7$ & $+2.7$ \\
                &       & 50  & $-52.6$ & $-14.1$ & $-34.8$ & $+6.3$ \\
                & Sybil & 20  & $-49.6$ & $+0.5$  & $-23.6$ & $+4.9$ \\
                &       & 100 & $-58.5$ & $+5.4$  & $-34.5$ & $+17.6$ \\
                &       & 400 & $-58.5$ & $+15.8$ & $-39.1$ & $+34.3$ \\
                & ring  & 20  & $-17.5$ & $-10.8$ & $-28.3$ & $+8.9$ \\
                &       & 100 & $-58.2$ & $-14.8$ & $-51.1$ & $+26.2$ \\
                &       & 400 & $-58.5$ & $-11.6$ & $-54.7$ & $+39.4$ \\
\addlinespace
T4 newcomer     & flood & 5   & $+3.1$  & $-10.4$ & $-21.9$ & $+3.0$ \\
                &       & 20  & $-51.4$ & $-8.0$  & $-33.7$ & $+9.4$ \\
                &       & 50  & $-61.4$ & $-2.3$  & $-33.1$ & $+18.2$ \\
                & Sybil & 20  & $-60.0$ & $+14.6$ & $-25.9$ & $+15.9$ \\
                &       & 100 & $-61.8$ & $+34.3$ & $-19.0$ & $+33.3$ \\
                &       & 400 & $-61.8$ & $+42.0$ & $-16.0$ & $+39.0$ \\
                & ring  & 20  & $-48.3$ & $-15.3$ & $-38.4$ & $+23.6$ \\
                &       & 100 & $-61.7$ & $-11.8$ & $-42.7$ & $+37.5$ \\
                &       & 400 & $-61.8$ & $-10.7$ & $-42.9$ & $+39.4$ \\
\addlinespace
T5 narrow field & flood & 5   & $+1.1$  & $-3.5$  & $-7.7$  & $+1.4$ \\
                &       & 20  & $-14.8$ & $-6.0$  & $-21.9$ & $+5.1$ \\
                &       & 50  & $-37.9$ & $-8.1$  & $-29.7$ & $+11.4$ \\
                & Sybil & 20  & $-40.5$ & $+8.5$  & $-19.3$ & $+8.7$ \\
                &       & 100 & $-40.6$ & $+20.0$ & $-26.9$ & $+28.7$ \\
                &       & 400 & $-40.6$ & $+29.7$ & $-27.8$ & $+52.4$ \\
                & ring  & 20  & $-27.8$ & $-6.5$  & $-31.7$ & $+15.6$ \\
                &       & 100 & $-40.6$ & $-2.8$  & $-43.7$ & $+41.3$ \\
                &       & 400 & $-40.6$ & $+2.1$  & $-45.0$ & $+58.9$ \\
\bottomrule
\end{tabular}
\end{table}

\paragraph{Rank gains, not net gains.} Table~\ref{tab:gains} reports changes of
rank, not net gains: the costs of the attacks are not deducted. Since costs are
non-negative, a negative rank gain implies a negative net gain for an attacker who
values only the rank of the attacking account; a positive rank gain says nothing about
profitability until costs are known; and an attacker may value other things, such as
the visibility of one paper or harm to a competitor, which the table does not measure.

\paragraph{Rank gains under each rule.} Under the minimum, every attack has a
negative rank gain beyond small intensities: from 20 generated papers or 20 added
recognitions, the attacking account loses 15 to 62 rank points, because a screen
collapses (yield for floods, endorser trust for Sybil farms, out-group share and
non-reciprocity for rings). Small floods still raise the rank, by $+0.8$ to $+3.1$
points for 5 papers: the yield screen tolerates a few unrecognized papers, as it must
for honest accounts. Under the arithmetic mean, Sybil farms raise the rank of every
profile, up to $+42$ points for the newcomer, because the gains on merit and
out-group share compensate the loss on endorser trust; floods lower it, and so do
rings except for the narrow-field profile at the highest intensity ($+2.1$).
Under merit only, every attack raises the rank, up to $+59$ points. Under the
geometric mean every attack tested has a negative rank gain, smaller in magnitude than
under the minimum, but the rule compensates: for the newcomer, the penalty of a Sybil
farm shrinks from $-26$ to $-16$ points as the farm grows from 20 to 400
recognitions, because the gain on merit keeps growing while the loss on endorser
trust saturates. This is the behavior that Proposition~\ref{prop:partial} predicts
for compensatory rules, and its absence under the minimum. The results do not
establish, however, that the minimum is preferable to the geometric mean, which ranks
honest accounts better (Table~\ref{tab:rules}) and gave every attack tested here a
negative rank gain: the comparison must be repeated against attackers who optimize a mix of actions,
in the planned simulation of Section~\ref{subsec:protocol}.

\subsection{A dynamic simulation with endogenous outcomes}
\label{subsec:num-dynmodel}

The analyses above treat recognitions and outcomes as exogenous. In MagMell they are
not: an assessment changes the visibility of its target, visibility changes reads and
recognitions, and recognitions determine both the outcome that settles the assessment
and the reputation of the authors, which in turn changes the visibility of their next
papers. Study~C simulates this loop.

\paragraph{Authors and papers.} There are 1500 incumbent authors in ten communities,
one of which is small (3\% of authors) and poorly connected, with
$\theta\sim\mathcal N(0,1)$. Time is counted in weekly epochs: a burn-in of 104
epochs is followed by 104 epochs of observation. An author publishes with probability
$0.03$ per epoch, and a paper has quality $q=\theta+\mathcal N(0,0.5^2)$. Papers stay
in the feed for 52 epochs.

\paragraph{Reading and recognition.} Every author is also a reader, with one read
per epoch, 90\% within her community and 10\% across communities. Reads are allocated
by a ranked feed with position bias (the expected number of reads of the paper of
rank $k$ is proportional to $1/k$) and an exploration share $\varepsilon_x$ of reads
spread over the 30 least-read papers (G6). The visibility score of a paper is the
mean of its authors' $R_{\mathrm{auth}}$ and of the percentile of its own evidence:
recognitions per read shrunk to the mean rate, an endorsement counting as two
pseudo-reads with recognition and a criticism as two pseudo-reads without. A read
produces a recognition with probability $\mathrm{logit}^{-1}(q-1.5)$, and an honest
recognition is reciprocated with probability $0.2$.

\paragraph{Reputation.} $R_{\mathrm{auth}}$ follows Section~\ref{subsec:profile}:
merit is the recognition per year of exposure, with in-community recognitions weighted
by the bridging coefficient of G3, shrunk towards the $\pi_0$-quantile ($q_0=0.25$)
with weight $n_0=5$ quarters, and ranked; the screens are the out-group share
(percentile) and non-reciprocity (percentile, or hybrid with $\theta_{43}=0.5$),
applied to authors with at least five recognitions.

\paragraph{Assessors.} A read may lead to an assessment. The reader observes the
quality of the paper relative to the merit rank of its authors,
$q-\Phi^{-1}(R_m)+\mathcal N(0,0.5^2)$, endorses above $0.9$ and criticizes below
$-1$. Two uninformative strategies are present: a \emph{coalition} of 5\% of the
authors, all in one community, spends half of its reads on coalition papers and
always recognizes, reciprocates and endorses them; \emph{random endorsers}, another
5\%, read the global feed and endorse 30\% of what they read, whatever its quality.

\paragraph{Settlement.} After $T_e=26$ epochs (shorter than the indicative value of
Table~\ref{tab:parameters}, to fit the horizon), each assessment is settled against the
recognitions its target received since the assessment from independent readers,
excluding the assessor's group and, under A5, her community. We compare three
measures of the outcome: counts, $\log(1+r)$; a rate, recognitions per independent
read shrunk to the mean; and the excess $r-\bar\rho\,d$ of recognitions over those
expected from the $d$ independent reads, with $\bar\rho$ the mean rate. $O$ is the
percentile of the measure within the batch of assessments settled at the same epoch;
the reference forecast $\hat O$ is the merit percentile of the authors at assessment
time, calibrated on $O$ by a linear fit over the resolved assessments of the batch
(A4). An assessment is unresolved if $d<d_{\min}=10$ or $|O-\hat O|<0.05$ (A6);
otherwise its return per unit of stake is $\beta a$, with $\beta=0.3$, multiplied by
$\gamma=1.5$ for a wrong criticism (A8).

\paragraph{Newcomers.} At the start of the observation period, 25 newcomers of each
of four types arrive and publish one paper, then publish at the common rate: an
\emph{excellent} researcher ($\theta=2$); an \emph{average} researcher ($\theta=0$),
to be compared with incumbents of average quality and long history ($|\theta|<0.25$);
a researcher in the \emph{poorly connected} field ($\theta=1$); and a
\emph{heterodox} researcher ($\theta=1.5$) whose papers are recognized late: for the
first 52 epochs, recognition is computed as if their quality were lower by $2$.

\paragraph{What is and is not endogenous.} Visibility, reads, recognitions,
outcomes, reference forecasts and reputations are endogenous. Strategies are not:
assessors follow fixed rules, and returns are measured but not fed back into their
behavior or their stakes. Study~C therefore tests whether fixed strategies, informative
or not, are profitable against endogenous outcomes; it does not compute an
equilibrium. Each condition is run with 10 seeds.

\subsection{Newcomers}
\label{subsec:num-newcomers}

For each type of newcomer, Table~\ref{tab:newcomers} reports the probability that a
paper is seen (at least five reads within 8 epochs) and assessed (within 26 epochs),
the median final rank, and the number of epochs before the median rank of the type
reaches the population median, without and with exploration.

\begin{table}[!htbp]
\centering
\caption{Newcomers in the dynamic simulation, $q_0=0.25$: probability that a paper
is seen (at least 5 reads within 8 epochs) and assessed (within 26 epochs), median
final rank of $R_{\mathrm{auth}}$ after 104 epochs, and epochs until the median rank
of the type reaches $0.5$; the incumbent has average quality and a long history.
Means across 10 seeds; standard deviations are at most $0.10$ for probabilities,
$0.06$ for ranks and $9$ epochs, except for the average newcomer, of median quality,
whose time to the median is censored by the horizon: ``$>104$'', not reached in any
seed. Script: \texttt{dynamics.py}, experiment C1.}
\label{tab:newcomers}
\footnotesize
\setlength{\tabcolsep}{4pt}
\renewcommand{\arraystretch}{1.1}
\begin{tabular}{lcccccccc}
\toprule
 & \multicolumn{2}{c}{P(seen)} & \multicolumn{2}{c}{P(assessed)} & \multicolumn{2}{c}{Final rank} & \multicolumn{2}{c}{To median} \\
\cmidrule(lr){2-3}\cmidrule(lr){4-5}\cmidrule(lr){6-7}\cmidrule(lr){8-9}
Type & $\varepsilon_x=0$ & $0.1$ & $0$ & $0.1$ & $0$ & $0.1$ & $0$ & $0.1$ \\
\midrule
Excellent ($\theta=2$)             & 0.54 & 0.86 & 0.98 & 0.98 & 0.94 & 0.95 & 14 & 10 \\
Average ($\theta=0$)               & 0.05 & 0.51 & 0.66 & 0.76 & 0.42 & 0.45 & $>104$ & $92^{a}$ \\
Average incumbent ($|\theta|<0.25$) & 0.09 & 0.54 & 0.66 & 0.77 & 0.47 & 0.47 & & \\
Poorly connected ($\theta=1$)      & 0.20 & 0.51 & 0.88 & 0.91 & 0.76 & 0.76 & 22 & 14 \\
Heterodox ($\theta=1.5$)           & 0.15 & 0.59 & 0.98 & 0.99 & 0.65 & 0.65 & 71 & 72 \\
\bottomrule
\end{tabular}
\par\smallskip{\footnotesize $^{a}$Mean over seeds, counting as 104 the seeds that do not reach the median.\par}
\end{table}

\paragraph{Effect of exploration.} Without exploration, the feed
concentrates reads on the most visible papers, and an average paper, new or not, is
seen with probability 5 to 9\%. With 10\% of reads spent on the least-read papers,
this probability is 51 to 54\%. Exploration does not favor newcomers over comparable
incumbents (the average newcomer and the average incumbent have similar visibility,
with 0.79 and 0.76 times the mean number of reads of their cohort) and in the long
run history is neither an advantage nor a handicap: both end at about the same rank.
The excellent newcomer reaches the median within 10 epochs with exploration and 14
without; the researcher in the poorly connected field within 14 and 22.

\paragraph{Effect of the prior.} With $q_0=0.5$, newcomers start
at the median, and the excellent, poorly connected and heterodox types reach it
within 1 to 6 epochs, while final ranks change by at most $0.07$. The quantile of the
prior thus sets the starting rank of newcomers and, through it, the incentive to
whitewash (Section~\ref{subsec:num-profiles}), much more than their long-run position.

\paragraph{Late-recognized work.} The heterodox newcomer is the most difficult case:
her papers are good but recognized only after a year. She reaches the median only
after about 70 epochs, some 20 epochs after recognition begins (the standard deviation
across seeds is about 7 epochs, the largest of the table), and at the end of the horizon she ranks at
$0.65$, far below the excellent newcomer: merit is a rate accumulated over the whole
exposure, and it keeps the memory of the year without recognition. Exploration
changes whether her papers are seen (from $0.15$ to $0.59$), not when she is
recognized. A ranking based on recognition cannot react faster than recognition
itself; at best, it can avoid delaying it further. Whether those who assess her work
early are rewarded is examined in Section~\ref{subsec:num-dynamics}.

\subsection{Reflexivity: assessments, visibility and outcomes}
\label{subsec:num-dynamics}

Table~\ref{tab:reflexivity} reports, for each type of assessment, the mean return
per assessment and the share of unresolved assessments, under the three measures of
the outcome, with and without the independence rule A5, and under variants of the
default configuration that test remedies.

\begin{table}[!htbp]
\centering
\caption{Mean return per assessment (percent of the stake) and, in parentheses, share
of unresolved assessments (percent), in the dynamic simulation; $\rho_S$: Spearman
correlation between $R_{\mathrm{auth}}$ and $\theta$ of honest incumbents. In the
coordinated variant, random endorsers all endorse the same third of the papers; with
a delayed effect, a criticism lowers visibility outside the critic's community only
after settlement; with independence by links, the outcome counts every reader not
linked to the assessor except the coalition, whose screens bind; the standardized
outcome divides the excess by its Poisson scale. Means
across 10 seeds; standard deviations of returns are at most $1.0$ point. The maximal
return is $\beta=30\%$ of the stake. Script: \texttt{dynamics.py}, experiments C2
and C3.}
\label{tab:reflexivity}
\footnotesize
\setlength{\tabcolsep}{3.5pt}
\renewcommand{\arraystretch}{1.1}
\begin{tabular}{lccccc}
\toprule
Configuration & \shortstack{Honest\\endorse} & \shortstack{Honest\\criticize} & \shortstack{Random\\endorse} & \shortstack{Coalition\\endorse} & $\rho_S$ \\
\midrule
Counts, A5 off                     & $0.0$ (78)  & $+0.1$ (97) & $+0.3$ (60) & $-0.2$ (90) & 0.85 \\
Counts, A5 on                      & $-0.1$ (82) & $+0.1$ (98) & $+0.2$ (65) & $0.0$ (98)  & 0.85 \\
Rate, A5 on                        & $+0.9$ (53) & $+0.1$ (98) & $-0.9$ (29) & $-0.2$ (95) & 0.85 \\
Excess, A5 off                     & $+1.4$ (50) & $-0.7$ (94) & $-1.6$ (27) & $-1.2$ (84) & 0.85 \\
Excess, A5 on (default)            & $+1.3$ (51) & $0.0$ (98)  & $-1.3$ (28) & $-0.3$ (95) & 0.85 \\
\midrule
Coordinated endorsers, counts & $-0.1$ (82) & $+0.1$ (98) & $+0.2$ (64) & $-0.1$ (97) & 0.85 \\
Coordinated endorsers, excess & $+1.4$ (52) & $+0.1$ (98) & $-0.9$ (27) & $-0.4$ (92) & 0.85 \\
Audit reads, 5\%                   & $+1.4$ (54) & $+0.1$ (95) & $-1.3$ (30) & $-0.6$ (95) & 0.81 \\
Audit reads, 15\%                  & $+2.0$ (56) & $+0.7$ (81) & $-0.9$ (32) & $-0.6$ (90) & 0.73 \\
No effect of criticism on visibility & $+1.4$ (50) & $+0.2$ (97) & $-1.1$ (28) & $-0.3$ (95) & 0.84 \\
Delayed effect of criticism        & $+1.3$ (52) & $+0.1$ (97) & $-1.0$ (28) & $-0.5$ (94) & 0.85 \\
Independence by links              & $+1.9$ (22) & $+1.1$ (84) & $+1.7$ (29) & $+0.3$ (95) & 0.85 \\
Independence by links, standardized & $+1.9$ (25) & $+1.6$ (83) & $+0.9$ (31) & $+0.1$ (95) & 0.85 \\
\bottomrule
\end{tabular}
\end{table}

\paragraph{Self-fulfilling outcomes.} When the outcome counts recognitions,
random endorsers earn a positive return and honest endorsers earn nothing: an
endorsement raises the visibility of its target, hence its reads and its
recognitions, whatever its quality, and the endorsement thereby confirms itself.
Measuring the outcome per independent read, as a rate or as an excess over the
expected recognitions, removes the effect in the model: honest endorsers gain
($+0.9$ to $+1.4\%$ of the stake per assessment) and random endorsers lose ($-0.9$ to
$-1.6\%$). This motivated the revision of A5. The effect is small here because an
endorsement weighs only two pseudo-reads in visibility; it should grow with that
weight, which we did not vary.
Coordinating uninformative endorsements, so that all random endorsers endorse the
same third of the papers, does not make them profitable: their return stays close to
zero under counts ($+0.2\%$) and negative under the excess measure ($-0.9\%$, against
$-1.3\%$ without coordination, a difference within the variation across seeds).

\paragraph{Calibration of the forecast.} A first version of the simulation used the
screened aggregate of the authors, without calibration, as reference forecast. Two
problems appeared. Colluders gained by endorsing coalition papers, whose authors the
screens rank low but whose papers the coalition recognizes; and under the
exposure-normalized measures honest assessors lost, because the forecast was on
another scale than the outcome and almost uncorrelated with it. The forecast is now
computed from merit and calibrated on the scale of $O$ (revised A4).

\paragraph{Effect of the independence rule.} Under A5, 95\% of the
endorsements of coalition papers by colluders are unresolved: few readers outside the
coalition's community read them, so that their outcome is rarely measurable. The
colluders lose slightly ($-0.3\%$) and gain nothing. Without A5 their endorsements are
resolved more often and lose ($-1.2\%$), but honest criticisms then lose as well
($-0.7\%$), which is why A5 is kept.

\paragraph{Settlement of criticism.} Under every configuration of
Table~\ref{tab:reflexivity}, 81 to 98\% of honest criticisms are unresolved, against
22 to 82\% of honest endorsements. In the default configuration, a criticized paper
has had on average 1.4 independent reads at settlement, against 79 for an endorsed
paper. The settlement rule therefore neither rewards nor punishes most criticisms,
whereas the decision analysis of Section~\ref{subsec:num-settlement} assumed 20\% of
unresolved assessments. In its current form, settlement is far more convincing for
endorsements than for criticisms.

\paragraph{Cause: availability of evidence.} A natural
hypothesis is that criticism censors its own outcome, since it lowers the visibility
of its target. The simulation refutes it as the main cause. Delaying the effect of a
criticism on visibility outside the critic's community until settlement leaves 97\% of
criticisms unresolved (1.7 independent reads on average), and removing the effect of
criticism on visibility altogether does no better (1.9 reads, 97\%). The cause lies
upstream. The papers that honest assessors criticize are papers whose recognition per
read is low, which the feed already demotes; and under A5 only readers from other
communities count, who read mostly the most visible papers. Evidence about poor papers
is scarce because few people read them, whether or not they have been criticized. In
the terms of a P2P system, this is a problem of evidence availability, discussed in
Section~\ref{subsec:num-distributed}.

\paragraph{Remedies acting on the supply of evidence.} Defining independence by links
rather than by communities (counting every reader not linked to the assessor, except
accounts whose screens bind) multiplies the independent reads of criticized papers by
six (8.6 on average), lowers the unresolved share of criticisms to 84\% and makes
honest criticism profitable ($+1.1\%$). In our model, however, it also makes random
endorsement profitable ($+1.7\%$); a standardized outcome, which removes the
dependence of the excess on the number of reads, reduces this gain without removing
it ($+0.9\%$). We have not identified the cause, and the rule is not adopted.
Reserving a share of every reader's budget for \emph{audit reads} of criticized papers
also restores part of the measurement: with 15\% of reads, the unresolved share of
criticisms falls to 81\% and honest criticism becomes profitable ($+0.7\%$). The cost
is high: attention diverted to criticized papers lowers the rank correlation of honest
authors from $0.85$ to $0.73$, the probability that a new average paper is seen from
$0.51$ to $0.30$, and the final rank of the heterodox newcomer from $0.65$ to $0.57$.
With 5\% of reads, the gain is negligible. Audit reads are therefore not part of the
default design either.

\paragraph{Research directions.} The settlement of criticism is the central open
problem of MagMell, and the diagnosis above points to two directions. The first is to
settle criticisms against evidence that does not depend on organic attention:
replication reports, which A5 already admits, and audit panels drawn at random among
qualified members not linked to the parties, triggered by the stake of the criticism
rather than by a fixed share of everyone's attention, so that their cost scales with
what is at stake. The second is to define independence at a finer grain than the
community, by links, with outcome measures that remain incentive-compatible when the
evidence available differs widely across assessments. Both are untested. The
unresolved shares also depend on the ratio of reads to papers, one read per author and
per epoch here, which is an assumption.

\paragraph{Early assessments of late-recognized work.} Assessors in the model
perceive the quality of heterodox papers before the community recognizes them, and
endorse them. Of these endorsements, 98\% of those settled before recognition begins
are unresolved, and those settled after gain little ($+0.2\%$). With
$T_e=26$ epochs, the settlement rule cannot reward foresight about work recognized
after a year; thanks to A6, it does not punish it either. When assessors share the
community's late recognition, the heterodox newcomer receives fewer endorsements and
reaches the median after 81 epochs instead of 72. Rewarding foresight would require
longer or rolling horizons, which conflicts with the role of the escrow cost in
participation (Section~\ref{subsec:num-settlement}).

\paragraph{Screens against a coalition.} We measure the advantage of the coalition
as the mean residual rank of its members with respect to the regression of rank on
$\theta$ among honest authors of the same community. Without screens it is $+45$
rank points; with percentile screens, $+34$; with the hybrid screen on
non-reciprocity, $-27$, the coalition then ranking below honest authors of the same
quality. The percentile screens fail because the coalition, with 5\% of the authors,
is larger than their tolerance $\tau_s=2\%$: most colluders are above the
$\tau_s$-quantile of a population that they partly make up
(Section~\ref{subsec:screens}). The hybrid screen does not measurably lower the rank
correlation of honest authors ($0.846$ against $0.847$). Its false positives fall on
authors with little evidence: with the honest reciprocation probability of $0.2$, an
honest author crosses the threshold $\theta_{43}=0.5$ with probability 5.8\% with 5
recognitions, 3.3\% with 10 and 0.3\% with 20.

\subsection{Distributed properties}
\label{subsec:num-distributed}

Study~D evaluates the distributed execution of Section~\ref{subsec:ledger}: the
dissemination of events among nodes, some of which are unavailable; the detection of
equivocation; the cost of tier-2 computation as the number of identities grows; and
the consistency of rankings computed on partial replicas. Consensus is neither
simulated nor benchmarked: MagMell relies on the known properties of BFT protocols,
and the corresponding measurements (latency, throughput, faulty validators, recovery)
remain to be done on an implementation.

\paragraph{Dissemination.} Events and blocks are disseminated by push gossip in its
infect-and-die form: a node that receives an item for the first time forwards it to
$\lceil\ln N\rceil+c$ peers chosen uniformly at random, a fanout for which the
probability that every node is reached tends to $e^{-e^{-c}}$ when all nodes are
available~\cite{kermarrec2003probabilistic}. A fraction $u$ of the nodes is
unavailable: messages sent to them are lost and they forward nothing.
Table~\ref{tab:gossip} reports, over 10 seeds, the share of available nodes
reached and the number of rounds. With $c=3$, gossip reaches on average at least 99.2\% of the
available nodes even when half of the nodes are unavailable (98.4\% in the worst
seed), and stops after at most 9.1 rounds on average, at a cost of $\lceil\ln N\rceil+3$ messages per node and per item (10 to 15 here). Nodes
that are missed obtain the item at their next synchronization with a full node:
gossip provides speed, the log provides completeness.

\begin{table}[!htbp]
\centering
\caption{Push gossip (infect-and-die, fanout $\lceil\ln N\rceil+3$): share of the
available nodes reached (mean and minimum over 10 seeds), rounds until the
dissemination stops, and rounds until 99\% of the available nodes are reached; last
column: rounds until a node holds two conflicting events signed by the same identity
(equivocation detected). Script:
\texttt{distributed.py}, D1.}
\label{tab:gossip}
\footnotesize
\setlength{\tabcolsep}{5pt}
\renewcommand{\arraystretch}{1.1}
\begin{tabular}{rcccccc}
\toprule
$N$ & Unavailable $u$ & Reached (mean) & Reached (min) & Rounds & Rounds to 99\% & Detection \\
\midrule
1\,000   & 0   & 1.0000 & 1.0000 & 5.7 & 4.0 & 2.0 \\
         & 0.3 & 0.9990 & 0.9956 & 6.2 & 5.0 & 2.0 \\
         & 0.5 & 0.9917 & 0.9842 & 7.3 & n.r.$^{a}$ & 2.6 \\
10\,000  & 0   & 1.0000 & 1.0000 & 6.0 & 5.0 & 2.0 \\
         & 0.3 & 0.9999 & 0.9999 & 7.0 & 5.1 & 2.3 \\
         & 0.5 & 0.9981 & 0.9976 & 8.0 & 6.0 & 2.7 \\
100\,000 & 0   & 1.0000 & 1.0000 & 7.0 & 5.0 & 2.4 \\
         & 0.3 & 1.0000 & 0.9999 & 8.0 & 6.0 & 2.8 \\
         & 0.5 & 0.9994 & 0.9993 & 9.1 & 7.0 & 2.9 \\
\bottomrule
\end{tabular}
\par\smallskip{\footnotesize $^{a}$Not reached in 2 of the 10 seeds.\par}
\end{table}

\paragraph{Inconsistent events.} An identity that equivocates (signs two different
events with the same sequence number and sends them to different nodes) is caught
when some node receives both. When the two events are injected at two random
available nodes and gossiped, detection takes 2 to 3 rounds (last column of
Table~\ref{tab:gossip}), and the proof of misbehavior, gossiped in turn by the nodes
that detected it, reaches 99\% of the available nodes within 5.0 to 9.2 rounds from the
injection. Within the log, total order prevents both events from being valid.

\paragraph{Scalability of computation and storage.} Table~\ref{tab:scaling} reports
two kinds of figures. The ledger, bandwidth and storage columns are the analytic
estimates of Section~\ref{subsec:costs}, extrapolated from assumed activity rates. The
last column is a measurement: the time of a full recomputation of the aggregate (merit,
the six screens, seed trust by 20 iterations of a propagation walk, percentiles and
minimum) with vectorized NumPy and SciPy on one processor core, from synthetic event
tables generated for $10^3$ to $10^6$ identities. It measures computation only, not a
deployed network. The time grows slightly faster than linearly with the
number of recognitions, from 0.01~s for $10^3$ identities to 1.6~s for $10^5$ and
24~s for $10^6$ on the machine of Table~\ref{tab:scaling}; a second machine, with
older library versions, took 1.0~s and 15~s. The benchmark generates 1.6 papers per identity and year without
sharing them among co-authors, and thus overestimates the number of papers and
recognitions.

\begin{table}[!htbp]
\centering
\caption{Costs as functions of the number of identities $N$: ledger growth,
transactions to order, ingest per replica, bandwidth of a node with gossip fanout
$\lceil\ln N\rceil+3$, content storage across replicas ($r=5$, 2~MB per paper), and
time of a full tier-2 recomputation on one core (recognitions in parentheses).
Analytic except the last column, which is measured with the versions of
Section~\ref{subsec:num-protocol}; wall-clock times vary between machines. Script: \texttt{distributed.py}, D2.}
\label{tab:scaling}
\footnotesize
\setlength{\tabcolsep}{4pt}
\renewcommand{\arraystretch}{1.1}
\begin{tabular}{rcccccc}
\toprule
 & \multicolumn{5}{c}{Analytic estimates} & Measured \\
\cmidrule(lr){2-6}\cmidrule(lr){7-7}
$N$ & Ledger/year & Tx/s & Ingest & Gossip & Contents/year & Recomputation \\
\midrule
$10^3$ & 0.03~GB & 0.002 & 0.01~kbit/s & 0.07~kbit/s & 0.01~TB & 0.01~s ($2.3\cdot10^4$) \\
$10^4$ & 0.26~GB & 0.02  & 0.07~kbit/s & 0.85~kbit/s & 0.05~TB & 0.13~s ($2.3\cdot10^5$) \\
$10^5$ & 2.6~GB  & 0.19  & 0.66~kbit/s & 9.9~kbit/s  & 0.53~TB & 1.6~s ($2.3\cdot10^6$) \\
$10^6$ & 26~GB   & 1.9   & 6.6~kbit/s  & 112~kbit/s  & 5.3~TB  & 24~s ($2.3\cdot10^7$) \\
\bottomrule
\end{tabular}
\end{table}

\paragraph{Partial replicas.} A light client that relies on gossip alone, or a full
node that is behind, computes rankings on a partial view of the log.
Table~\ref{tab:partial} reports how the tier-2 ranking of Study~A changes when the
node has observed each event independently with probability $v$, for 10 populations
of 3000 identities. Rankings are robust to small gaps: with 1\% of events missing, the
rank correlation with the full view is $0.997$ and 0.4\% of the accounts change
examination status. With 20\% missing, the correlation is still $0.95$, but one
account in ten moves by more than 10 rank points and 3\% change examination status,
because missing events shift both the account's statistics and the reference
population. Decisions with consequences (settlement, minting, examination) are
therefore taken on the full log, at checkpointed positions, never on partial views.

\begin{table}[!htbp]
\centering
\caption{Tier-2 ranking (minimum rule) on a partial replica that observed each event
with probability $v$, against the full log: Spearman correlation of ranks (mean
$\pm$ standard deviation over 10 populations), mean absolute rank difference
(points), share of accounts moving by more than 10 points, share whose examination
status (Section~\ref{subsec:num-robustness}) differs, and share whose binding
sub-indicator differs. Script: \texttt{distributed.py}, D3.}
\label{tab:partial}
\footnotesize
\setlength{\tabcolsep}{4pt}
\renewcommand{\arraystretch}{1.1}
\begin{tabular}{cccccc}
\toprule
$v$ & Spearman & Mean $|\Delta|$ & $|\Delta|>10$ (\%) & Examination differs (\%) & Binding differs (\%) \\
\midrule
0.99 & $0.997\pm0.001$ & 0.5 & 0.6  & 0.4 & 0.5 \\
0.95 & $0.988\pm0.002$ & 1.6 & 2.9  & 1.2 & 1.6 \\
0.90 & $0.976\pm0.004$ & 2.6 & 5.6  & 1.9 & 2.6 \\
0.80 & $0.953\pm0.005$ & 4.2 & 10.3 & 3.0 & 4.3 \\
\bottomrule
\end{tabular}
\end{table}

\paragraph{Evidence availability.} The results of
Section~\ref{subsec:num-dynamics} can be read in the same terms. Settlement is a
distributed measurement whose input, independent accesses to the target, is produced
only by the attention of other nodes. For a poor paper this input is rarely
produced, whether or not the paper has been criticized. Unlike the availability of
contents, the availability of evidence cannot be obtained by replication: evidence
must be produced, not copied.

\subsection{Design corrections derived from the analyses}
\label{subsec:num-corrections}

The analyses led to the following changes, all included in
Section~\ref{sec:multirep}.
\begin{enumerate}[label=(\roman*),leftmargin=2.2em]
  \item Integrity sub-indicators are screens, $\min(1,p/\tau_s)$, not percentiles,
        and missing values of merit are imputed on the scale of the aggregate
        (Section~\ref{subsec:screens}; Study~A).
  \item The evidence of $z_{31}$ is exposure time, not the number of publications,
        so that attackers cannot inflate it (Section~\ref{subsec:screens}).
  \item Screens whose honest values are bounded by human behavior are hybrid,
        percentile and absolute, so as to catch groups larger than $\tau_s$ (P6,
        Section~\ref{subsec:screens}; Study~C).
  \item Human examination is triggered only when a binding screen costs at least
        $\delta_r$ rank points (Section~\ref{subsec:profile}; Study~A).
  \item The outcome of A5 counts recognitions in excess of those expected from
        independent exposure (Study~C).
  \item The reference forecast of A4 is computed from merit and calibrated on the
        scale of the outcome (Study~C).
  \item An assessment is unresolved below $d_{\min}$ independent accesses (A6;
        Study~C).
  \item The penalty factor for wrong criticisms is lowered to $\gamma=1.5$, and rule
        G7 addresses the selection of assessors by risk and capital
        (Section~\ref{subsec:governance}; settlement analysis).
\end{enumerate}
One weakness found by the analyses is not corrected: criticism is rarely settled,
because the papers that deserve it are rarely read by independent readers, and none of
the remedies tested is satisfactory (Section~\ref{subsec:num-dynamics}).

\section{Discussion}
\label{sec:discussion}

This section states what the evaluation establishes and what it cannot
(Section~\ref{subsec:num-establish}), discusses the limitations of this work and the
problems that remain open (Section~\ref{subsec:evaluation}), and specifies the first
full simulation (Section~\ref{subsec:protocol}).

\subsection{What the evaluation establishes}
\label{subsec:num-establish}

\paragraph{Established within the model.} The following hold under the stated
assumptions, across the scenarios and seeds tested, and are logical consequences of
the design that the analyses check rather than discover. The naive pipeline ranks
honest accounts largely by noise and creates a strong incentive to whitewash. The
revised pipeline corrects the former problem and makes the remaining whitewashing
incentive explicit, transferring its mitigation to the identity and onboarding
mechanisms. With percentile screens, the false-positive funnel is set by $\tau_s$, the
number of screens and $\delta_r$, not by the honest population. The minimum ranks
honest accounts less well than the arithmetic or geometric mean in every scenario:
robustness has a price, which grows as the merit signal weakens. Under the minimum,
the three attacks modeled lower the rank of the attacking account beyond small
intensities, while Sybil farms raise it under the arithmetic mean and every attack
raises it under merit only; costs are not deducted. Outcomes measured by raw
counts make the settlement of endorsements self-fulfilling. Percentile screens cannot
catch a group larger than their tolerance. On the distributed side, push gossip with
fanout $\lceil\ln N\rceil+3$ reaches almost all available nodes in a few rounds even
when half of the nodes are unavailable, and equivocation is detected within three
rounds, as known results on epidemic dissemination
predict~\cite{karp2000randomized,kermarrec2003probabilistic}.

\paragraph{Suggested.} The following depend on parameter values that are
assumptions and would change with them: the entries of the estimated gaming matrix,
which are local to three profiles and to small intensities; the orders of magnitude of
the required costs (0.15 to 1.6 rank points per unit of attack) and the fact that
they are highest for newcomers; the residual profitability of small floods; that exploration sets the
visibility of newcomers while the prior sets their starting rank and the incentive to
whitewash; that the escrow cost drives participation and that $\gamma>1$ suppresses
criticism among risk-averse assessors; that the independence rule neutralizes
collusive endorsement rather than punishing it; that criticism is rarely settled for
lack of independent attention, its own effect on visibility being a minor cause; that late-recognized work is neither rewarded nor punished at settlement;
that hybrid screens reverse the advantage of a coalition at a small cost for
honest authors; the costs of the protocol (an analytic estimate of about 26~kB of
log per identity and year, dominated by access events) and of tier-2 computation
(1.0 to 1.6~s per full recomputation for $10^5$ identities, measured on synthetic
event tables on two machines); and the sensitivity of rankings to partial replicas.

\paragraph{Cannot be concluded.} The analyses say nothing about real populations,
real attackers or real costs: no parameter is calibrated on data, and the static and
dynamic models are not calibrated on each other, so that their rank correlations
($0.51$ and $0.85$) are not comparable. They do not establish robustness against adaptive
attackers who optimize a mix of actions, since the attacks tested are few and fixed
and the evolutionary simulation that would test them has not been run.
They do not establish any equilibrium property of the assessment mechanism, since
strategies are fixed and returns are not fed back. They do not establish that the
minimum is preferable to the geometric mean. And they do not establish the fairness of
the governance rules: seed selection, community detection, earned entry and the
capture of exploration slots by Sybils were not simulated, and bridging and
exploration were simulated only in simplified form. Finally, no part of MagMell has
been implemented. Consensus was neither simulated nor benchmarked, so that nothing is
established about confirmation latency, throughput, replication cost, faulty or
malicious validators, or recovery. The costs are analytic estimates that ignore
protocol overheads other than signatures, keys and hashes, and the dissemination model
ignores network latency, bandwidth heterogeneity and network-level attacks.

\subsection{Limitations and open problems}
\label{subsec:evaluation}

\paragraph{Scope of the theory.} The exact result holds in an idealized linear
model, and the result for realistic aggregates only locally
(Section~\ref{subsec:limits}). The gaming matrix is unknown and drifts as attackers
innovate, and we do not claim that the admissible sets of MagMell are non-empty: the
estimates of Section~\ref{subsec:num-gaming} are local to three profiles and to small
intensities.

\paragraph{Settlement of criticism.} The weakest part of the design is the
settlement of criticism. In the dynamic simulation, 98\% of honest criticisms remain
unresolved under the default rules, so that settlement is far more convincing for
endorsements than for criticisms. The natural explanation (criticism censors its own
outcome by lowering the visibility of its target) is refuted as the main cause:
removing or delaying that effect changes almost nothing. The cause is a problem of
evidence availability in the network: the input of settlement, independent accesses,
is produced only by the attention of other nodes, and poor papers receive little of
it. Unlike contents, evidence cannot be replicated; it must be produced. Two
directions follow:
settling criticism against evidence that does not depend on organic attention, such
as replication reports and audit panels drawn at random and triggered by the stake;
and defining independence by links rather than by communities, with outcome measures
that remain incentive-compatible when evidence is unevenly available.

\paragraph{Fairness and side effects.} The governance rules G1--G7 are proposals,
whose parameters trade robustness against fairness, and none of them is a guarantee
of fairness. The analyses also made several costs explicit rather than removing them:
whitewashing still tempts 23\% of honest accounts, and its deterrence rests on
onboarding; screens send about 5\% of honest accounts to examination, atypical ones
nearly three times as often; and the settlement rule selects assessors by their
tolerance to risk and their capital. The side effects of any reputation system on
honest but atypical researchers (newcomers, small fields, late-recognized work) may
matter as much as its efficacy against attackers. The permanent record of assessments
deters manipulation but also chills criticism, which pseudonymous but accountable
assessments (A10) are meant to mitigate once a cryptographic construction exists.

\paragraph{Distributed design.} The specification of
Section~\ref{subsec:sysmodel} is conceptual: the protocol is specified, not implemented
or benchmarked. Ordering by a committee of validators is a trust assumption that suits
institutions but departs from fully permissionless operation. It exposes the network to
the capture of the committee, to its dependence on governance for renewal, and to
institutional concentration, and it guarantees neither safety, nor liveness, nor
freedom from censorship if more than $f$ validators are Byzantine. Access events, which
are needed to check admissibility and independence, make reading behavior public, and
privacy-preserving access records remain to be designed. Finally, the pseudonymous
assessments of A10 lack a cryptographic construction, so that protection against
retaliation is not guaranteed (Table~\ref{tab:guarantees}).

\paragraph{Evaluation.} The evaluation relies on synthetic assumptions, fixed
attacks and fixed assessor strategies, and establishes no equilibrium; an equilibrium
analysis of the assessment mechanism remains to be done. The empirical effectiveness
of MagMell can only be established through deployment and adoption. On the positive
side, a two-tier
system allows detection algorithms to evolve without changing the protocol: new
signals can be added when new attacks appear, and weights can be varied within the
admissible set.

\subsection{Protocol of the first full simulation (planned)}
\label{subsec:protocol}

The first full simulation has not been run. Its protocol is fixed in advance, so that
its results can falsify parts of the design; it extends Study~C along the lines left
open above.

\paragraph{Populations.} $10\,000$ identities in communities whose sizes follow a
power law, the smallest holding 1\% of the identities; 30\% of honest identities
with one atypical trait (Study~A); assessors with heterogeneous attitudes to risk
(risk neutral, CARA with $A=2$ and $A=10$, loss averse, in equal shares) and with the
reputational capital accumulated during a two-year burn-in; newcomers of the four
types of Study~C arriving every quarter; a three-year horizon.

\paragraph{Attackers.} 5\% of the identities, in five families: Sybil farm,
collusion ring, flood of generated papers, coordinated uninformative assessment, and
mixed strategies. A strategy is a vector of intensities and timings of actions, of
targets, of allocations of budgets and of onboarding paths. Strategies evolve by a
genetic algorithm (population 200, 100 generations) against each rule, with fitness
equal to the rank gain minus the costs incurred (budgets, burned stakes, identity
costs); the best responses of Study~B serve as initial strategies.
Appendix~\ref{app:ga} details the representation of strategies, the conversion of
costs into fitness, and the checks against premature convergence and insufficient
exploration.

\paragraph{Rules compared.} Minimum, arithmetic mean, geometric mean and merit only,
each with and without the revised A4--A6, hybrid screens, exploration and G7.

\paragraph{Metrics.} Gain of the best evolved strategy and its trajectory over
generations; Spearman correlation between rank and quality of honest identities; the
false-positive funnel FP1--FP4, overall and by type of identity; for newcomers, the
probabilities of being seen and assessed and the time to the median rank; returns and
unresolved shares per type of assessor; share of criticisms, and of informative
criticisms, by risk class and capital tier; share of exploration slots captured by
attackers; share of honest identities that would gain by whitewashing.

\paragraph{Repetitions.} 30 seeds per condition, paired across rules; 95\% bootstrap
confidence intervals over seeds.

\paragraph{Success criteria.} The design is considered internally consistent if:
\begin{enumerate}[label=(\alph*),leftmargin=2.2em]
  \item the best evolved gain under the minimum is below 5 rank points and at most
        a third of that under the arithmetic mean;
  \item the rank correlation of honest identities under the minimum is at least 90\%
        of that under the arithmetic mean;
  \item fewer than 1\% of honest identities are wrongly sanctioned, and atypical
        identities are sent to examination at most twice as often as typical ones;
  \item the excellent newcomer reaches the median within 26 epochs in at least 90\%
        of runs, and the heterodox newcomer within 26 epochs after recognition
        begins;
  \item random endorsers and colluders have negative expected returns, honest
        endorsers and critics positive ones, and fewer than half of honest
        criticisms are unresolved;
  \item criticisms make up at least 40\% of assessments in every risk class, and at
        least 30\% of criticisms come from the lower half of the capital
        distribution;
  \item attackers capture a share of exploration slots no larger than their share
        of the population.
\end{enumerate}
The analyses of this section suggest that criteria (b), (c), (d) for the heterodox
newcomer, and (e) for criticisms are at risk with the current design. A failed
criterion is a result: it identifies which part of the design must be revised. The
simulation will be followed by replays on public data (the Steem
blockchain~\cite{li2021steemops}, Stack Exchange data dumps, and open citation graphs
such as OpenAlex~\cite{priem2022openalex}, on which citation cartels can be searched
for~\cite{fister2016toward}) to estimate entries of the gaming matrix with
confidence intervals, to apply the robust non-emptiness test of
Section~\ref{subsec:nonempty} to their upper bounds, and to evaluate the governance
rules with the metrics of Table~\ref{tab:governance}.

\section{Conclusion}
\label{sec:conclusion}

We asked how the signals of a reputation system should be chosen and combined so
that, in an open P2P network where identities are cheap, manipulation does not pay
while genuine contribution does. For RQ1, we stated six design constraints and showed
their interdependencies: rate limits and the asymmetry between gains and losses are
only effective if identities have a cost, rewards for assessments call for escrow,
and locality is required by impossibility results on Sybil-proofness. For RQ2, we
developed a combination-therapy theory of reputation signals. In a global linear
model, conjunctive aggregation leaves the attacker with two possibilities: either a
mix of gaming actions raises every signal by more than it costs, or the net gain of
gaming is bounded by the imbalance of the attacker's honest profile. Locally, gaming
brings no first-order gain exactly when an admissible weighting is supported on the
attacker's weakest signals. Conjunctive aggregation exploits admissible weightings
without knowing them, but their existence is a property of the signals, which must be
designed and tested. For RQ3, MagMell instantiates these principles as a two-tier
P2P system: a protocol that orders signed events (by a BFT committee of identifiable
institutions in the instantiation analyzed), enforces budgets and
executes a deterministic reference settlement, with guarantees that hold whatever the
reputation algorithm; and an open reputation layer with ten sub-indicators in which
merit is ranked and integrity is screened, costly identities, and escrowed, asymmetric
settlement of staked assessments against independent outcomes. The protocol is
specified but not implemented; its estimated costs are modest (an analytic estimate of
about 26~kB of log per identity and year), and a full reputation recomputation
takes 1.0 to 1.6~s for $10^5$ identities on synthetic data. The evaluation on synthetic data led to eight corrections of
the design. On the attacks modeled, minimum aggregation gave Sybil farms, rings and
floods a negative rank gain beyond small intensities, at the price of a lower
correlation between rank and quality for honest accounts; the geometric mean remains
a serious alternative; and the settlement of criticism, limited by the
availability of evidence in the network, is the main open problem: MagMell is not yet
a fully operational mechanism of post-publication evaluation, since its settlement is
convincing for endorsements but not for criticisms. The theory is the
main contribution of this paper; MagMell is one instantiation of it in a P2P
scientific network, and the evaluation delimits the conditions under which this
instantiation works and what it costs.

Future work will follow the protocol of Section~\ref{subsec:protocol} and
Appendix~\ref{app:ga}: in silico evolution of attacker strategies against each
aggregation rule, with success criteria fixed in advance; settlement of criticism
against evidence that does not depend on organic attention; an implementation of
the protocol, with measurements of confirmation latency, throughput and recovery under
faulty validators; replay on public data to
estimate the gaming matrix and the false-positive rate; an equilibrium analysis of the
assessment mechanism; and the cryptographic design of pseudonymous but accountable
assessments.

\section*{Statements}

\paragraph{Code and data availability.} All data analysed in this study are
synthetic. The scripts, parameters, seeds and outputs that reproduce every table of
Section~\ref{sec:numerics} are provided as ancillary files of this preprint
(directory \texttt{anc/}), with instructions and dependency versions (Python~3.13,
NumPy~2.5, SciPy~1.18); a single script reproduces every table in about twenty
minutes on one processor core.

\paragraph{Funding and competing interests.} No funding was received for conducting
this study. The author has no relevant financial or non-financial interests to
disclose.

\paragraph{Use of generative artificial intelligence.} The use of a large language
model is documented in Section~\ref{subsec:num-protocol}.

\begin{appendices}

\section{Methodology of the planned evolutionary simulation}
\label{app:ga}

This appendix specifies in advance the methodology of the first full simulation
(Section~\ref{subsec:protocol}), which has not been run: no result reported in this
paper comes from it. It details how attackers will be represented, evaluated and
evolved, and how the results will be protected against the two failures of any
search-based evaluation: an optimizer that converges too early, and a rule that looks
robust only because the search did not explore enough.

\subsection{Representation of strategies}

A strategy is a genome with five groups of genes.
\begin{itemize}
  \item \emph{Allocation}: the shares of the attackers' resources (reading and
        publication budgets, identities, free Rep available for stakes) given to
        five families of actions: Sybil farm, collusion ring, flood of generated
        papers, coordinated assessment, and honest activity. Honest activity is
        included because the theory treats it as the zero action: an attacker may mix
        gaming with genuine contribution. Shares are the softmax of real genes.
  \item \emph{Intensities}, per family: number of Sybil identities and their rate of
        recognitions; size of the ring, rate of reciprocation, share of recent
        recognitions, spread across communities; generated papers per quarter and
        their target communities; assessments per epoch, polarity mix and rule for
        choosing targets (random, top of the feed, coalition papers).
  \item \emph{Timing}: start, duration, and on/off periods, which allow oscillation
        attacks.
  \item \emph{Beneficiaries}: one account, one ally, or a set of coalition members.
  \item \emph{Onboarding path} of new identities: vouching, earned entry, or purchase
        of an existing identity.
\end{itemize}
Real genes are encoded on $[0,1]$ and mapped to their ranges, on a logarithmic scale
for intensities; categorical genes are integers.

\paragraph{Variation.} Parents are selected by tournaments of size 3, with an
elitism of 5\%. Uniform crossover, with probability $0.7$, exchanges whole groups of
genes, so that coherent sub-strategies are recombined rather than cut. Real genes
undergo Gaussian mutation with a self-adaptive step size carried by each individual
(log-normal update); categorical genes mutate with probability $1/L$, $L$ being the
number of genes.

\subsection{Fitness and conversion of costs}

The fitness of a strategy $s$ against a rule $r$ is
\[
  F_r(s)=\bar G_r(s)-\kappa_c\,K(s).
\]
$\bar G_r(s)$ is the mean, over the horizon and over evaluation seeds, of the rank
of the beneficiaries minus their rank in a counterfactual run with the same seed in
which the attackers' identities behave honestly (or, for identities created by the
attack, do not exist); common random numbers make the difference due to the strategy.
$K(s)$ collects the costs in natural units, each converted into rank points:
budgets diverted from honest activity are accounted for by the counterfactual itself,
which uses them honestly; burned stakes and lost Rep are converted at the marginal
rank value of Rep at the attacker's profile, estimated numerically; identity costs
(stakes forfeited by vouchers, probation time, the effort of earned entry) are
expressed in Rep equivalents and converted in the same way. Since how attackers value
these costs is unknown, $\kappa_c$ is a sensitivity parameter, $\kappa_c\in\{0, 0.5, 1,
2\}$, and results will be reported as functions of $\kappa_c$; $\kappa_c=0$ is the worst
case of an attacker for whom costs do not matter. Attackers whose objective is not
their own rank (promoting one paper, harming a competitor) are evaluated with the
corresponding objective in place of $\bar G_r$.

\paragraph{Noise.} During evolution, fitness is the mean over $E=3$ seeds. Because
selecting the best of many noisy evaluations overestimates its value (the winner's
curse), the final best strategies will be re-evaluated on 30 held-out seeds, and only
these re-evaluations will be reported.

\subsection{Avoiding premature convergence}

Four devices keep the search from settling early on a local optimum.
\begin{itemize}
  \item An island model: four islands of 50 individuals, with migration of the two
        best individuals along a ring every 10 generations.
  \item Fitness sharing by family: individuals whose allocations are close share
        their fitness, so that one family of attacks cannot take over the population
        in the first generations.
  \item Five independent restarts per rule, with different seeds, and a diverse
        initial population: 20\% best responses of Study~B, 20\% single-family
        strategies with random intensities, 60\% uniformly random genomes.
  \item A second optimizer of a different nature: the covariance matrix adaptation evolution strategy (CMA-ES) on the real genes, with the
        categorical genes fixed at their values in the best strategy of the genetic
        algorithm, from five starting points.
\end{itemize}

\subsection{Sufficiency of the search}

Evolution will run on a reduced simulation (2000 identities, a one-year horizon) in
which one evaluation should take about a second, by extrapolation from Study~C; the
best strategies will then be re-evaluated on the full simulation. With a population of
200, 100 generations, $E=3$, four rules and five restarts, a campaign would cost about
$1.2$ million reduced runs, a few hours on a 64-core machine. A run stops after 100 generations unless the best fitness improved by
more than $0.5$ rank point over the last 20, in which case it continues by steps of 50
generations, up to 300. Best-so-far curves will be reported for every run. As a check of
the budget, the minimum and the arithmetic mean are also evolved against with a
population four times larger and three times more generations; if the best fitness of
these runs exceeds that of the standard runs by more than 2 rank points, the standard
budget is declared insufficient and every campaign is repeated with the larger one.

\subsection{Robust rule or poorly explored search?}

A low best gain against a rule may mean that the rule is robust, or that the search
failed. Four checks separate the two.
\begin{enumerate}[label=(\roman*),leftmargin=2.2em]
  \item \emph{Positive controls.} The search must rediscover the gains that Study~B
        found where they exist: Sybil farms against the arithmetic mean (at least
        $+40$ rank points for the newcomer at $\kappa_c=0$), and every attack against
        merit only. If it does not, the search is declared inadequate, whatever it
        finds against the minimum.
  \item \emph{Equal budgets.} Every rule is evolved against with the same search
        budget, so that comparisons between rules do not depend on it.
  \item \emph{Local exhaustive checks.} Around the best strategy found against each
        rule, a $20\times20$ grid search is run on the two-dimensional slices spanned
        by the most influential pairs of genes.
  \item \emph{Cross-evaluation.} Strategies evolved against each rule are evaluated
        under every other rule. A rule is reported as robust only if no strategy
        evolved against any rule exceeds, under it, the threshold of criterion (a) of
        Section~\ref{subsec:protocol}.
\end{enumerate}
Even with these checks, results will be statements about the best attack found with
a given search budget: they are lower bounds on the capability of attackers, and will
be reported as such.

\end{appendices}

\bibliographystyle{plainnat}
\bibliography{biblio}

\end{document}